%% file: 20260713_threshold.tex
\documentclass[bj,authoryear,noshowframe]{imsart}

\RequirePackage{amsthm,amsmath,amsfonts,amssymb}
\RequirePackage[authoryear]{natbib}
\RequirePackage{graphicx}
\RequirePackage{multirow, array}

\usepackage{mathrsfs}
\usepackage{epsfig}
\usepackage{ifthen,latexsym,syntonly}
\usepackage{soul,color}
\usepackage{bm}
\usepackage{pstricks}
\usepackage{mdwlist}
\usepackage{verbatim}
\usepackage{enumitem}
\usepackage{subfigure}
\usepackage{float}
\usepackage{colortbl}
\usepackage{inputenc}
\usepackage{xcolor,lipsum}
\usepackage{algorithm}
\usepackage{algpseudocode}
\usepackage{lineno} 
\usepackage[toc,page,header]{appendix}
\usepackage{titletoc}

\startlocaldefs
\theoremstyle{plain}

\newtheorem{theorem}{Theorem}[section]
\newtheorem{lemma}[theorem]{Lemma}
\newtheorem{proposition}{Proposition}
\newtheorem{corollary}{Corollary}
\theoremstyle{definition}
\rmfamily
\newtheorem{Remark}{\sc Remark}
\newenvironment{remark}{\begin{Remark}\rm}{\hskip
1cm\hfill\quad$\Box$\end{Remark}}

\newtheorem{assumption}{Assumption}
\def\A{\mathcal A}

\def\F{\mathcal F}
\def\C{\mathcal C}

\def\Q{\mathcal Q}

\def\I{\mathrm{I}}

\def\ZZ{\mathbb{Z}}

\def\EE{\mathbb E}
\def\PP{\mathbb P}
\def\log{\mathrm {log}}
\def\RR{\mathbb{R}}
\def\d{\mathrm{d}}

\def\1{\textbf{1}}

\def\var{\mathrm{Var}}
\def\cov{\mathrm{Cov}}

\theoremstyle{definition}

\def\Var{\mbox{Var}}
\def\A{\mathcal A}

\def\F{\mathcal F}
\def\C{\mathcal C}

\def\I{\mathrm{I}}

\def\ZZ{\mathbb{Z}}
\def\EE{\mathbb E}
\def\PP{\mathbb P}
\def\log{\mathrm {log}}
\def\RR{\mathbb{R}}

\def\d{\mathrm{d}}

\def\1{\textbf{1}}

\def\var{\mathrm{Var}}
\def\cov{\mathrm{Cov}}

\def\E{\mathbf{E}}

\newcommand\DoToC{%
  \startcontents
  \printcontents{}{1}{\textbf{\Large Contents}\vskip3pt\hrule\vskip5pt}
  \vskip3pt\hrule\vskip5pt
}

\endlocaldefs

\begin{document}

\begin{frontmatter}
\title{High-dimensional inference on jumps in nonparametric time series regression models}
\runtitle{High-dimensional inference in nonparametric time series regression models}

\begin{aug}
\author[A]{\fnms{Likai}~\snm{Chen}\ead[label=e1]{likai.chen@wustl.edu}}
\author[B]{\fnms{Georg }~\snm{Keilbar}\ead[label=e2]{georg.keilbar@hu-berlin.de}}
\author[C]{\fnms{Liangjun}~\snm{Su}\ead[label=e3]{sulj@sem.tsinghua.edu.cn}}
\author[D]{\fnms{Weining}~\snm{Wang}\ead[label=e4]{weining.wang@bristol.ac.uk}}

\address[A]{Department of Statistics and Data Science,
Washington University in Saint Louis \printead[presep={ ,\ }]{e1}}

\address[B]{Chair of Statistics,
Humboldt-Universit\"at zu Berlin \printead[presep={,\ }]{e2}}

\address[C]{School of Economics and Management,
Tsinghua University \printead[presep={,\ }]{e3}}

\address[D]{Department of Economics,
University of Bristol \printead[presep={,\ }]{e4}}
\end{aug}

\begin{abstract}
We study simultaneous inference on jumps in the conditional
mean functions of a high-dimensional collection of heterogeneous
nonparametric time series, where the number of series may exceed the sample size and the data may exhibit strong cross-sectional dependence. The jump depends on one specific covariate, and we
allow the regression function to vary with additional latent variables. We
propose two uniform tests: one for the existence of jumps and one for their
homogeneity across series. We derive a
simple closed-form approximation to the covariance structure of the jump
estimators and establish a high-dimensional Gaussian approximation showing
that, owing to the localized construction of the statistics, the maximum of
the studentized jumps is approximated by the maximum of independent
Gaussians. The cross-sectional dependence is thus asymptotically negligible
for critical values, even under strong (e.g., factor) dependence, and the
approximation requires estimating only the variance for each series. For pronounced
cross-sectional dependence, a dependence-aware refinement restores the
off-diagonal covariances, improving finite-sample size and power.
Simulations show accurate size and reasonable power under both
cross-sectional and serial dependence, and two empirical applications
reveal significant non-smooth effects.
\end{abstract}



\end{frontmatter}


\section{Introduction}

  Recently, there has been a notable surge in the study and application of change-point analysis in high-dimensional time series data (see, e.g., \cite{cho2015multiple}, \cite{cao2018multi}, \cite{chen2022inference}, and \cite{li2024}, along with the references cited therein). Detecting change-points in regression models is important and has wide applications in the field of biology \citep{pastor1998use}, agricultural sciences \citep{freeman1998credit}, economics \citep{li2024}, etc. The purpose of these studies is to estimate not only where the threshold effect exists but also the magnitude of the effects.

{\normalsize 
\label{lin:test1}
There is extensive literature on parametric high-dimensional change-point regression analysis. \cite{li2016panel}, \cite{manner2019testing}, \cite{wang2021statistically}, \cite{rinaldo2021localizing}, and \cite{cho2024detection} investigate change-point analysis of coefficients in high-dimensional linear regression models using penalization methods. \cite{lee2016lasso} examine a high-dimensional regression model with a change-point due to a covariate threshold in a cross-sectional study. These studies mainly focus on parametric models with high-dimensional covariates. Similar to \cite{lee2016lasso}, this paper also considers change-point
regression due to a covariate threshold. However, our focus is on
nonparametric regression with a high-dimensional change-point effect driven
by cross-sectional heterogeneity across a large number of
series.
Specifically, we investigate a high-dimensional collection
of heterogeneous nonparametric time series with jumps in the conditional
mean functions:
\begin{align}
\label{modelstructural}
Y_{jt}=h_j(X_{jt},U_{jt})+\tau _{j}(X_{jt},U_{jt})\mathbf{1}_{\{X_{jt}\geq
c_{0j}\}}+\sigma _{j}(X_{jt},U_{jt})\epsilon _{jt},
\end{align}
where {$X_{jt}$ is a one-dimensional threshold variable,}  $Y_{jt}\in
\mathbb{R}$ and {$U_{jt}\in \mathbb{R}^d$ ($d\geq 1$ is an integer)} are the
outcome and a vector of additional covariates for unit
$j\in \lbrack N]:=\{1,\ldots ,N\}$ at time point $t\in \lbrack T],$
$h_{j}(\cdot,\cdot )$, $\tau_j(\cdot,\cdot)$ and $\sigma _{j}(\cdot,\cdot)$
are continuous functions, $c_{0j}$ is a threshold value that
is either known or unknown and $\epsilon_{jt}$ is the error term.
The additional variables $U_{jt}$ can be either observed or unobserved. However, our inferential procedure is identical in both cases,
since information about these variables is not used in the estimation.
Throughout the paper, we therefore refer to $U_{jt}$ as latent variables.
This specification is a natural generalization of the classical linear
threshold regression model (see Remark~\ref{comparison} in Section
\ref{sec:model} for details).
{Further, our model also extends nonparametric threshold models to understand threshold effects and break points, see e.g., \cite{porter2015regression}.}
Our model captures regime-switching behavior through a structural break at an unknown threshold location \( c_{0j}\), consistent with the core mechanism of threshold models widely used in  high-dimensional time series settings.

Our main focus is to test for the existence of change-point effects associated with $\tau_{j}(X_{jt},U_{jt})$. Specifically, we develop a uniform testing procedure to detect potential discontinuities in the conditional expectations of $Y_{jt}$ given $X_{jt}$. Furthermore, we examine whether the jump sizes are uniformly identical across units $j$ under general assumptions of spatial and temporal dependence. We provide Gaussian approximation (GA) results for both the known and unknown threshold cases.
For the unknown case, the unit-specific thresholds $c_{0j}$ are searched over a bounded grid; this boundedness is used in the union-bound step underlying our Gaussian approximation result and is therefore retained as a condition throughout.

The foundation of this model lies in the nonparametric literature on testing for jumps in regression models, as studied in \cite{muller1999discontinuous} and \cite{spokoiny1998estimation}. However, these approaches typically consider only a single-dimensional covariate. In practical applications, multivariate regressors are common, and the focus is often on testing jumps in a specific covariate, as illustrated in \cite{griffin2007investors}. While these additional covariates do not affect the presence of a jump, they can still influence the response, posing challenges for inference. For instance, in financial applications, the response variable could denote the stock volatility of firm $j$, the threshold variable might represent the lagged return, and the latent factors $U_{jt}$ could be unobserved risk factors or market sentiment (see Section \ref{sec:application} for details).
Similarly, in biostatistics, discontinuities may arise in dose-response relationships, where a sudden change in treatment effect can indicate a threshold beyond which efficacy or toxicity dramatically shifts. Here, latent patient-specific factors $U_{jt}$  such as genetic predisposition or underlying health conditions can influence observed responses.

Our main contributions are twofold. First, we establish a
novel whitening phenomenon in localization, which distinguishes our work from prior results
for MOSUM change-point analysis that mainly focus on temporal localization (e.g., \cite{chen2022inference} and \cite{li2024}). By leveraging the localization effect in the covariates, we generalize the \textit{whitening-by-window principle} introduced in \cite{hart1996some} to high-dimensional time series with very general forms of spatio-temporal dependence. The existing literature focuses exclusively on temporal dependence \citep{neumann1998strong, kreiss1998regression, fan2003nonlinear}, leaving the effect of cross-sectional dependence undeveloped. Recent work has shown that the maxima of many nonlocalized statistics admit Gaussian process approximations \citep{chernozhukov2013gaussian, zhang2017gaussian}; these approaches, however, do not incorporate spatial localization, so their limiting distributions inherit the full dependence structure of the data generating process and force practitioners to model spatial dependence in detail to obtain accurate critical values. In contrast, by extending the whitening principle to the spatial domain, we show that under mild regularity conditions the cross-sectional covariances between the localized statistics are of smaller order than their variances, so that the \emph{max-norm} of the studentized statistics is approximated at leading order by the maximum of i.i.d.\ Gaussian variables, even though the joint law of the full vector need not be close to i.i.d. This yields a simple framework for practical inference. When one wishes to exploit residual cross-sectional correlation to improve finite-sample performance, we further provide a dependence-aware version of the procedure that draws critical values from a correlated Gaussian maximum with an estimated, shrinkage-regularized correlation matrix.

Second, we introduce a high-dimensional GA result that incorporates latent variables. Applying the standard central limit theorem to high-dimensional time series is challenging in our high-dimensional time series data setup when the dimension $N$ significantly exceeds the effective sample size. Previous works, including \cite{chernozhukov2015comparison, MR3693963}, \cite{zhang2017gaussian}, and \cite{chernozhukov2019inference}, have addressed this issue by developing high-dimensional GA methods. However, these existing results are not directly applicable to our setting. The major challenge lies in accommodating non-identically distributed time series under general spatial and temporal dependence structures, as well as incorporating latent variables, which may arise in scenarios involving missing covariates or model misspecification. In this paper, we propose a novel GA result that resolves all these challenges. As a methodological by-product, the resulting tests for the existence and homogeneity of threshold effects are easy to implement and remarkably simple to calibrate: our GA limit requires only the estimation of the variance for each coordinate, with neither long-run covariance estimation nor resampling.

Our paper is related to three strands of literature. First, our approach can be viewed as an extension of the MOSUM change-point methodology, see for example \cite{eichinger2018mosum}, to the nonparametric regression case with latent covariates. The whitening phenomenon in our work arises from \emph{covariate localization}, in contrast to the temporal localization in the previous MOSUM literature, and it operates on data with strong cross-sectional dependence. More broadly, our work belongs to the nonparametric literature on detecting jumps in regression and trend functions \citep{muller1999discontinuous,muller1999multivariate,qiu1998discontinuous,spokoiny1998estimation}, which mostly focuses on independent and identically distributed settings; we generalize this line of work to heterogeneous, high-dimensional time series with general spatio-temporal dependence. Also related is the literature on inference for smooth nonparametric trends in time series \citep{chen2018testing,gao2024time,karmakar2022simultaneous,zhang2012inference}, which by contrast does not address discontinuities.

Second, our results contribute to the literature on high-dimensional Gaussian approximation. The maxima of high-dimensional sums and their bootstrap counterparts have been studied by \cite{chernozhukov2013gaussian}, \cite{chernozhukov2015comparison}, \cite{zhang2017gaussian} and \cite{chernozhukov2019inference}, among others. These results are not directly applicable in our setting, which features non-identically distributed and spatio-temporally dependent localized statistics together with latent variables; a central methodological contribution of our paper is a Gaussian approximation result tailored to this setting, in which the limiting covariance structure is asymptotically diagonal owing to the localization.

Third, our framework relates to threshold and homogeneity analysis for high-dimensional time series data, see e.g.\ \citet{massacci2022high}, \citet{barassi2023threshold} and \citet{su2018identifying}. This literature is largely parametric, whereas our nonparametric approach is robust to misspecification of the mean and jump functions; in particular, our test for homogeneity provides, to our knowledge, the first uniform test for the homogeneity of nonparametric jump effects across a high-dimensional time series.

 The rest of the paper is organized as follows. Section \ref{sec:model} provides the model setup and the practical steps of the testing
procedure. Section~\ref{sec:theorem} presents the assumptions and the
main theorems: Theorem~\ref{thm:ga} establishes the Gaussian approximation
for the existence test, Theorem~\ref{thm:homotesting} the analogue for the
homogeneity test, and Theorem~\ref{thm:unknown} the extension to unknown
threshold locations, with Theorem~\ref{consistency} giving consistency of
the estimated thresholds. In Section \ref{sec:simulation} we provide simulation results.\footnote{The code for our method is available in the R package \texttt{hdthreshold}, with an illustration accessible at \url{https://rpubs.com/Lk1110/1259698}.} The paper concludes in Section \ref{sec:conclusion}. The Supplementary Materials include additional theorems, proofs of all results, extra simulation outcomes, and two empirical applications.

 \textit{Notation.} For a vector $v=(v_{1},\ldots,v_{d})\in
\mathbb{R}^{d}$ and $q>0$, we denote $|v|_{q}=(%
\sum_{i=1}^{d}|v_{i}|^{q})^{1/q}$ and $|v|_{\infty }=\max_{1\leq i\leq
d}|v_{i}|$. For a matrix $A=(a_{i,j})_{1\leq i\leq m,1\leq j\leq n}$, we
define the max norm $|A|_{\text{max}}=\max_{i,j}|a_{i,j}|$. For $s>0$ and a
random vector $X$, we say $X\in \mathcal{L}^{s}$ if $\lVert X\rVert _{s}=[%
\mathbb{E}(|X|^{s})]^{1/s}<\infty $. For two sequences of positive numbers $(a_{n})$ and $(b_{n})$, we say $a_{n}=O(b_{n})$ or $a_{n}\lesssim b_{n}$
(resp. $a_{n}\asymp b_{n}$) if there exists $C>0$ such that $a_{n}/b_{n}\leq
C$ (resp. $1/C\leq a_{n}/b_{n}\leq C$) for all large $n$, and say $%
a_{n}=o(b_{n})$ if $a_{n}/b_{n}\rightarrow 0$ as $n\rightarrow \infty $. We
set $(X_{n})$ and $(Y_{n})$ to be two sequences of random variables. Write $%
X_{n}=O_{\mathbb{P}}(Y_{n})$ if for any $\epsilon >0$, there exists $C>0$
such that $\mathbb{P}(|X_{n}/Y_{n}|\leq C)>1-\epsilon $ for all large $n$, and
say $X_{n}=o_{\mathbb{P}}(Y_{n})$ if $X_{n}/Y_{n}\rightarrow 0$ in
probability as $n\rightarrow \infty $.
}

\section{Model Setup and Test Procedure}\label{sec:model}
In this section, we present the model, the hypotheses and
estimators, and the test procedure. Subsections \ref{subsec:model} and \ref{subsec:hypotheses} are concerned with the known threshold case. In Subsection \ref{unknown_c1} we focus on the case in which the threshold locations $c_{0j}$ are unknown. The testing procedures are described in Subsection \ref{eq:algo} and in the Supplementary Materials.

\subsection{Model Setup}
\label{subsec:model}
\label{modelsetup} In this subsection, we formulate our model.

Recall the full model in (\ref{modelstructural})
where $U_{jt} \in \mathbb{R}^{d}$ is a latent random vector that can be arbitrarily
correlated with $X_{jt},$ innovations $\epsilon _{jt}$ are mean zero terms that are
independent of $\left( X_{jt},U_{jt}\right) $.

Without loss of generality set $c_{0j} = 0$ for the known case.
The full model in \eqref{modelstructural} can be rewritten as the following reduced form model:
\begin{align} 
&Y_{jt}=\tilde{h}_{j}(X_{jt})+\tilde{\tau}_{j}(X_{jt})\mathbf{1}_{\{X_{jt}\geq c_{0j}\}}+e_{jt},
\label{mainmodel}
\end{align}
where $\tilde{\tau}_{j}(X_{jt})=\mathbb{E}[\tau _{j}(X_{jt},U_{jt})|X_{jt}]$ and  $\tilde{h}_{j}(X_{jt})=\mathbb{E}[h_{j}(X_{jt},U_{jt})|X_{jt}]$, and the reduced form error
\begin{align}
\label{eq:deffjvarepj} 
e_{jt}=\varepsilon _{j}(X_{jt},U_{jt})+\sigma _{j}(X_{jt},U_{jt})\epsilon
_{jt},
\end{align}
with
\begin{align*}
 \varepsilon _{j}(X_{jt},U_{jt})=h_{j}(X_{jt},U_{jt})-\tilde{h}%
_{j}(X_{jt})+[\tau _{j}(X_{jt},U_{jt})-\tilde{\tau}_{j}(X_{jt})]\mathbf{1}%
_{\{X_{jt}\geq c_{0j}\}}.   
\end{align*}
The noise term $e_{jt}$ has a complex structure, influenced by both the latent variable and by the heterogeneous volatility of the observations. This intricate structure, combined with the temporal and cross-sectional dependencies of the noise, makes the inference process particularly challenging. In our application on modeling stock volatility (Section~\ref{sec:application} in the supplement), $X_{jt}$ is the lagged stock return. Functions $h_{j}(\cdot ,\cdot
) $, {$\tau _{j}(\cdot ,\cdot )$} and $\sigma _{j}(\cdot ,\cdot )$ represent baseline effect, the
jump effect and the volatility function, respectively. 
The $U_{jt}$ are unobserved risk factors or other variables.
Allowing for latent variables $U_{jt}$ is important in this circumstance as the volatility
of a single stock is most likely affected by other risk factors such as news events or unobserved market factors.
{Note that under this reduced model, we have $\mathbb{E}(e_{jt}| X_{jt})=0$ and 
the conditional expectation of the observed outcome given $X_{jt}$ is $%
\mathbb{E}(Y_{jt}|X_{jt})=\tilde{h}_{j}(X_{jt})+\tilde{\tau}_{j}(X_{jt})%
\mathbf{1}_{\{X_{jt}\geq c_{0j}\}}.$ 
In Section~\ref{addtionalcovariate} in the Supplementary Materials, we discuss possible extensions of our model to add more covariates and incorporating fixed effects.}

\begin{remark}\label{comparison}
Consider the following linear threshold model, originating from the threshold autoregressive framework of \cite{tong1978threshold} and further developed by \cite{hansen2000sample}, and \cite{barassi2023threshold} for panel threshold models,
\begin{equation}
\label{eq:yjtlinearform}
Y_{jt}=V_{jt}^{\top}\theta_{j,1}+V_{jt}^{\top}\theta_{j,2}\mathbf{1}_{\{X_{jt}\geq
c_{0j}\}}+e _{jt},
\end{equation}
where $V_{jt}$ is a vector of $d$ regressors, $X_{jt}$ is a threshold variable, and $\theta_{j,1}$ and $\theta_{j,2}$ are parameters, and $e _{jt}$s are the error terms.
For each \( l = 1, \ldots, d \), note that
$V_{jt,l} = \mathbb{E}(V_{jt,l} \mid X_{jt}) + \left( V_{jt,l} - \mathbb{E}(V_{jt,l} \mid X_{jt}) \right)$.
Define the conditional mean function:
$f_l(X_{jt}) = \mathbb{E}(V_{jt,l} \mid X_{jt})$
and $f(X_{jt})=(f_1(X_{jt}),\ldots,f_d(X_{jt}))^\top.$
Let $U_{jt}=V_{jt}-f(X_{jt}).$
Then, we can rewrite the linear model \eqref{eq:yjtlinearform} as model \eqref{modelstructural} with
$h_j(X_{jt}, U_{jt}) = \left( f(X_{jt}) + U_{jt} \right)^{\top} \theta_{j,1}$ and $
\tau_j(X_{jt}, U_{jt}) = \left( f(X_{jt}) + U_{jt} \right)^{\top} \theta_{j,2}$.
For the more general nonlinear model,
$Y_{jt} = h_{j}(V_{jt}) + \tau_{j}(V_{jt}) \, \mathbf{1}_{\{X_{jt} \geq c_{0j}\}} + e_{jt}$,
a similar argument applies, and the model can still be rewritten within our framework.

An important special case is the self-exciting threshold autoregressive model of order one (SETAR(1)):
\begin{align*}
Y_{jt} = \left(1, Y_{j,t-1}\right)^\top \beta + \left(1, Y_{j,t-1}\right)^\top \delta \mathbf{1}_{\{Y_{j,t-1} \geq c_{0j}\}} + e_{jt},
\end{align*}
where the regime shift is triggered by the lagged dependent variable crossing a unit-specific threshold $c_{0j}$. This model can be extended to a multivariate setting as threshold vector autoregressions (TVAR), as in \cite{tsay1998testing}, and is related to recent developments in panel threshold VARs with stochastic volatility-in-mean, enabling regime-dependent volatility and dynamic interactions across multiple units (\cite{soave2023panel}). In this formulation, both $\tilde{h}_j(\cdot)$ and $\tilde{\tau}_j(\cdot)$ in the reduced form \eqref{mainmodel} are linear, with $X_{jt} = Y_{j,t-1}$ as the threshold variable.
\end{remark}

This paper aims to explore methods for conducting simultaneous inference on threshold effect estimators in a large $N$ and large $T$
setup. In particular, we allow $N\gg T.$ 
{\normalsize Assuming that $X_{jt}$ is a continuous random variable, the
jump effects can be identified as a \textquotedblleft gap" in the
conditional expectations and 
their derivatives at the threshold. Let $\mu _{j}(x):=\mathbb{E}(Y_{jt}|X_{jt}=x)$ and $\mu
_{j}(c_{0j}-)=\lim_{x\uparrow c_{0j}}\mu _{j}(x)$ and $\mu
_{j}(c_{0j}+)=\lim_{x\downarrow c_{0j}}\mu _{j}(x).$ Consider $\partial
_{+}\mu _{j}(\cdot )$ (resp. $\partial _{-}\mu _{j}(\cdot )$) as the right
(resp. left) derivative of function $\mu _{j}(\cdot ).$  The threshold effect 
for the individual $j$ is defined as
\begin{eqnarray}
\gamma _{j}&=&\mu
_{j}(c_{0j}+)-%
\mu
_{j}(c_{0j}-),\label{gammaj1}\\
\gamma _{j}^{[1]}&=&\partial _{+}\mu _{j}(c_{0j} )-
\partial _{-}\mu _{j}(c_{0j}). \label{gammaj}
\end{eqnarray}
}  
By continuity of $h_j(\cdot,\cdot)$ and $\tau_j(\cdot,\cdot)$, we have
\[
\gamma_j=\tilde\tau_j(c_{0j}).
\]
A simple expression for $\gamma_j^{[1]}$ requires additional differentiability assumptions, so we treat the derivative case separately in the Supplementary Materials.
Our focus is on the threshold effect, starting with the case where the threshold value $c_{0j}$
is known and then extending the analysis to the unknown case in Subsections \ref%
{unknown_c1} and \ref{unknown_c2}.

\subsection{\protect\normalsize Hypotheses and Estimators}\label{subsec:hypotheses}
In this section, we introduce the hypotheses and estimators. We
would like to conduct simultaneous inference on $\gamma _{j}$ and $\gamma_j^{[1]}$ for $%
j\in \left[ N\right] $. The importance of studying individual/group-specific
threshold effects is discussed in \cite{zimmert2019nonparametric}.

{\normalsize First, we are interested in testing the existence of
the threshold effects ($\gamma_j$ or $\gamma_j^{[1]}$). 
The null and alternative hypotheses are defined as follows:
\begin{eqnarray}
&&H_{0}^{\left( 1\right) }:\gamma_{j} =0\ \forall \text{ }j\in \left[ N\right]
,\quad \text{and }H_{a}^{\left( 1\right) }:\gamma _{j}\neq 0\ \text{for some 
}j\in \left[ N\right] .  \label{null1}
\end{eqnarray}

Such a test is well motivated. In Figure \ref{figure:comparison}, we present a simple comparison with the pooled threshold test in the literature, for example in \cite{fong2017chngpt}. It is clear that the conventional pooled tests may exhibit very low power due to possible signal cancellation, with performance only comparable in cases of strictly positive signals.

\begin{figure}[tbp]
\centering
\includegraphics[width=.49\linewidth]{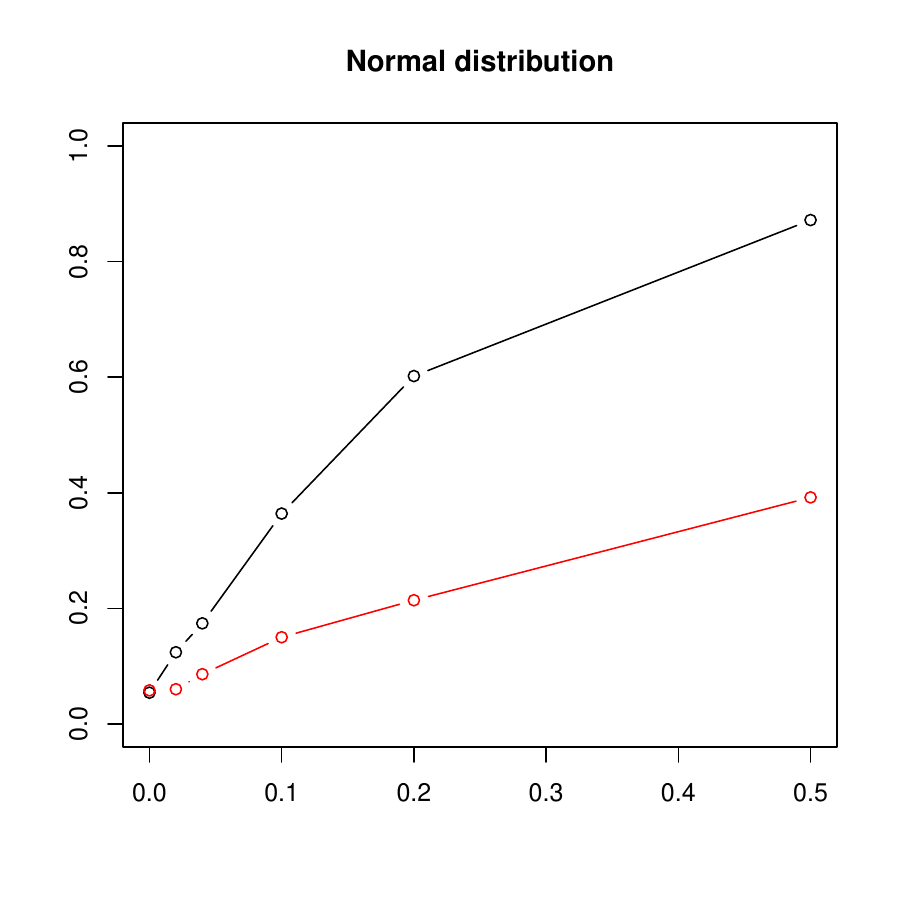}\hfill
\includegraphics[width=.49\linewidth]{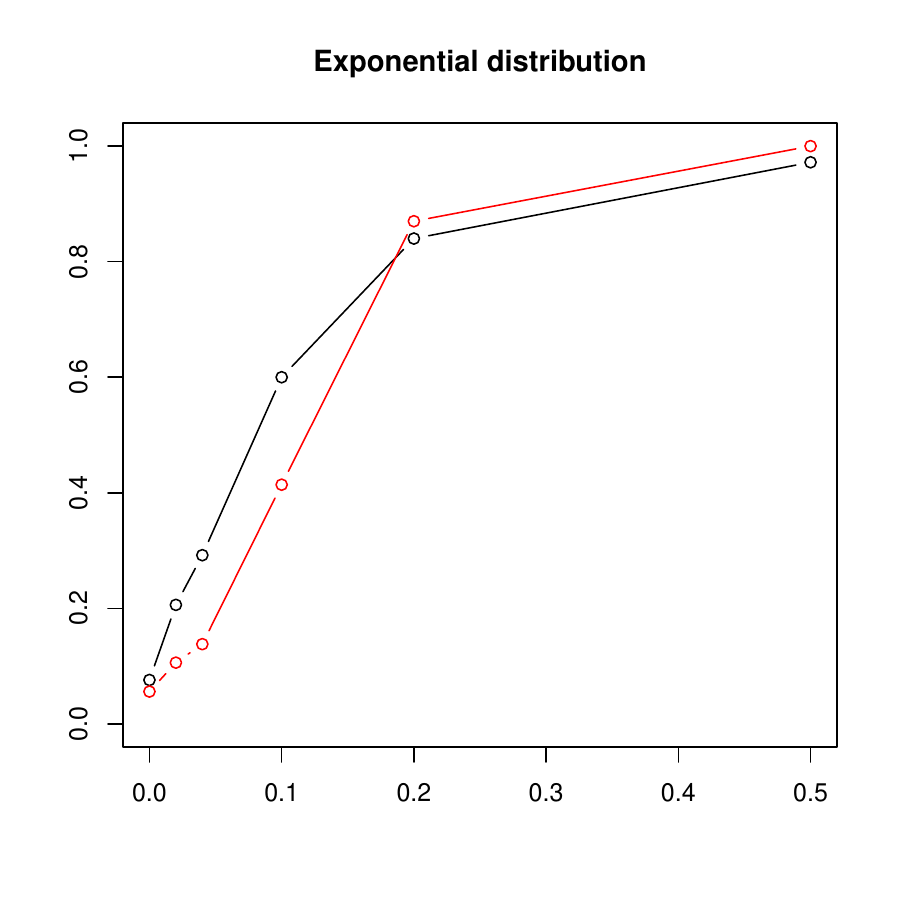}
\caption{Empirical power comparison between our uniform testing procedure
(black) for the existence of threshold effects and the test based on the pooled linear threshold model (red) at $5\%$ significance level for different proportions of non-zero coefficients ($\gamma_j$) on the x-axis. The non-zero $\gamma_j$s are drawn from a standard normal distribution (left panel) and an exponential distribution (right panel). The true process is linear,  $\EE(Y_{jt}|X_{jt}=x)=0.2x+\mathbf{1}_{\{x\geq0\}}\gamma_j$ (for details see DGP7 in Section \ref{sec:simulation_appendix} of the Supplementary Materials). }
\label{figure:comparison}
\end{figure}

}

When $N$ is large, assuming homogeneous threshold effects 
$\gamma_j$ across all $j$ is restrictive, and inferences based on this assumption can be misleading if the threshold effects are heterogeneous.  Therefore, it is important to test for heterogeneous threshold effects. In this case, the null and alternative hypotheses are:
\begin{eqnarray*}
&&H_{0}^{\left( 2\right) }:\gamma _{j}=\gamma _{0}\ \text{for some }\gamma _{0},
\text{ }\forall \text{ }j\in \left[ N\right],\quad
\\
\text{and}\quad&&
H_{a}^{\left( 2\right) }:\text{there is no constant }
\gamma_0 \text{ such that }\gamma
_{j}=\gamma_0, \text{ }\forall \ j\in \left[ N\right] .  \label{test}
\end{eqnarray*}
{It should be noted that the above hypothesis can be extended to other forms, such as testing for specific parametric structures, partial homogeneity across subgroups, or functional relationships in $\gamma_j$.
}

We can similarly define $H_0^{(1)[1]}$ and $H_0^{(2)[1]}$ for the existence and homogeneity tests of the derivative, respectively. Rejecting $H_{0}^{\left( 2\right) }$ suggests the
presence of heterogeneous threshold effects. Moreover, providing simultaneous confidence intervals for the threshold
effects is also tempting. If we want to study the asymptotics of $\hat{%
\gamma}=(\hat{\gamma}_{1},\ldots,\hat{\gamma}_{N})^{\top }$, the standard
central limit theorem may fail due to the high dimensionality $N$. One of the key theoretical contributions of this paper
is to provide a general framework for making uniform inferences on
heterogeneous threshold effects.

{\normalsize

To estimate $\gamma
_{j} $, we need to estimate both $\mu _{j}(c_{0j}+)$ and $\mu _{j}(c_{0j}-).$
To this end, it is natural to adopt local linear estimators, {see for example  \cite{fan2018local}}. Denote the
estimators: 
\begin{align}
\big(\hat{\mu}_{j}(c_{0j}+),{\partial _{+}\hat{\mu}_{j}}(c_{0j})\big)=& 
\mathrm{argmin}_{\beta _{j,0},\beta _{j,1}}\sum_{t=1}^{T}(Y_{jt}-\beta
_{j,0}-\beta _{j,1}X_{jt})^{2}K_{jt}\mathbf{1}_{\{X_{jt}\geq c_{0j}\}}, \\
\big(\hat{\mu}_{j}(c_{0j}-),{\partial _{-}\hat{\mu}_{j}}(c_{0j})\big)=& 
\mathrm{argmin}_{\beta _{j,0},\beta _{j,1}}\sum_{t=1}^{T}(Y_{jt}-\beta
_{j,0}-\beta _{j,1}X_{jt})^{2}K_{jt}\mathbf{1}_{\{X_{jt}<c_{0j}\}},
\end{align}
where $K_{jt}:=K((X_{jt}-c_{0j})/b_{j}),${\normalsize \ $b_{j}$ is a
bandwidth parameter, and $K(\cdot )$ is a kernel function. 
For $l=0,1,2$,
define the weights as%
\begin{align}
w_{j,b}^{+}(u,v)& =\frac{K((v-u)/b_{j})[S_{j2,b}^{+}(u)-(v-u)S_{j1,b}^{+}(u)]%
}{S_{j2,b}^{+}(u)S_{j0,b}^{+}(u)-S_{j1,b}^{+2}(u)}\mathbf{1}_{\{v\geq u\}},  \label{eq:wjbuv} \\
w_{j,b}^{+[1]}(u,v)& =\frac{K((v-u)/b_{j})[S_{j1,b}^{+}(u)-(v-u)S_{j0,b}^{+}(u)]%
}{-S_{j2,b}^{+}(u)S_{j0,b}^{+}(u)+S_{j1,b}^{+2}(u)}\mathbf{1}_{\{v\geq u\}},
\label{eq:wjbuv[1]}
\end{align}
where
\begin{align}
S_{jl,b}^{+}(u)& =\sum_{t=1}^{T}(X_{jt}-u)^{l}K((X_{jt}-u)/b_{j})\mathbf{1}%
_{\{X_{jt}\geq u\}},
\end{align}
and $w_{j,b}^{-}(u,v)$ (respectively, $S_{jl,b}^{-}(u)$, $w_{j,b}^{-[1]}(u,v)$) is the same as $%
w_{j,b}^{+}(u,v)$ (respectively, $S_{jl,b}^{+}(u)$, $w_{j,b}^{+[1]}(u,v)$) with $+$ and $\geq $ therein
replaced by $-$ and $<$. 

We can thus estimate the one-sided conditional mean and derivative at the cut-off point and define
\begin{align}
\hat{\mu}_{j}(c_{0j}+)
=\sum_{t=1}^{T}w_{jt,b}^{+}Y_{jt}
\quad\textrm{and}\quad {\partial _{+}\hat{\mu}_{j}}(c_{0j})=\sum_{t=1}^{T}w_{jt,b}^{+[1]}Y_{jt},
\end{align}
where the weights take the following form $%
w_{jt,b}^{+}=w_{j,b}^{+}(c_{0j},X_{jt})$ and $%
w_{jt,b}^{-}=w_{j,b}^{-}(c_{0j},X_{jt})$.  

Accordingly, the local linear estimators of the threshold effect and derivative jump are
\begin{eqnarray}
&\hat{\gamma}_{j}=
\sum_{t=1}^{T}w_{jt,b}^{+}Y_{jt}-\sum_{t=1}^{T}w_{jt,b}^{-}Y_{jt} \quad \mbox{and} \quad\hat{\gamma}_{j}^{[1]}=%
\sum_{t=1}^{T}w_{jt,b}^{+[1]}Y_{jt}-\sum_{t=1}^{T}w_{jt,b}^{-[1]}Y_{jt}.
\label{estimate}
\end{eqnarray}
We shall focus on $\hat{\gamma}_{j}$ from now on. The algorithms and theorems concerning the statistical properties of $\hat{\gamma}_{j}^{[1]}$ are developed in Section~\ref{derivatives} in the Supplementary Materials, where we also consider a stacked uniform testing procedure designed to detect jumps in both the level and the slope of the regression function.

Under $H_{0}^{\left( 1\right) }:$ $\gamma _{j}=0$ for all $j\in %
\left[ N\right] $, we are exclusively testing for the overall significance
of threshold effects. Since the functions $\tilde{h}_{j}(\cdot)$ and $\tilde{\tau}_{j}(\cdot)$ are continuous, by \eqref{mainmodel}, the difference between our estimator and the
parameter can be approximately expressed in terms of a weighted average of the
error sequence, 
\begin{equation}
\hat{\gamma}_{j}\approx \gamma
_{j}+\sum_{t=1}^{T}(w_{jt,b}^{+}e_{jt}-w_{jt,b}^{-}e_{jt}),
\label{eq:hatgammajgammaj}
\end{equation}
where $e_{jt}$ is the composite error in (\ref{mainmodel}%
). Under $H_{0}^{\left( 1\right) }$, one would expect $\left\vert \hat{%
\gamma}_{j}\right\vert $ to be small for all $j\in \left[ N\right] $. {This
motivates us to consider the test statistic 
\begin{equation}
\mathcal{I}=\max_{1\leq j\leq N}(Tb_{j})^{1/2}|\hat{\gamma}_{j}/v_{j}|,
\label{I1}
\end{equation}%
where 
\begin{equation}
v_{j}^{2}=v_{j}^2(c_{0j})=(Tb_{j})%
\sum_{t=1}^{T}(w_{jt,b}^{+}-w_{jt,b}^{-})^{2}\mathrm{Var}\big(e_{jt}|\mathcal F_T\big)
\label{eq:def_sigma}
\end{equation}%
}is an approximation of the conditional variance of $(Tb_{j})^{1/2}\hat{\gamma}_{j}$ for
standardization purposes. For now, we assume $v_{j}^{2}$ is given. In
general, it is unknown and has to be replaced by its estimate. We discuss a
consistent estimator of $v_{j}^{2}$ in \eqref{eq:hatvj2} in Subsection \ref{eq:algo}. It can be shown that all
asymptotic results based on $v_{j}^{2}$ remain valid when it is replaced by
the consistent estimate $\hat{v}_{j}^{2}$, see Proposition \ref{thm:hatv} and the discussion afterwards for details.
{It is worth noting that using the above maximum-type statistic is particularly efficient when the signal is sparse. In contrast, for dense signals, it may be advantageous to further employ an additional aggregation in an $\ell_2$-norm fashion, as suggested by \cite{chen2022inference}.
}

Similarly, under $H_{0}^{\left( 2\right) }$, one would expect 
$\big|\hat{\gamma}_{j}-\overline{\hat{\gamma}}\big|$ to be
small for all $j\in \left[ N\right] ,$ where $\overline{\hat{\gamma}}=\frac{1%
}{N}\sum_{j=1}^{N}\hat{\gamma}_{j}$. This motivates us to consider the test
statistic 
\begin{equation}
\mathcal{Q}=\max_{1\leq j\leq N}(Tb_{j})^{1/2}|\hat{\gamma}_{j}-\overline{%
\hat{\gamma}}|/\tilde{v}_j,\quad
\textrm{where}\ \tilde{v}_{j}^{2}=(1-1/N)^{2}v_{j}^{2}+\sum_{i\neq
j}^{N}v_{i}^{2}/N^{2}.
\label{Q1}
\end{equation}
For practical implementation, we introduce a feasible version
of $\mathcal{Q}$ in Algorithm \ref{alg2} in Section \ref{sec:algo} of the Supplementary Materials.

\subsection{{\protect\normalsize The Unknown Threshold Case \label{unknown_c1}}}
When the threshold locations are unknown, we generalize our methodology by searching
over a grid $[c_1, c_2, \ldots, c_K]$ for each individual $j \in [N]$. Let
\begin{align}
\label{eq:wjtbci}
w_{jt,b}(c_{i})=w_{j,b}^{+}(c_{i},X_{jt})-w_{j,b}^{-}(c_{i},X_{jt}) \quad \mbox{and}\quad
\hat{\gamma}_{j}(c_{i})=\sum_{t=1}^{T}w_{jt,b}(c_{i})Y_{jt}.
\end{align}
Further, let $v_{j}^{2}(c_{i})=(Tb_{j})\sum_{t=1}^{T}w_{jt,b}^{2}(c_{i})\mathrm{Var}%
\big(e_{jt}|X_{jt}\big)$ and define 
\begin{equation}
\mathcal{I}^{C}=\max_{1\leq i\leq K}\max_{1\leq j\leq N}(Tb_{j})^{1/2}|\hat{%
\gamma}_{j}(c_{i})/v_{j}(c_{i})|.
\end{equation}
We propose to estimate $c_{0j}$ by $\hat{c}_{j}:=\arg \max_{1\leq i\leq
K}(Tb_{j})^{1/2}|\hat{\gamma}_{j}(c_{i})/\hat{v}_{j}(c_{i})|\ $for $j\in %
\left[ N\right] ,${\normalsize \ where }$\hat{v}_{j}(c_{i})${\normalsize \
is a consistent estimator of }$v_{j}(c_{i})$ which will be introduced in Algorithm %
 \ref{algunknown} in Section \ref{sec:algo} of the Supplementary Materials.

\subsection{{\protect\normalsize Test Procedure \label{eq:algo} }}

In this subsection, we present the steps of the test procedures. To
proceed, we begin by discussing the estimation of the error variance $\sigma
_{e,j}^{2}$ of $e_{jt}$ when $c_{0j}$'s are known.

\begin{enumerate}
\item Recall the definition of $\hat{\gamma}_j$ as in Equation \eqref{estimate}. Regress $Y_{jt}-\hat{\gamma}_{j}\mathbf{1}_{\{X_{jt}\geq c_{0j}\}}$ on $X_{jt}$ using the
local linear method with kernel $K\left( \cdot \right) $ and bandwidth $%
b_{j} $. 

\item Calculate the regression residuals $\hat{e}_{jt}$. 

\item  Estimate $\sigma_{e,j}^{2}$ by
\begin{align}
\label{eq:sigmaej2}
\hat{\sigma }_{e,j}^{2}=\frac{\sum_{t=1}^{T}%
\mathbf{1}_{\{|X_{jt}-c_{0j}|\leq b_{j}\}}\hat{e}_{jt}^{2}}{\sum_{t=1}^{T}%
\mathbf{1}_{\{|X_{jt}-c_{0j}|\leq b_{j}\}}} .   
\end{align}  
\end{enumerate}
The estimator $\hat{\sigma}_{e,j}^2$ is used to form $\hat{v}_{j}^{2}$ in the Algorithms \ref%
{alg1} and \ref{alg2} in the case of known $c_{0j}$, which are
applied to conduct simultaneous tests for $H_{0}^{\left( 1\right) }$ and $%
H_{0}^{\left( 2\right) }$. The algorithms are supported by
the theoretical GA results in Section \ref{sec:theorem}.
The following Algorithm \ref{alg1} is for testing $H_0^{(1)}$ and the counterpart for the derivative case is presented in the parentheses.
\begin{algorithm}
	\caption{Testing for existence of jumps in conditional means (in derivative): i.e., $\gamma_j =0$ {( $\gamma_j^{[1]} =0$}) for all $j$.}\label{alg1}
	\begin{algorithmic}[1]
		\For {$j=1,\ldots,N$}
			\State Estimate $\gamma_j$ for each group/individual $j$ using equation \eqref{estimate}.
			\State Estimate the variance by 
            \begin{align}
            \label{eq:hatvj2}
			\hat{v}_j^2=Tb_j\sum_{t=1}^{T}\left(w_{jt,b}^{+}-w_{jt,b}^-\right)^2\hat{\sigma}_{e,j}^2,\quad
            \Big(\hat{v}_j^{2[1]}=Tb_j^3\sum_{t=1}^{T}\left(w_{jt,b}^{+[1]}-w_{jt,b}^{-[1]}\right)^2\hat{\sigma}_{e,j}^2\Big) 
			\end{align} 
            where $\hat{\sigma}_{e,j}^2$ is defined in \eqref{eq:sigmaej2}.
		\EndFor
    \State Calculate the feasible test statistic $\hat{\mathcal{I}}:=\mbox{max}_{1\leq j\leq N}(Tb_j)^{1/2}|\hat{\gamma}_j|/\hat{v}_j$ 
    {( $\hat{\mathcal{I}}^{[1]}:=\max_{1\leq j\leq N}(Tb_j^3)^{1/2}|\hat{\gamma}_j^{[1]}|/\hat{v}_j^{[1]}$})
    .
    \State Reject the null hypothesis if $\hat{\mathcal{I}}>q_{\alpha}$ ( $\hat{\mathcal{I}}^{[1]}>q_{\alpha}$), where $q_{\alpha}$ is the $(1-\alpha)$ quantile of $\max_{1\leq j\leq N}|Z_j|$ with $Z_j$'s being independent standard normal variables.
	\end{algorithmic} 
\end{algorithm}

Furthermore, Algorithm \ref{alg2} in Section \ref{sec:algo} in the Supplementary Materials is used for testing the homogeneity hypothesis $H_0^{(2)}$. For the case where the threshold is unknown, the corresponding algorithm is detailed in Algorithm \ref{algunknown} in the same section.   
Note that $\mathcal{\hat{I}}$ in Algorithm \ref{alg1} (resp. $\mathcal{\hat{Q}}$ in Algorithm \ref{alg2}, $\mathcal{\hat{I}}^{C}$ in Algorithm {\ref{algunknown}}) is a feasible version of $\mathcal{I}$ (resp. $\mathcal{Q}$, $\mathcal{{I}}^{C}$). In Section \ref{sec:theorem}, we will study the asymptotic properties of 
$\mathcal{\hat{I}}$, $\mathcal{\hat{Q}}$ and $
\mathcal{\hat{I}}^{C}.$ 

It is surprising that, according to the above algorithm, $q_{\alpha}$ can be
obtained by simulating only i.i.d.\ Gaussian random variables, even though the
reduced-form error $e_{jt}$ in \eqref{eq:deffjvarepj} inherits a nontrivial,
possibly strong (e.g., factor) cross-sectional dependence structure. This is a
consequence of the localized nature of the estimators $\hat{\gamma}_{j}$, which
renders the off-diagonal covariance terms negligible relative to the variances.
We develop this intuition in Section \ref{intuition} and state the underlying
condition in Assumption \ref{asmp:bddxj1j2density}. For a discussion on the issue of bandwidth selection, we refer to
Remark \ref{remark_bandwith} in Section \ref{sec:theorem}. We additionally consider a modified testing procedure that takes finite-sample correlation into account in the case of very strong cross-sectional dependence. See Remark \ref{rmk:rmkvj1j2} in Section~\ref{sec:theorem} and
Algorithm \ref{algdepaware} in Section \ref{sec:algo} of the Supplementary
Materials.

\section{{\protect\normalsize Main Theorems \label{sec:theorem}}}

In this section, we present our main results, with a particular focus on Theorem \ref{thm:ga}, which provides a Gaussian approximation result that determines the critical values for testing the existence of threshold effects. Theorem \ref{thm:homotesting} shows GA for testing homogeneity. 
Theorem \ref{thm:unknown} extends the result to the unknown threshold case and Theorem \ref{consistency} shows the consistency of the estimated threshold locations. Propositions \ref{thm:hatv} and \ref{thm:hatvap} derive the consistency of variance estimators to ensure the validity of our test statistics. Corollary \ref{thm:power} analyzes the power of the tests for both the threshold effects and homogeneity. Results regarding the derivative case are presented in Section \ref{derivatives} of the Supplementary Materials, where we also analyze the behavior of a stacked testing procedure for detecting jumps in either the level or slope.

\subsection{\protect\normalsize Assumptions}
\label{assum}
In this subsection, we state the assumptions used in our asymptotic analysis.  First of all, we pose a general spatial and
temporal dependence assumption on the data generating processes. As for the threshold variable $X_{jt}$ and the
latent variables $U_{jt} \in \mathbb{R}^d$, we assume that they take the form 
\begin{equation}
X_{jt}=H_{j}(\iota_{t}, \iota_{t-1},\ldots )\quad \text{and}\quad U_{jt}=\tilde{H}%
_{j}(X_{jt},\tilde{\iota}_{t}), \label{eq:xjtujtdef}
\end{equation}%
where 
$\tilde{H}_{j}=(\tilde{H}_{j1},\ldots, \tilde{H}_{jd})^\top$,  $\{\tilde{H}_{ji}\}_{j\in \lbrack N], i\in \lbrack d]}$ and
 $\left\{ H_{j}\right\} _{j\in \lbrack N]}$ are functions
so that $X_{jt}$ and $U_{jt}$ are well defined. Here both $\{\iota_{t}\}_{t\in\ZZ}$ and $\{\tilde{\iota}_{t}\}_{t\in\ZZ}$ are sequences of i.i.d.\ innovations, and they are mutually independent.
We emphasize that $\iota_{t}$ may be a vector of arbitrary dimension whose components are i.i.d.\ over $t$ but may have any contemporaneous cross-component dependence; through the unit-specific filters $H_{j}$, this allows the cross-sectional dependence in $\{X_{jt}\}_{j\in[N]}$ to range from full independence (when each $H_{j}$ depends on a distinct independent component of $\iota_{t}$) to strong dependence (when the $H_{j}$ share common components), so that no restriction on the cross-sectional dependence structure is imposed at this stage.

Denote 
\begin{align}
\label{eq:FttildeFt}
\mathcal{F}_{t}=(\iota_{t}, \iota_{t-1},\ldots )\quad \textrm{and}\quad \tilde{\mathcal{F}}_{t}=(\tilde{\iota}_{t},\tilde{\iota}_{t-1},\ldots
).
\end{align}

\begin{remark}(Dependence between $X_{jt}$ and $U_{jt}$)
 Note that Equation (\ref{eq:xjtujtdef}) is a fairly general setup, where
almost all kinds of dependence structures between $X_{jt}$ and $U_{jt}$ can
be fulfilled. 
To see this, consider $U_{jt}$ as a univariate random variable for simplicity in this illustration.
For any two continuous
random variables $X$ and $U$ with the conditional cumulative distribution
function of $U$ given $X$ as $F_{U|X}(\cdot)$. Let $\tilde \iota=F_{U|X}(U)$. Then by Lemma F.1 in Appendix F in \cite{rio2017asymptotic}, $\tilde\iota$ is a uniformly distributed random variable on $[0,1]$ and is
independent of $X$. Moreover, $U=F_{U|X}^{-1}(\tilde\iota).$ 
\end{remark}

{
For the innovation part $\epsilon _{t}=(\epsilon_{1t},\ldots,\epsilon_{Nt})^\top$ as in (\ref{modelstructural}), the dependence
is allowed to be strong (e.g., with a factor structure) along the cross-sectional dimension and it has to be weak along the time
dimension, as our procedure aggregates over time while relying only on maxima across the cross section for estimating the threshold effects.} Specifically, we
assume that $\epsilon _{t}$ follows an MA($\infty $) process as follows: 
\begin{equation}
\epsilon _{t}=\sum_{k\geq 0}A_{k}\eta _{t-k},
\end{equation}%
where $\eta _{t}=(\eta _{1t},\ldots ,\eta _{\tilde{N}t})^{\top }\in \mathbb{R%
}^{\tilde{N}},$ $t\in \mathbb{Z},$ are i.i.d.\ random vectors with zero mean
and identity covariance matrix $I_{\tilde{N}}$, and $A_{k}\in \mathbb{R}%
^{N\times \tilde{N}}$ are real-valued matrices with $\tilde{N}\leq cN$ for
some constant $c>0$. In particular, we allow $\tilde{N}=N$ so that the $A_{k}$'s
become square matrices. We assume that $\left\{ \iota_{t},\tilde{\iota}
_{t}\right\} _{t\in \mathbb{Z}}$ are independent of $\left\{ \eta
_{t}\right\} _{t\in \mathbb{Z}}.$
Next, we impose some conditions on the elements of $\eta _{t}.$ 

\begin{assumption}
{\normalsize (Moment) \label{asmp:moment} $\eta _{jt}$ are i.i.d.\ with
zero mean and unit variance for $j\in[N]$ and $t\in[T]$. One of the following
two conditions is satisfied: }

\begin{enumerate}
\item[(i)] {\normalsize (Finite moments) For some constant $q>2,$ $\eta
_{11} $ has finite $q$th moment. }

\item[(ii)] {\normalsize (Sub-exponential) For some constant $\lambda _{0}>0$%
, $\eta _{1}=(\eta _{11},\ldots ,\eta _{\tilde{N}1})^{\top }$ is
sub-exponential with $a_{0}:=\sup_{|v|_{2}\leq \lambda _{0}}\mathbb{E}%
(e^{|v^{\top }\eta _{1}|})<\infty .$ }
\end{enumerate}
\end{assumption}

{\normalsize The following assumption imposes some conditions on the
temporal dependence for the processes $(X_{jt})_{t\in \mathbb{Z}}$ and $%
(\epsilon _{jt})_{t\in \mathbb{Z}}.$ }
{Denote $\mathcal{C}_j$ as a set of search values for the time series $j$ containing at least the threshold values $c_{0j}$. If the true threshold is given, then $\mathcal{C}_j = \{c_{0j}\}$ is a single point in Subsections \ref{known} and \ref{known2}. We will specify $\mathcal{C}_j$ later in the theorems for the unknown case.}
Let $g_{j}(\cdot )$ be the density function of $X_{jt}.$
Recall $\F_s$ in \eqref{eq:FttildeFt}, for $t\geq s+1,$ let $$g_{j,t}(x|\mathcal{F}_{s}):=d\mathbb{P}(X_{jt}\leq x|%
\mathcal{F}_{s})/dx.$$ Let $\mathcal{F}_{t,\{s\}}$ be $\F_t$ with $\iota_s$ therein replaced by $\iota_s'$ where $(\iota_s')_{s\in\ZZ}$ are independent copies of $(\iota_s)_{s\in\ZZ}$.
That is, for $s\leq t$, $\mathcal{F}_{t,\{s\}}=(\iota_t,\iota_{t-1},\ldots, \iota_{s+1},\iota_s',\iota_{s-1},\ldots).$
Following the idea in \cite{wu2005nonlinear}, we define the following dependence measure.
\begin{assumption}
(Dependence) \label{asmp:dependence} 
(i) For any $x\in\C_j,$ assume $g_{j,t}(x|%
\mathcal{F}_{t-1})$ has finite $p$th moment for some $p>2$. 
Denote 
\begin{equation*}
\theta _{k,p}=\Big\|\sup_{1\leq j\leq N,1\leq t\leq T,x\in\C_j}\big|g_{j,t}(x|
\mathcal{F}_{t-1})-g_{j,t}(x|\mathcal{F}_{t-1,\{t-1-k\}})\big|\Big\|
_{p}.
\end{equation*}
Assume $\sup_{m\geq 0}m^{\alpha }\sum_{k\geq m}\theta _{k,p}<\infty $ for
some $\alpha >0.$ 

{\normalsize (ii) For some $C>0$ and $\beta >0,$ we have $\max_{1\leq j\leq
N}\sum_{k\geq m}|A_{k,j,\cdot }|_{2}\leq C(m\vee 1)^{-\beta }, $where $%
m\geq 0$ and $A_{k,j,\cdot }\in \mathbb{R}^{\tilde{N}}$ is the $j$th row of
matrix $A_{k}.$ }
\end{assumption}
 Assumption \ref{asmp:dependence} $(i)$-$(ii)$ essentially assumes
weak temporal dependence for the processes $(X_{jt})_{t\in \mathbb{Z}}$
and $(\epsilon _{jt})_{t\in \mathbb{Z}}$ for any $j\in \left[ N\right]$. We allow strong cross-sectional
dependence over the $j$ dimension. We shall also note that $\beta $
indicates the strength of the dependence structure of the error processes.

Let $g_{j_{1},j_{2},t}(\cdot ,\cdot |\mathcal{F}_{t-1})$ be
the conditional joint density of $X_{j_{1}t}$ and $X_{j_{2}t}$ given $\mathcal{F}_{t-1}.$ Finally, the following assumption imposes a condition on $%
g_{j_{1},j_{2},t}(\cdot ,\cdot |\mathcal{F}_{t-1}).$

\begin{assumption}(Joint density) \label{asmp:bddxj1j2density} There is no
perfect or asymptotically perfect collinearity in $X_{j_{1}t}$ and $%
X_{j_{2}t}$ for any $j_{1}\neq j_{2}.$ The conditional densities $
g_{j_{1},j_{2},t}(x_{1},x_{2}|\mathcal{F}_{t-1})$ are uniformly upper bounded,
that is, $\max_{1\leq j_{1},j_{2}\leq N}\sup_{x_{1},x_{2}\in \mathbb{R}%
}|g_{j_{1},j_{2},t}(x_{1},x_{2}|\mathcal{F}_{t-1})|<\infty .$
\end{assumption}

We provide examples for Assumptions \ref{asmp:dependence} and \ref{asmp:bddxj1j2density} in Section~\ref{addtionalcovariate} of the Supplementary Materials.

{\normalsize 
}

{\normalsize 
}

{\normalsize 
}
Let
\begin{align}
\label{eq:defbarbunderb}
\bar{b}=\max_{1\leq j\leq N}b_{j}\quad\text{and}\quad \underline{b}%
=\min_{1\leq j\leq N}b_{j}.    
\end{align}

Assumptions \ref{boundedness}-\ref{kernel} are standard assumptions on boundedness, smoothness and kernel functions.

\begin{assumption}
\label{boundedness}
Assume that \(h_j(x,y)\) and \(\tau_j(x,y)\) are uniformly Lipschitz continuous with respect to \(y\). That is, there exists a constant \(L>0\), independent of \(j\), such that for all \(j\in[N]\),
\begin{align*}
\sup_{x\in\mathbb R}
|h_j(x,y)-h_j(x,y')|
\leq L|y-y'|_2
\quad\textrm{and}\quad\sup_{x\in\mathbb R}
|\tau_j(x,y)-\tau_j(x,y')|
\leq L|y-y'|_2.
\end{align*}
Moreover, for some \(q>2\), assume that
\[
\max_{1\leq j\leq N}\max_{1\leq t\leq T}
\mathbb E\bigl(|U_{jt}|_2^q\mid \mathcal F_T\bigr)
\leq C_U
\]
almost surely, where \(C_U>0\) is a constant.
\end{assumption}
It is not hard to see that the above assumption implies that $\mathbb E\left(
|\varepsilon_j(X_{jt},U_{jt})|^q
\mid \mathcal F_T
\right)
<\infty$. We refer to Section~\ref{addtionalcovariate} in the Supplementary Materials for a detailed derivation.

Recall the model specified in \eqref{mainmodel}. Let $f_j(\cdot)$ denote the smooth trend component of the model, defined as follows:
\begin{align} 
\label{eq:defoffjx}
&f_j(x)=\tilde{h}_{j}(x)+[\tilde{\tau}_{j}
(x)-\tilde{\tau}_{j}
(c_{0j})]\mathbf{1}_{\{x\geq c_{0j}\}}.
\end{align}
Consequently, the observed response $Y_{jt}$ can be expressed as:
\begin{align*} 
&Y_{jt}=f_j(X_{jt})+\tilde{\tau}_{j}
(c_{0j})\mathbf{1}_{\{X_{jt}\geq c_{0j}\}}+e_{jt}.
\end{align*}
This formulation captures the smooth trend and the discontinuity at the threshold $c_{0j}$.
The following assumption imposes smoothness conditions on functions.
\begin{assumption}
{\normalsize (Smoothness) \label{asmp:smooth}(i) The density functions $%
g_{j}(\cdot )$ of $X_{jt}$ are lower bounded on $\C_j,$ that is for any $x\in\C_j,$ assume $\min_{1\leq j\leq
N}g_{j}(x)\geq c$ for some constant $c>0$. The conditional densities $%
g_{j,t}(\cdot |\mathcal{F}_{s})$ satisfy that $\max_{j,t}|g_{j,t}(\cdot|%
\mathcal{F}_{t-1})|$ $=O(1)$ and that
\begin{align*}
\max_{1\leq j\leq N,1\leq t\leq
T}\sup_{x\in\C_j}\sup_{|y-x|\leq b_{j}}|g_{j,t}(y|\mathcal{F}_{t-1})-g_{j,t}(x|%
\mathcal{F}_{t-1})|=O(\bar{b}).    
\end{align*} 
}

{\normalsize (ii) For all $j\in \lbrack N],$ assume that for any $x\in\C_j$, function $f_{j}(\cdot )$ satisfies
\begin{equation*}
f_{j}(y)=%
\begin{cases}
f_{j}(x)+\partial _{+}f_{j}(x)(y-x)+O((y-x)^{2})\quad & 
\text{if}\ y\in \lbrack x,x+b_{j}], \\ 
f_{j}(x)+\partial _{-}f_{j}(x)(y-x)+O((y-x)^{2}) & \text{%
if}\ y\in \lbrack x-b_{j},x],%
\end{cases}%
\end{equation*}%
where $\partial _{+}f_{j}(x)$ (resp. $\partial _{-}f_{j}(x)$) is
the right (resp. left) derivative of function $f_{j}\left( \cdot \right) $
at point $x$, and the constants in $O(\cdot )$ are independent of $%
j,b_{j}$. }

(iii) The standard deviation functions $\sigma _{j}(\cdot
,\cdot )$ satisfy $\inf_{u\in\RR^d,x\in\C_j}\sigma _{j}(x,u)>0$ with 
\begin{align*}
\max_{1\leq j\leq
N}\sup_{u\in\RR^d, x\in\C_j, |y-x|\leq b_j}|\sigma _{j}(y,u)/\sigma
_{j}(x,u)-1|=O(\bar{b}).   
\end{align*}

\end{assumption}

Assumption \ref{asmp:smooth} $(i)$ imposes a lower bound condition
for the density functions $g_{j}(\cdot )$, as well as an upper bound for the conditional density $g_{j,t}(\cdot|\F_{t-1})$ and smoothness. Condition $(ii)$
imposes some smoothness conditions of the mean function $f_{j}(\cdot ).$
Note that we do not require $f_{j}(\cdot )$ to be differentiable at point $x$. It suffices for it to have bounded left and right
derivatives near $x$. That is we do not impose any differentiability assumptions on $\tilde{%
h}_{j}(\cdot)$ and $\tilde{\tau}_{j}(\cdot)$. Furthermore, we do not require the left and right derivatives of 
$f_j(\cdot)$ to be the same, allowing the slope to differ on either side of the threshold. Assumption $(iii)$ is for
the smoothness of the variance function $\sigma _{j}^{2}(\cdot ,\cdot)$. If $%
\sigma _{j}(x ,u)$ has bounded partial left and right derivatives with respect to $x$, for any $x\in\C_j$ uniformly over $u,$ then assumption $(iii)$ is fulfilled. 

\begin{assumption}\label{kernel}
{\normalsize (Kernel and bandwidth) \label{asmp:kernel} (i) The kernel
function $K\left( \cdot \right) $ is nonnegative and has bounded support $%
[-1,1].$ Also assume $|K|_{\infty }<\infty ,|K^{\prime }|_{\infty }<\infty $
and $\int_{-1}^{1}K(x)dx=1,$ where $K^{\prime }$ denotes the first
derivative of $K.$ For $l=0,1,2,$ denote $K_{l}^{+}=\int_{0}^{1}x^{l}K(x)dx$ and $%
K_{l}^{-}=\int_{-1}^{0}x^{l}K(x)dx.$ Assume 
\begin{equation}
1<\frac{K_{2}^{+}K_{0}^{+}}{K_{1}^{+2}},\frac{K_{2}^{-}K_{0}^{-}}{K_{1}^{-2}}%
\leq c  \label{eq:kernelcond}
\end{equation}%
for some constant $c>1.$ }

(ii) As $\left( N,T\right) \rightarrow \infty ,$ assume $\bar{b}%
\rightarrow 0$ and $\underline{b}T/\mathrm{log}(NT)\rightarrow \infty .$  Additionally, assume that $b\asymp 
\underline{b}\asymp \bar{b}$.

\end{assumption}

Assumption \ref{asmp:kernel}$(ii)$ imposes the standard rate conditions on the bandwidth parameter in local linear kernel estimation, up to logarithmic factors.

\subsection{\protect\normalsize Gaussian Approximation Results}

{\normalsize \label{known} }

{If the true threshold is given, then $\mathcal{C}_j = \{c_{0j}\}$ is a single point in Subsections \ref{known} and \ref{known2}.}
{We first consider the GA result for $\mathcal{I}$ defined in (%
\ref{I1}). Define 
\begin{align*}
d_{j}=(Tb_j)^{1/2}\gamma _{j}/v_{j}\quad \textrm{and}\quad\underline{d}%
=(d_{1},d_{2},\ldots ,d_{N})^{\top }.    
\end{align*}
Further define $Z$ as a mean zero Gaussian random vector with identity covariance matrix.  Under the null }$H_{0}^{\left(
1\right) },$ the bias term $\underline{d}$
becomes a zero vector. The following theorem states that the infeasible statistic $\mathcal{I}$ can be approximated well by the maximum of Gaussian random
variables. Denote 
\begin{equation*}
\Delta =\mathrm{log}^{7/6}(NT)(\underline{b}T)^{-1/6}+\mathrm{log}%
^{1/2}(N)T^{1/2}\bar{b}^{5/2}+T^{-(\alpha p)\wedge (p/2-1)}+\bar{b}^{1/3}%
\mathrm{log}^{2/3}(N).
\end{equation*}%
Define $\mathcal{R}_{NT}=\Delta +(\underline{b}T^{1-2/q})^{-1/3}\mathrm{log}
(NT)+\mathrm{log}(NT)N^{1/q}T^{-\beta }$ under Assumption \ref{asmp:moment} $(i)$, and $
\mathcal{R}_{NT}=\Delta +\mathrm{log}(NT)T^{-\beta }$ under Assumption
\ref{asmp:moment} $(ii)$.

\begin{theorem}[Gaussian Approximation]
\label{thm:ga} Let Assumptions \ref{asmp:moment}-\ref{kernel} hold. Then we have
\begin{equation}
\sup_{u\in \mathbb{R}}\big|\mathbb{P}(\mathcal{I}\leq u)-\mathbb{P}(|Z+%
\underline{d}|_{\infty }\leq u)\big|\lesssim \mathcal{R}_{NT}.
\label{eq:ga1}
\end{equation}%
\end{theorem}
Now we list the related rate requirement on $T$, $\bar b$ and $N$ under different moment conditions.
Under Assumption \ref{asmp:moment}$(i)$,
if $T\bar{b}^{5}\mathrm{log}(N)\rightarrow 0,$ $\underline{b}%
^{-1}T^{2/q-1}\mathrm{log}^{3}(NT)\rightarrow 0$ and $\mathrm{log}%
(NT)N^{1/q}T^{-\beta }\rightarrow 0,$ then $\Delta \rightarrow 0$ and $\mathcal{R}_{NT}\to 0$.
Under Assumption \ref{asmp:moment}$(ii)$, if $T\bar{b}^{5}\mathrm{log}(N)\rightarrow 0,$ $(T\underline{b})^{-1}
\mathrm{log}^{7}(NT)\rightarrow 0,$ $\bar b\log^3(N)\rightarrow0$ and $\mathrm{log}(NT)T^{-\beta
}\rightarrow 0,$ then $\Delta \rightarrow 0$ and $\mathcal{R}_{NT}\to 0$.

{\normalsize 
}

Theorem \ref{thm:ga} lays down the foundation for testing the
null hypothesis $H_{0}^{\left( 1\right) }.$ The proof of Theorem \ref{thm:ga}
is technically challenging and involved for two reasons. First, we allow for
both temporal and cross-sectional dependence among $\left\{ \left(
X_{jt},\epsilon _{jt}\right) \right\} .$ Second, the error term \(e_{jt}\) in \eqref{eq:deffjvarepj} is highly complex, 
involving both the heterogeneous component and the latent variable. Such a complicated structure makes
it impossible to apply any existing GA results or their proof strategies
directly. In Section \ref{SecA.1} of the Supplementary Materials, we outline the main idea used in the proof of the above theorem. We note that the term $\log^{7/6}(NT)(\underline{b} T)^{-1/6} + \log^{1/2}(N)T^{1/2}\bar{b}^{5/2}$ arises from extending the original GA results in \cite{MR3693963}. The term $\log(NT)N^{1/q}T^{-\beta }$ is due to the temporal dependence and the moment condition of the {innovation term $\eta _{jt}$}. The term $\bar{b}^{1/3} \log^{2/3}(N)$ results from the covariance approximation and Gaussian comparison.
 

Next, we consider the GA result for {$\mathcal{Q}$} defined in (\ref{Q1}). Note that 
\begin{align}
\mathcal{Q}& =\max_{1\leq j\leq N}(Tb_{j})^{1/2}\Big|%
\sum_{t=1}^{T}\Big((w_{jt,b}^{+}-w_{jt,b}^{-})Y_{jt}-\sum_{j=1}^N(w_{jt,b}^{+}-w_{jt,b}^{-})Y_{jt}/N\Big)\Big|/%
\tilde{v}_{j}\nonumber\\
&\approx \max_{1\leq j\leq N}\Big|(Tb_{j})^{1/2}
\sum_{t=1}^{T}\Big((w_{jt,b}^{+}-w_{jt,b}^{-})e_{jt}-\sum_{j=1}^N(w_{jt,b}^{+}-w_{jt,b}^{-})e_{jt}/N\Big)/
\tilde{v}_{j}+\tilde d_j\Big|,\label{Q2}
\end{align}%
where $\tilde{d}_{j}=(Tb_{j})^{1/2}(\gamma _{j}-\frac{1}{N}\sum_{i=1}^{N}\gamma _{i})/%
\tilde{v}_{j}.$ Let $\tilde{\underline{d}}=(\tilde{d}_{1},\tilde{d}%
_{2},\ldots ,\tilde{d}_{N})^{\top }.$  The approximation is valid due to the smoothness of $\tilde h_j(\cdot)$ and $\tilde \tau_j(\cdot)$; the remainder is addressed in Subsection \ref{SecB.3}. 
We then deliver the GA results for $\mathcal{Q}$ and the proofs in Section \ref{proofhomo}  in the Supplementary Materials. Furthermore, the theoretical results regarding the derivatives and their proofs are given in Section \ref{derivatives} in the Supplementary Materials. In addition,  in Section \ref{known2}, we establish the validity of the test statistics when variance estimators are utilized.

The most interesting finding is that in the GA result \eqref{eq:ga1}, neither temporal nor cross-sectional dependencies appear in $|Z+\underline{d}|_\infty$, despite allowing  strong cross-sectional dependence for the underlying data generating process $X_{jt}, U_{jt}, \epsilon_{jt}$ within $\mathcal{I}$. The theoretical intuition behind this result is that local smoothing effectively whitens the noise, see Section \ref{intuition}. In practice, this observation suggests that, when the Assumption \ref{asmp:bddxj1j2density} holds for cross-sectional dependence, the correlation in the estimated statistics can be safely ignored, requiring only adjustments for heteroskedasticity.

\begin{theorem}[Gaussian Approximation for homogeneity]
{\normalsize \label{thm:homotesting} Suppose that Assumptions \ref{asmp:moment}-\ref{kernel} hold. Then we have
}

\begin{equation}\label{eq:ga_homo}
\sup_{u\in \mathbb{R}}\Big|\mathbb{P}\Big(\mathcal{Q}\leq u\Big)-\mathbb{P}%
(|Z+\tilde{\underline{d}}|_{\infty }\leq u)\Big|\lesssim \mathcal{R}_{NT}.
\end{equation}
\end{theorem}
Theorem \ref{thm:homotesting} establishes the foundation for testing the null hypothesis $H_{0}^{\left( 2\right) }.$ Its proof does not directly follow from that of Theorem \ref{thm:ga} due to the additional cross-sectional averaging term in $Q$, which might not be of smaller order. A simpler
argument would be available under weak cross-sectional dependence, since in that case
the additional averaging term becomes higher order.


\begin{remark}
(Relative rates of $N$ and $T$, and bandwidth selection) 
\label{remark_bandwith} 
The relative rate of $N$ and $T$ is determined by the moment assumptions on the
error $\eta _{11}$ (Assumption \ref{asmp:moment}) and the temporal dependence (Assumption \ref{asmp:dependence}). When we
assume exponential moment conditions (i.e., Assumption \ref{asmp:moment}
$(ii)$), $N$ can be of exponential order of $T$. Specifically, if we assume
that $b\propto T^{-c}$ with $c>1/5$, then $\mathcal{R}_{NT}\rightarrow 0$ when  $\mathrm{log}N\propto
T^{r} $, where $0<r<(1-c)/7\wedge (5c-1)\wedge \beta\wedge c/2$. When we assume finite moment conditions (i.e., Assumption \ref{asmp:moment} $(i)$), we require $N$ to be of polynomial order
of $T$. If we assume
that $b\propto T^{-c}$ with $c\in(1/5,1-2/q),$ then $\mathcal{R}_{NT}\rightarrow 0$ when $N\propto T^{\tilde{r}}$ for  $0<\tilde{r}<\beta q$.
To address the bandwidth selection, we can calculate $\hat{b}%
_{j} $ for each individual, following the procedure of \cite{calonico2014robust}. {Sometimes, it is
also recommended to aggregate the bandwidths to avoid the computational burden of
estimating the bandwidths for each individual/group.} 
\end{remark}

\begin{remark}(Critical values accounting for higher order terms, the dependence aware scheme.)
\label{rmk:rmkvj1j2}
Although Theorem~\ref{thm:ga} shows that the cross-sectional dependence of the localized statistics is asymptotically
negligible, these off-diagonal covariances are not exactly zero in finite samples.
Adding them back for the calculation of critical values can improve power. Define
\begin{equation}
v_{j_1j_2}
=
T(b_{j_1}b_{j_2})^{1/2}
\sum_{t=1}^{T}
\bigl(w_{j_1t,b}^{+}-w_{j_1t,b}^{-}\bigr)
\bigl(w_{j_2t,b}^{+}-w_{j_2t,b}^{-}\bigr)
\operatorname{Cov}
\bigl(e_{j_1t},e_{j_2t}\mid \mathcal F_T\bigr).
\label{eq:def_sigmaj12}
\end{equation}
Then \(v_{jj}=v_j^2\), where \(v_j^2\) is defined in \eqref{eq:def_sigma}. Let the correlation matrix be
\[
\Sigma=(\Sigma_{j_1j_2})_{1\leq j_1,j_2\leq N},
\qquad
\Sigma_{j_1j_2}
=
\frac{v_{j_1j_2}}{v_{j_1}v_{j_2}} .
\]
If Assumption \ref{asmp:bddxj1j2density} is removed from Theorem \ref{thm:ga},
then the following more general Gaussian approximation holds:
\begin{equation}
\sup_{u\in \mathbb R}
\left|
\mathbb P(\mathcal I\leq u)
-
\mathbb P\left(
\big|\Sigma^{1/2}Z+\underline d\big|_\infty\leq u
\right)
\right|
\lesssim
\mathcal R_{NT},
\end{equation}
where \(Z\sim N(0,I_N)\). This result is more general than Theorem \ref{thm:ga}. Without Assumption
\ref{asmp:bddxj1j2density}, the localization argument no longer necessarily
whitens the cross-sectional dependence, and the limiting Gaussian vector may retain
a nontrivial covariance structure. The matrix \(\Sigma\) captures this remaining
cross-sectional dependence, including cases with extremely strong or even perfect
cross-sectional dependence. The price for this additional generality is that one
needs to estimate the full correlation matrix \(\Sigma\). In contrast, under
Assumption \ref{asmp:bddxj1j2density}, Theorem \ref{thm:ga} avoids this step
because the limiting correlation matrix is asymptotically diagonal. We refer to Algorithm~\ref{algdepaware} in Section~\ref{sec:algo} of the Supplementary Materials for a feasible implementation of the modified testing procedure.
\end{remark}

\subsection{Asymptotic Validity of the Feasible Testing Procedures}
\label{known2} 
Since $\mathcal{I}$ is infeasible, it is replaced by a
feasible version $\mathcal{\hat{I}}$ in Algorithm \ref{alg1}, where the variance $v_{j}^2$ is substituted by its consistent estimator $\hat{v}_{j}^2.$ Let $\sigma _{e,j}^{2}=%
\mathrm{Var}\big(e_{jt}|X_{jt}\big).$ Recall that $\hat{\sigma}%
_{e,j}^{2}$ is defined in \eqref{eq:sigmaej2} and $\hat v_j^2$ in Algorithm \ref{alg1}. The following theorem establishes the  consistency of these estimators.

\begin{proposition}[Consistency of the variance estimator]
{\normalsize \label{thm:hatv} Suppose that Assumptions {\ref{asmp:moment}}-\ref{kernel} hold and $N(\underline{b}T)^{-q/4}\rightarrow 0.$ Then we have $
\max_{j\in \lbrack N]}|\hat{\sigma}_{e,j}^{2}-\sigma _{e,j}^{2}|$ $=O_{%
\mathbb{P}}(\bar b+(\underline{b}T)^{-1/2}\mathrm{log}(N))$ and $
\max_{j\in \lbrack N]}|\hat{v}_{j}^{2}-v_{j}^{2}|$ $=O_{\mathbb{P}}(
\bar b+(\underline{b}T)^{-1/2}\mathrm{log}(N)).$}
\end{proposition}

Proposition \ref{thm:hatv} implies that $\hat{\sigma}_{e,j}^{2}$
and $\hat{v}_{j}^{2}$ are uniformly consistent estimators of $\sigma
_{e,j}^{2}$ and $v_{j}^{2},$ respectively. Combining the results in Theorem %
\ref{thm:ga} and Proposition \ref{thm:hatv}, we observe that the estimation of $v_j^2$
does not change the accuracy of the GA of $\mathcal{\hat{I}}$ in Algorithm %
\ref{alg1} under $H_{0}^{(1)}$. Analogously, combining the results in
Theorem \ref{thm:homotesting} and Proposition \ref{thm:hatv}, we observe that the
feasible version $\mathcal{\hat{Q}}$ of $\mathcal{Q}$ shares the same first
order limit distribution as $\max_{j\in \left[ N\right] }\left\vert
Z_{j}\right\vert $ under $H_{0}^{(2)}$, where $Z_{j}$'s are i.i.d.\ $N(
0,1) .$ As a result, one can obtain analytical critical values from $%
\max_{j\in \left[ N\right] }\left\vert Z_{j}\right\vert $ for both $\mathcal{%
\hat{I}}$ and $\mathcal{\hat{Q}}$ and the tests in Algorithms \ref{alg1} and %
\ref{alg2} have well-controlled size asymptotically.

Recall that $q_{\alpha}$ is the $1-\alpha $ quantile of the
maximum Gaussian random variable $\max_{j\in \left[ N\right] }\left\vert
Z_{j}\right\vert $ in Algorithm \ref{alg1}. The following theorem studies the asymptotic power
properties of the two tests. 

\begin{corollary}[Power]
\label{thm:power}Suppose that Assumptions \ref%
{asmp:moment}-\ref{kernel} hold. Let $\bar{\gamma}=%
\sum_{j=1}^{N}\gamma _{j}/N.$ 

{\normalsize $(i)$ If $\mathcal{R}_{NT}\to 0$ in Theorem \ref{thm:ga} and $\max_{j\in \left[ N\right] }(Tb_{j})^{1/2}|\gamma
_{j}|\gg (\mathrm{log}N)^{1/2},$ then we have \\$\lim_{\left( N,T\right) \rightarrow
\infty }\mathbb{P}(\mathcal{\hat{I}}>q_{\alpha})=1;$ }

{\normalsize $(ii)$ If $\mathcal{R}_{NT}\to 0$ in Theorem \ref{thm:homotesting} and $\max_{j\in \left[ N\right] }(Tb_{j})^{1/2}|%
\gamma _{j}-\bar{\gamma}|\gg (\mathrm{log}N)^{1/2},$ then \\$\lim_{\left(
N,T\right) \rightarrow \infty }\mathbb{P}(\mathcal{\hat{Q}}>q_{\alpha})=1. $ }
\end{corollary}

Corollary \ref{thm:power} implies that the tests based on $%
\mathcal{\hat{I}}$ and $\mathcal{\hat{Q}}$ and the simulated critical values
are consistent against global alternatives and have nontrivial power against
local alternatives. Specifically, if $(\mathrm{log}N)^{-1/2}
\max_{j\in \left[ N\right] }(Tb_{j})^{1/2}|\gamma _{j}|$ $\asymp 1,$ $%
\mathcal{\hat{I}}$ has non-trivial local power even though it is not easy to
present the explicit local power function. The power of $\mathcal{\hat{I}%
}$ approaches 1 provided $(\mathrm{log}N)^{-1/2}\max_{j\in \left[ N%
\right] }(Tb_{j})^{1/2}|\gamma _{j}|\rightarrow \infty $ as $\left(
N,T\right) \rightarrow \infty .$ Similar remarks apply to $\mathcal{\hat{Q}}$. 

\subsection{Key Intuition on the Dependence Structure}
\label{intuition}

The goal of this subsection is to provide intuition on why the critical values of our testing procedure can be computed as if the time series were independent, even if we allow for strong cross-sectional dependence across them. For this purpose, we distinguish between dependence on the level of the \emph{data-generating process}, and the dependence that survives in the \emph{localized statistics} $\hat{\gamma}_j$ after kernel smoothing. The former can be extremely strong, while the latter is asymptotically negligible, which ultimately governs the critical values.

At the level of our data-generating process, the covariates $X_{jt}$ and the error process $\epsilon_t$ are both allowed to exhibit temporal as well as cross-sectional dependence. We impose only mild restrictions on the dependence structure, so the cross-sectional dependence is permitted to be extremely strong.
At the level of the localized statistics, by contrast, this dependence largely
washes out in the sup-norm at the leading order. We will show that for
$i \ne j$ the cross-sectional covariance between $\hat\gamma_i$ and
$\hat\gamma_j$ is of strictly smaller order than the variance of
$\hat\gamma_j$. This is a \emph{consequence} of kernel localization rather than
a restriction on the DGP, and it is exactly what is needed for our test: the
maximum of the studentized statistics behaves like the maximum of $N$
approximately independent Gaussians, so the critical values can be simulated as
if the panels were independent. We stress that this concerns only the
\emph{maximum} of the statistics; the joint distribution of the full vector
need not be close to i.i.d.\ Gaussian, and we do not require it to be.

This is in contrast to the standard Gaussian-approximation route to
critical values. In that approach, one typically needs to simulate from an estimated variance-covariance matrix \citep{chernozhukov2019inference}. To estimate the covariance between $\hat\gamma_i$ and $\hat\gamma_j$, one must control the long-run covariance structure of the error process. This is challenging because temporal and cross-sectional dependence have to be handled at the same time. The difficulty becomes even more pronounced in high-dimensional settings. Our framework
sidesteps this entirely: under mild conditions, and provided the joint density
of $(X_{it}, X_{jt})$ is non-degenerate (Assumption~\ref{asmp:bddxj1j2density}),
the off-diagonal covariances are of smaller order than the variances, so
standardising by the marginal standard deviation (or a consistent estimate
thereof) suffices asymptotically, with no covariance estimation at all.

To see why localization produces this effect, consider a simplified setting in
which the latent variable $U_{jt}$ is absent from \eqref{modelstructural} and
$\sigma_j(X_{jt}, U_{jt}) = 1$, so that $e_{jt} = \epsilon_{jt}$, and in which
$(X_{it}, \epsilon_{it})$ is i.i.d.\ over $t$ with $X_{it}$ independent of
$\epsilon_{it}$ within each panel. Heuristically, the weighted statistic
\[
\max_j (bT)^{1/2} \sum_{t=1}^T (w_{jt,b}^+ \epsilon_{jt} - w_{jt,b}^- \epsilon_{jt})
\]
is of the same order as the corresponding uniform-kernel statistic
\[
\max_j (bT)^{-1/2} \sum_{t=1}^T \mathbf{1}_{\{|X_{jt}| \leq b\}}\, \epsilon_{jt},
\]
and the high-dimensional Gaussian approximation lets us replace the latter by
the maximum of a Gaussian vector with the same covariance structure as
$\{ b^{-1/2} \mathbf{1}_{\{|X_{jt}| \leq b\}}\, \epsilon_{jt} \}_{j=1}^N$. The
population covariance of this vector is
\[
  b^{-1} \E\!\left[ \mathbf{1}_{\{|X_{it}| \leq b\}} \mathbf{1}_{\{|X_{jt}| \leq b\}}
                    \epsilon_{it} \epsilon_{jt}\right]
  = b^{-1} \E(\epsilon_{it} \epsilon_{jt})\,
    \E\!\left[ \mathbf{1}_{\{|X_{it}| \leq b\}} \mathbf{1}_{\{|X_{jt}| \leq b\}}\right],
\]
where we take $\E(\epsilon_{it} \epsilon_{jt}) \neq 0$ since cross-sectional
dependence is the case of interest. The key observation is that the indicator
expectation is of a different order on and off the diagonal:
\[
  \E\!\left[ \mathbf{1}_{\{|X_{it}| \leq b\}}\mathbf{1}_{\{|X_{jt}| \leq b\}}\right]
  = \E\!\left[ \mathbf{1}_{\{|X_{it}| \leq b\}}\right] = O(b)
  \quad \text{if } i = j,
  \qquad
  = O(b^2) \quad \text{otherwise.}
\]
Hence the $i \neq j$ covariances are of strictly smaller order than the $i = j$
variances and are negligible asymptotically. Crucially, this holds even under
strong cross-sectional correlation among $\{(X_{jt}, \epsilon_{jt})\}_{j=1}^N$,
as long as the joint density of $(X_{it}, X_{jt})$ is non-degenerate --- ruling
out exact linear dependence ($X_{jt} = a_0 + a_1 X_{it}$) and asymptotic linear
dependence ($X_{jt} = a_1 X_{it} + \eta_{it}$ with $\Var(\eta_{it}) \to 0$),
but otherwise leaving the cross-sectional dependence structure unrestricted.

Two remarks qualify this calculation. First, the negligibility of the
off-diagonal terms is a \emph{leading-order} statement: at first order the
localized statistics behave like independent Gaussians, which is what justifies
simulating critical values from $\max_{j}|Z_j|$ with i.i.d.\ $Z_j$. The
off-diagonal covariances are of smaller order but not exactly zero, so in finite
samples one may recover additional accuracy and power by adding them back.
That is, by simulating from $\max_{j}|Z_j|$ with $Z \sim N(0, \hat R)$ for
an estimated correlation matrix $\hat R$, rather than from the identity. We
develop this dependence-aware refinement in Remark~\ref{rmk:rmkvj1j2} and study
its finite-sample behavior in the Supplementary Materials. Second, the whitening
we exploit is a statement about the \emph{maximum} of the localized statistics,
i.e.\ about their sup-norm; it does not assert that the joint law of the vector
is close to i.i.d.\ Gaussian. 

These observations complement what is known in the temporal direction. In the
classical nonparametric literature, the asymptotic variance of the kernel
estimator coincides with the one obtained by ignoring the temporal
dependence among the observations; see, e.g., Theorem~5.1 in
\cite{fan2003nonlinear} and Theorems~18.2 and~18.4 in
\cite{li2007nonparametric}. Our setting differs in two respects. First, we
do not sum over the cross-sectional dimension $j$, so no decay condition
along $j$ is needed. Second, we treat temporal and cross-sectional
dependence simultaneously: the covariance between $\hat\gamma_i$ and
$\hat\gamma_j$ is strictly dominated by the variance of $\hat\gamma_j$,
regardless of whether the cross-sectional dependence is strong or weak,
provided the temporal dependence is weak in the usual sense.

In the simulation study of Section~\ref{sec:simulation}, we generate strong
cross-sectional dependence in $\{(X_{jt}, \epsilon_{jt})\}_{j=1}^N$ via factor
structures to illustrate these calculations. As expected, stronger
cross-sectional dependence may cost some finite-sample accuracy relative to the
independent case, so that a larger sample size is needed to attain comparable
performance.

\subsection{Theoretical Results for the Unknown
Threshold Case\label{unknown_c2}}
\label{unknown} In this section, we present theoretical results in the case of unknown thresholds. 
For simplicity of notation, we define $\mathcal{C}_j$ as the unified interval $\mathcal{C}_j = [c_{\min}, c_{\max}]$, where $c_{\min}$ and $c_{\max}$ are constants in $\RR$. The theorem still holds even when $\C_j$ differs for different $j$. Recall that $[c_1,\ldots, c_K]$ is the grid we search over. Let $c_1=c_{\min}$ and  $c_{K}=c_{\max}.$ As our statistics aggregate over different threshold grid points, the problem becomes more demanding.
Let $\Delta_{c,\min }=\min_{1\leq i\leq {K-1}}|c_{i+1}-c_{i}|$ and $\Delta_{c,\max}=\max_{1\leq i\leq {K-1}}|c_{i+1}-c_{i}|$.

Recall $w_{jt,b}(c_{i})$ in \eqref{eq:wjtbci} and define ${\gamma }_{j}(c_{i})=\sum_{t=1}^{T}w_{jt,b}(c_{i})%
\gamma _{j}$. Let $$d_{jK}=(Tb_j)^{1/2}(\gamma _{j}(c_{1})/v_{j}(c_{1}),\ldots ,\gamma
_{j}(c_{K})/v_{j}(c_{K}))^{\top }$$ and $\underline{d}%
^{C}=(d_{1K}^{\top },d_{2K}^{\top },\ldots ,d_{NK}^{\top })^{\top }.$
Define $Z^{C}=\left( {{Z_{1}^{C\top },\ldots,Z_{N}^{C\top }}}\right) ^{\top }$
as an $NK$-vector of mean zero Gaussian distributed random vector with
covariance matrix $\Sigma^C$
specified below. For each $j\in \lbrack N]$ and $1\leq i_{1},i_{2}\leq K$,
the covariance between the 
$i_{1}$- and $i_{2}$-th elements of $Z_{j}^C$  {{is
{given} by{%
\begin{equation*}
\cov(Z_{j,i_1}^C, Z_{j,i_2}^C)=(v_{j}(c_{i_{1}})v_{j}(c_{i_{2}}))^{-1}(Tb_{j})%
\sum_{t=1}^{T}w_{jt,b}(c_{i_{1}})w_{jt,b}(c_{i_{2}})\mathrm{Var}\big(%
e_{jt}|X_{jt}\big).
\end{equation*}%
}}}${\normalsize {{Z_{j_1}^C}}}$ and $Z_{j_2}^C$ are
independent for $ j_1\neq j_2$,  so that the matrix $\Sigma^C$ is block diagonal. Note that $
\underline{d}^{C}$ denotes a bias term, which is a zero vector under the null hypothesis. In addition, when $2\max_{1\leq j\leq
N}b_{j}<\Delta_{c,\min },$ one can ensure the matrix to be diagonal. The following theorem establishes the
asymptotic property of our test statistics with unknown thresholds. 

\begin{theorem}\label{thm:unknown}
{\normalsize (Gaussian Approximation) Suppose that Assumptions \ref%
{asmp:moment}-\ref{kernel} hold and that $\Delta_{c,\max }\rightarrow 0.$} Then 
\begin{equation}
\sup_{u\in \mathbb{R}}\big|\mathbb{P}(\mathcal{I}^{C}\leq u)-\mathbb{P}%
(|Z^C+\underline{d}^C|_{\infty }\leq u)\big|\lesssim \mathcal{R}_{(NK)T},
\label{eq:ga}
\end{equation}
where $\mathcal{R}_{(NK)T}$ is $\mathcal{R}_{NT} $ in Theorem
3.1 with $N$ replaced by $NK$.
\end{theorem}

The proof of the above theorem is a generalization 
of Theorem \ref{thm:ga} with a different variance-covariance structure. It should be noted that the conclusion of the above theorem follows the same pattern as Theorem \ref{thm:ga}, except that $N$ is replaced with $NK$.
The following theorem provides the theoretical support for the consistency
of the threshold estimator. Note that we only focus on the location $j$ where $\gamma_j$ does not equal $0$, and we require those breaks to be significant.

\begin{theorem}[Consistency of the estimation of thresholds]
{\normalsize \label{consistency} Suppose that Assumptions \ref{asmp:moment}-\ref{kernel}  hold. Assume $\min_{1 \leq j\leq N: \gamma_j \neq 0}(Tb_{j})^{1/2}|\gamma _{j}|\gg 1\ $ and }$\Delta_{c,\max }\rightarrow 0.$
Suppose $c_{0j}\in\{c_i, 1\leq i\leq K\}$, then 
\begin{equation}
\max_{1\leq j\leq N,\gamma_j\neq 0} \gamma _{j}^{2}|\hat{c}_{j}-c_{0j}|=O_{\PP}((\log N)/T).
\end{equation}

\end{theorem}

If the true break $c_{0j}$ does not fall in the grid, then the estimation error is bounded by $\max\{\Delta_{c,\max},\allowbreak  \log(N)/(T\gamma_j^2)\}.$ Under the assumption of a minimum break signal, we can achieve
consistency in detecting the break. The rate is expected to depend
on both the signal strength $\gamma_j$ and the available sample size $T$.
Note that the precision of the  threshold estimation is not affected by the bandwidth $b_j$. This rate is in line with the high-dimensional change-point literature, see for example Theorem 2 in \cite{li2024}.

{
To further evaluate the feasibility of the test statistic $\hat{\mathcal{I}}^{C}$, we also examine the consistency of the variance estimator $\hat{\sigma}_{e,j}^2(c_i)$. Estimating the variance in the presence of an unknown threshold requires additional caution, particularly when accounting for jumps, see Section \ref{sec:algo} in the Supplementary Materials. We will consider a truncation operation to eliminate the contamination from jumps. The consistency of the estimator will be established in a similar manner.  For completeness, we provide Proposition \ref{thm:hatvap} as a counterpart to Proposition \ref{thm:hatv}, specifically addressing the case with an unknown threshold.
}

Let $\hat{e}_{jt}$ be the residual by regressing $Y_{jt}$ on $X_{jt}$ after a local linear regression with a bandwidth $b^*\ll \underline{b} $. 
Let $\phi_{\acute{a}}(x)=\min(\acute{a},|x|)$ with $\acute{a}=c\log^{1/2}(NK).$ Define
\begin{align}
\label{eq:sigmaej2_unknown}
\hat{\sigma }_{e,j}^{2}(c_i)=\frac{\sum_{t=1}^{T}%
\mathbf{1}_{\{|X_{jt}-c_{i}|\leq b_{j}\}}\phi_{\acute{a}}(\hat{e}_{jt}^{2})}{\sum_{t=1}^{T}%
\mathbf{1}_{\{|X_{jt}-c_{i}|\leq b_{j}\}}} .  
\end{align}

\begin{proposition}[Consistency of the variance estimator]
\label{thm:hatvap} Suppose that Assumptions {\ref{asmp:moment}}-\ref{kernel} hold and $NK(bT)^{-q/4}\rightarrow 0.$ Then we have\\ $
\max_{j\in \lbrack N]}\max_{i\in[K]}|\hat{\sigma}_{e,j}^{2}(c_i)-\sigma _{e,j}^{2}(c_i)|$ $=O_{\PP}(\sqrt{\log(NKT)/(T\underline b)}+\bar b+\acute{a}b^*/\bar b+\acute{a}^{-p/2})$ and\\ $
\max_{j\in \lbrack N]}\max_{i\in[K]}|\hat{v}_{j}^{2}(c_i)-v_{j}^{2}(c_i)|$ $=O_{\PP}(\sqrt{\log(NKT)/(T\underline b)}+\bar b+\acute{a}b^*/\bar b+\acute{a}^{-p/2})$.
\end{proposition}
{The estimator accounts for robust variance estimation to address potential breaks. Here, we use the truncation function $\phi_{\acute{a}}(x)$, other choices are also possible, see for example the robust M-estimation method proposed in \cite{chen2022inference}.
}

\section{{\protect\normalsize Monte Carlo Simulations \label{sec:simulation}}%
}

In this section, we analyze the finite sample performance of
our GA results. In particular, we study the size and
power of our uniform testing procedures in different settings. {The explicit focus of this simulation study is to study the impact of cross-sectional dependence on the size and power of our test. While the effect of weak temporal dependence in nonparametric regression models is well studied \citep{fan2003nonlinear}, we demonstrate that the `whitening by windowing' principle also applies to weak and even strong forms of cross-sectional dependence in our model setting.}

\subsection{\protect\normalsize Data Generating Processes}

{\normalsize Throughout the simulation study, we consider the following
setup: 
\begin{equation}
Y_{jt}=\cos (X_{jt})+\sin (U_{jt})+\gamma _{j}\mathbf{1}_{\{X_{jt}\geq
0\}}+\sigma _{j}(X_{jt},U_{jt})\epsilon _{jt},  \label{sim1}
\end{equation}%
where the latent variable is generated by $U_{jt}=\tilde{U}%
_{jt}X_{jt}$ and $\tilde{U}_{jt}\sim i.i.d.\ U[-1,1]$. To study the
performance of our uniform testing procedure, we consider six different data generating processes (DGPs) with
different specifications of the threshold variable $X_{jt}$, the error term  $\epsilon _{jt}$ and the volatility function $\sigma _{j}(\cdot ,\cdot
)$.
For the first one, we consider the independent setting. 
For the other settings, we include
cross-sectional and temporal dependences. We consider either 
homoskedastic settings (DGP2) in which $\sigma _{j}(\cdot ,\cdot )=1$ and
heteroskedastic settings (DGP1, DGP3--6) in which we set 
\begin{equation}
\sigma _{j}(x,u)=1+\left( \frac{3}{8}-\frac{\left\vert x\right\vert }{4}%
\right) \left( \frac{3}{2}\right) ^{2u}.  \label{sim2}
\end{equation}
}

\begin{itemize}
\item[DGP1:] {\normalsize i.i.d.\emph{\ structure with heteroskedasticity}. $%
X_{jt}\sim i.i.d.\ U[-1,1]$, $\epsilon _{jt}\sim i.i.d.\ N(0,1)$. }

\item[DGP2:] {\normalsize \emph{Factor structure and homoskedasticity}. Both
the running variable and error term have a factor structure,
\begin{align*}
    \epsilon _{jt}&=\lambda _{\epsilon ,j}f_{\epsilon ,t}+u_{\epsilon ,jt},\\
    X_{jt}&=1/4\left( \lambda _{x,j}f_{x,t}+u_{x,jt}\right),
\end{align*}
where $f_{\epsilon ,t}$, $%
u_{\epsilon ,jt}$, $f_{x,t}$ and $u_{x,jt}$ are each generated from an MA($%
\infty $) process with algebraic decay parameter, $\beta =1.5$. The factor
loadings, $\lambda _{\epsilon ,j}$ and $\lambda _{x,j}$, are i.i.d. draws
from a standard normal distribution. }

\item[DGP3:] {\normalsize \emph{Factor structure and heteroskedasticity}.
The setting is the same as in DGP2 except that the error term is
heteroskedastic with the conditional standard deviation described in (\ref%
{sim2}). }
\end{itemize}

We examine three additional cases, referred to as DGP4--6 in Section \ref{sec:simulation_appendix} of the Supplementary Materials, which account for varying degrees of cross-sectional dependence and a possible time effect.
All DGPs except for the first one exhibit strong cross-sectional
and weak temporal correlation in the threshold variable and the error term.
We are interested in testing the existence of the jump effect,
namely, $H_{0}^{\left( 1\right) }:\gamma _{j}=0$ for all $j\in \lbrack N]$.
For each DGP, we study both the size and local power performance of our
testing procedure. When studying the size performance, we set all $\gamma
_{j}=0$. When examining the local power, according to Corollary \ref{thm:power} and to the selection of the bandwidth, we set a fraction of the threshold
effects to be $\gamma _{j}=T^{-2/5}(\mathrm{log}N)^{1/2}B_{j}$, where $%
B_{j}\sim i.i.d.\ U[2,10]$, and the remaining $\gamma _{j}$'s to be zero. We
consider two different choices for the cross-sectional dimension: $N=10,$
100. When $N=10$, the above fraction is set to $20\%$, whereas in the
high-dimensional setting with $N=100,$ the fraction is decreased to $10\%$. Additional simulation setups and results are presented in Section \ref{sec:simulation_appendix} of the Supplementary Materials.

\subsection{\protect\normalsize Simulation Results}
Throughout the study, we use a local linear estimator with a uniform kernel and select the bandwidth according to the procedure described in \cite{calonico2014robust}. We consider three
significance levels, namely, $\alpha =0.1,$ $0.05,$ and $0.01$. The results
are based on $1000$ Monte Carlo iterations. 

\begin{table}[tbp]
\centering
\begin{tabular}{rrr|rrr|rrr}
\hline
\multicolumn{3}{r}{} & \multicolumn{3}{c}{Size} & \multicolumn{3}{c}{Power}
\\ 
& $N$ & $T$ & $\alpha=0.1$ & $\alpha=0.05$ & $\alpha=0.01$ & $\alpha=0.1$ & $%
\alpha=0.05$ & $\alpha=0.01$ \\ \hline
\multirow{6}{*}{DGP1} & 10 & 200 & 0.113 & 0.059 & 0.008 & 0.395 & 0.293 & 
0.142 \\ 
& 10 & 400 & 0.133 & 0.060 & 0.014 & 0.428 & 0.316 & 0.153 \\ 
& 10 & 800 & 0.115 & 0.054 & 0.005 & 0.466 & 0.333 & 0.172 \\ 
& 100 & 200 & 0.099 & 0.051 & 0.014 & 0.884 & 0.797 & 0.596 \\ 
& 100 & 400 & 0.115 & 0.054 & 0.011 & 0.926 & 0.874 & 0.699 \\ 
& 100 & 800 & 0.099 & 0.050 & 0.008 & 0.948 & 0.922 & 0.786 \\ \hline
\multirow{6}{*}{DGP2} & 10 & 200 & 0.105 & 0.054 & 0.013 & 0.539 & 0.443 & 
0.269 \\ 
& 10 & 400 & 0.122 & 0.061 & 0.012 & 0.583 & 0.489 & 0.316 \\ 
& 10 & 800 & 0.115 & 0.062 & 0.010 & 0.647 & 0.548 & 0.377 \\ 
& 100 & 200 & 0.101 & 0.042 & 0.007 & 0.981 & 0.962 & 0.892 \\ 
& 100 & 400 & 0.107 & 0.043 & 0.010 & 0.982 & 0.976 & 0.944 \\ 
& 100 & 800 & 0.096 & 0.053 & 0.010 & 0.996 & 0.991 & 0.964 \\ \hline
\multirow{6}{*}{DGP3} & 10 & 200 & 0.123 & 0.055 & 0.007 & 0.371 & 0.275 & 
0.105 \\ 
& 10 & 400 & 0.121 & 0.058 & 0.017 & 0.393 & 0.299 & 0.152 \\ 
& 10 & 800 & 0.122 & 0.066 & 0.014 & 0.465 & 0.380 & 0.220 \\ 
& 100 & 200 & 0.115 & 0.056 & 0.019 & 0.848 & 0.780 & 0.594 \\ 
& 100 & 400 & 0.111 & 0.061 & 0.011 & 0.909 & 0.849 & 0.689 \\ 
& 100 & 800 & 0.116 & 0.064 & 0.012 & 0.952 & 0.902 & 0.775 \\ \hline
\end{tabular}%
\caption{Empirical size and power results for the uniform test for the existence of threshold effects for DGP1--3.}
\label{table:sim1}
\end{table}

{\normalsize 
The results of simultaneous jump effects testing for DGP1--DGP3 are displayed in Table \ref{table:sim1}. 
First, the empirical size closely matches the nominal size across all $N$ values and for large $T$ in all DGPs. For DGP3, we observe a slightly
oversized test when $N=10\ $ which may be caused by the presence of
heteroskedasticity in comparison with results with DGP2. Overall, these
simulation results confirm our theoretical findings in Theorem \ref{thm:ga}.
In particular, they show that the test procedure, which is based on
critical values neglecting the dependence in the covariance structure, has fairly accurate size control
even in the case of strong cross-sectional dependence in the running variables and innovations. Second, the local
power properties of the test are reasonably good for all DGPs under investigation, and
the power increases as either $N$ or $T$ increases. Later we will compare the power of our test with the power of the test based on pooling the observations along the cross-sectional dimension. This result confirms that our
testing procedure is well-suited to detect sparse alternatives.
In addition, we achieve comparable simulation performance in terms of size and power results for the test of \emph{homogeneity} of the threshold effects in Table \ref{table:sim2} in Section \ref{sec:simulation_appendix} of the Supplementary Materials. However, for DGP3, the test procedure shows a slight tendency to over-reject the null, potentially due to the presence of heteroskedasticity in the error terms. The local power results align with the earlier findings from the significance test for the existence of jump effects. The results confirm our Gaussian approximation result in Theorem \ref{thm:homotesting}.

For robustness, we study the size and power properties of the test for the existence of jump
effects for DGP4 -- DGP6 in Table \ref{table:sim3} of the Supplementary Materials. These DGPs are the ones
with the strong cross-sectional dependence we consider in our simulation
study. For DGP4, we observe an empirical size slightly below the nominal
size for the high-dimensional setting with $N=100$ which is due to the
increase in the cross-sectional dependence in the covariate and the
error term. The sizes for DGP5 are almost perfect, while the
inclusion of an additional time effect in DGP6 increases the empirical size
slightly. This time effect adds more noise to the system, so it is not
surprising that the empirical power of DGP6 is lower than that of the other
DGPs we consider. The results show that size and power are indeed robust to strong cross-sectional dependence.

While the preceding results are concerned with the known threshold case, we now examine the size and power properties of our uniform testing procedure in the case of unknown thresholds. We consider the same model specification as in the known threshold case. Therefore, the true threshold is $c_{0j}=0$ for all $j$. For the practical implementation of our test procedure, we consider the following grids of threshold locations. Due to different ranges in the covariate distribution, we consider the grid $[-0.6,-0.3,0,0.3,0.6]$ for DGP1 and $[-0.3,-0.15,0,0.15,0.3]$ for DGP2 and 3. The results in Table \ref{table:sim4} reveal that the test has still good size properties, with the empirical size being slightly smaller than nominal in some settings and slightly larger in others. As expected, the power of the test decreases compared to the case with a known threshold. However, as $T$ increases, the power still approaches one, confirming our theoretical results from Theorem \ref{thm:unknown}. We additionally show the good finite sample performance for the estimation accuracy of the threshold location estimator in Table \ref{table:sim5}, providing Monte Carlo evidence supporting the consistency result of the threshold in Theorem \ref{consistency}.}

\begin{table}[tbp]
\centering
\begin{tabular}{rrr|rrr|rrr}
\hline
\multicolumn{3}{r}{} & \multicolumn{3}{c}{Size} & \multicolumn{3}{c}{Power}
\\ 
& $N$ & $T$ & $\alpha=0.1$ & $\alpha=0.05$ & $\alpha=0.01$ & $\alpha=0.1$ & $%
\alpha=0.05$ & $\alpha=0.01$ \\ \hline
\multirow{6}{*}{DGP1} 
& 10 & 400 & 0.091 & 0.038 & 0.005 & 0.184 & 0.110 & 0.031 \\ 
& 10 & 800 & 0.120 & 0.060 & 0.008 & 0.323 & 0.221 & 0.098 \\ 
& 100 & 400 & 0.086 & 0.039 & 0.006 & 0.463 & 0.339 & 0.186 \\ 
& 100 & 800 & 0.099 & 0.048 & 0.012 & 0.883 & 0.813 & 0.649 \\   \hline
\multirow{6}{*}{DGP2} 
& 10 & 400 & 0.118 & 0.058 & 0.013 & 0.444 & 0.349 & 0.195 \\ 
& 10 & 800 & 0.102 & 0.055 & 0.008 & 0.503 & 0.414 & 0.263 \\
& 100 & 400 & 0.107 & 0.045 & 0.006 & 0.955 & 0.941 & 0.869 \\ 
& 100 & 800 & 0.117 & 0.057 & 0.011 & 0.986 & 0.972 & 0.935 \\   \hline
\multirow{6}{*}{DGP3} 
& 10 & 400 & 0.100 & 0.045 & 0.006 & 0.316 & 0.231 & 0.134 \\ 
& 10 & 800 & 0.116 & 0.071 & 0.017 & 0.534 & 0.458 & 0.355 \\ 
& 100 & 400 & 0.124 & 0.058 & 0.009 & 0.624 & 0.535 & 0.388 \\ 
& 100 & 800 & 0.124 & 0.068 & 0.010 & 0.877 & 0.842 & 0.751 \\   \hline
\end{tabular}%
\caption{Empirical size and power results for the uniform test of the existence of
jump effects under unknown threshold locations for DGP1--3.}
\label{table:sim4}
\end{table}

For comparison, one direct method is the sup-Wald or sup-likelihood ratio test from the pooled threshold regression literature, neither of which accounts for nonlinearity or heterogeneity (see, for example, \cite{andrews1993tests}, \cite{hansen2000sample} and \cite{fong2017chngpt}). Notably, recent studies, such as \cite{barassi2023threshold}, propose linear models with heterogeneous threshold effects in panel settings. However, they do not provide uniform testing methods.
To facilitate a fair comparison, we first restrict ourselves to a linear threshold model. See DGP7 in Section \ref{sec:simulation_appendix} in the Supplementary Materials for a detailed description of the setup.
Under the alternative we sample a fraction of coefficients $\gamma_j$ from a standard normal distribution. As shown in Table \ref{table:sim6} in the Supplementary Materials, our method has similar performance to the sup-Wald type statistic under the null hypothesis, while it has much better power under the alternative. See also Figure \ref{figure:comparison} in Section \ref{sec:model}. The reasons for this are twofold. First, our uniform testing procedure is robust towards settings with sparse signals, in which a pooled test suffers from dilution of signals. Second, our uniform testing procedure does not have the problem of signal cancellation in case the signals for different $j$'s have opposite signs and cancel each other out. As a second comparison, we consider the nonparametric pooled test statistic under DGP3. Again, we draw the non-zero threshold coefficients from a standard normal. Such a setup without accounting for parameter heterogeneity is considered in \cite{spokoiny1998estimation}, \cite{yang2014jump} and \cite{chiou2018nonparametric}. As shown in Table \ref{table:sim7} in the Supplementary Materials, similar to the linear case, our method works with better power. The above superior performance is mainly due to the sparsity of the signal under the alternative hypothesis. In addition, our method has better power even for dense alternatives compared with the pooled type test statistic when there might exist signal cancellation as illustrated in Figure \ref{figure:comparison}.

In the Supplementary Materials, we also consider the test of existence of a threshold effect in the derivative under DGP1--3.  As shown in Table \ref{table:sim8}, oversmoothing is needed to achieve similar size and power performance. We also study the size and power properties of our modified testing procedure that incorporates the higher order covariance terms (see Remark~\ref{rmk:rmkvj1j2}). The results in Table \ref{table:sim9} show that incorporating dependence in the calculation of critical values can indeed lead to a significant increase in power, in the case of extreme cross-sectional dependence, compared with the testing procedure based on critical values from i.i.d.\ Gaussian variables.

{To demonstrate the practical relevance of our testing procedure, we present two real
data applications in Section \ref{sec:application} of the Supplementary Materials. In Subsection \ref{application1}, we detect
significant threshold effects in the news impact curve \citep{engle1993measuring}, i.e., in a model
where past returns influence present volatility of stocks. It is an illustration of our testing
procedure under unknown thresholds. As a second application, we revisit the study of the
incumbency effect in US House elections in Subsection \ref{application2}, and find significant jump effects, as well as heterogeneity across different states. Since the threshold is naturally given in
this problem, it functions as an illustration of our testing procedures for the existence of
jump effects and the test for homogeneity in the known case.}

\section{Conclusion}
\label{sec:conclusion}
This paper focuses on the estimation and inference of heterogeneous threshold effects in high-dimensional nonparametric regression models, accounting for both cross-sectional and temporal dependencies in the covariates and error terms. We propose a test to assess the significance of the threshold effect and another to evaluate whether the threshold effects are homogeneous across individuals or groups. Additionally, we introduce a consistent method for estimating unknown threshold locations. Our tests are based on high-dimensional Gaussian approximation results. Despite the complexity of the underlying dependence structures, we show that the variance-covariance structure of the threshold effect estimators has a simple analytical expression under general dependence conditions. Unlike existing change-point analysis in time series, our approach considers a setup involving latent variables, leading to a substantially different development of the GA results. Simulations demonstrate considerable improvement in power compared to existing methods under various settings. 

\bibliographystyle{chicago}
\bibliography{literature.bib}

\newpage

\appendix

\bigskip
\begin{center}
{\Large\bf Supplementary Materials for ``High-dimensional inference on jumps in nonparametric time series regression models"}
\end{center}

\DoToC

\linespread{1.25}\setcounter{section}{0} 

\input{20260713_theorem}

\section{Additional Algorithms}\label{sec:algo}

We shall list three additional algorithms below. I.e., Algorithm \ref{alg2} for the test for homogeneity and Algorithm \ref{algunknown} for the test of existence of jumps under unknown threshold locations. 
The derivative case follows a similar approach to the mean case. Thus, we include the algorithm for the derivative case below, presented in a bracket format.

\begin{algorithm}[H]
	\caption{Testing for homogeneity, i.e., $\gamma_1=\ldots=\gamma_N$ ( $\gamma_1^{[1]}=\ldots=\gamma_N^{[1]}$).\label{alg2}}
	\begin{algorithmic}[1]
		\For {$j=1,\ldots,N$}
			\State Estimate $\gamma_j$ for each group/individual $j$ using equation (\ref{estimate}).
			\State Estimate the variance $v_j^2$ ($v_j^{[1]2}$) of $(Tb_j)^{1/2}\hat{\gamma}_j$($(Tb_j^3)^{1/2}\hat{\gamma}_j^{[1]}$) by $\hat{v}_j^2=Tb_j\sum_{t=1}^{T}\left(w_{jt,b}^{+}-w_{jt,b}^-\right)^2\hat{\sigma}_{e,j}^2$ (  $\hat{v}_j^{[1]2}=Tb_j^3\sum_{t=1}^{T}\left(w_{jt,b}^{+[1]}-w_{jt,b}^{-[1]}\right)^2\hat{\sigma}_{e,j}^2$), where $\hat{\sigma}_{e,j}^2$ is a consistent estimator for the error variance using the observations close to the threshold $c_{0j}$.
		\EndFor
    \State Calculate the feasible test statistic $\hat{\mathcal{Q}}:=  \mbox{max}_{1\leq j\leq N}(Tb_j)^{1/2} |\hat{\gamma}_j-\frac{1}{N}\sum_{i=1}^N\hat{\gamma}_i|/\hat{\tilde{v}}_j$ ($\hat{\mathcal{Q}}^{[1]}:=  \mbox{max}_{1\leq j\leq N}(Tb_j^3)^{1/2} |\hat{\gamma}_j^{[1]}-\frac{1}{N}\sum_{i=1}^N\hat{\gamma}_i^{[1]}|/\hat{\tilde{v}}_j^{[1]}$), where $\hat{\tilde{v}}_j=\left(\hat{v}_j^2\left(1-\frac{1}{N}\right)^2+\frac{1}{N^2}\sum_{i\neq j}^N\hat{v}_i^2\right)^{1/2}$ ($\hat{\tilde{v}}_j^{[1]}=\left(\hat{v}_j^{[1]2}\left(1-\frac{1}{N}\right)^2+\frac{1}{N^2}\sum_{i\neq j}^N\hat{v}_i^{[1]2}\right)^{1/2}$).
    \State Reject the null hypothesis if $\hat{\mathcal{Q}}( \hat{\mathcal{Q}}^{[1]})>q_{\alpha}$, where $q_{\alpha}$ is the $(1-\alpha)$ quantile of the Gaussian random variable $\max_{1\leq j\leq N}|Z_j|$ with $Z_j$'s being independent standard normal variables.
	\end{algorithmic} 
\end{algorithm}

Similarly for the algorithm below in derivative, we shall replace $\hat{\gamma}_j(c_i)$ by $\hat{\gamma}_j^{[1]}(c_i)$, $\hat{v}_j^2(c_i)$ by $\hat{v}_j^{[1]2}(c_i)$ and $Tb_j$ by $Tb_j^3$. The newly estimated threshold estimator can be denoted as $\hat{c}_j^{[1]}$. 

{\normalsize 
\begin{algorithm}[H]
{
	\caption{Testing for existence of jumps: i.e., $\gamma_j =0$ for all $j$ {with an unknown threshold}. }\label{algunknown}
	\begin{algorithmic}[1]
		\For {$j=1,\ldots,N$}
			\State For a given grid of $[c_1, c_2, \cdots, c_K]$, estimate $\gamma_j(c_i)$, for each threshold location $c_i$, ($i=1,\cdots, K$), using equation (\ref{estimate}).
			\State Estimate the variance $v_j^2(c_i)$ of $(Tb_j)^{1/2}\hat{\gamma}_j(c_i)$ by $\hat{v}_j^2(c_i)=Tb_j\sum_{t=1}^{T}\left(w_{jt,b}^{+}(c_i)-w_{jt,b}^-(c_i)\right)^2\hat{\sigma}_{e,j}^2(c_i)$ for each $i$, where $\hat{\sigma}_{e,j}^2(c_i)$ is estimated by (\ref{eq:sigmaej2_unknown}).
		\EndFor

      \State Calculate the feasible test statistic $\hat{\mathcal{I}}^C:=\max_{1\leq i\leq K}\max_{1\leq j\leq N}(Tb_{j})^{1/2}|\hat{\gamma}_{j}(c_i)/\hat{v}_{j}(c_i)|$.
    \State Reject the null hypothesis if $\hat{\mathcal{I}}^C>q^{C}_{\alpha}$, where $q^{C}_{\alpha}$ is the $(1-\alpha)$ quantile of the Gaussian random variable $\max_{1\leq j\leq N}\max_{1\leq l\leq K}|Z_{jl}|$ with $Z_{jl}$'s being independent standard normal variables.

    \State Return the $\hat{c}_{j}$ which corresponds to the maximum absolute value of $\max_{1\leq i\leq K}(Tb_{j})^{1/2}|\hat{\gamma}_{j}(c_i)/\hat{v}_{j}(c_i)|$.
	\end{algorithmic} }
\end{algorithm}}

We also provide an algorithm for the \emph{dependence-aware} version of the existence test referenced in Remark~\ref{rmk:rmkvj1j2}. It relies on the same statistic $\hat{\mathcal I}$ as
Algorithm~\ref{alg1}, but draws critical values from a correlated Gaussian
maximum whose correlation matrix is estimated from the localized residuals and
regularized by Ledoit--Wolf shrinkage toward the identity.

\begin{algorithm}[H]
	\caption{Dependence-aware testing for existence of jumps in conditional means: i.e., $\gamma_j=0$ for all $j$.}\label{algdepaware}
	\begin{algorithmic}[1]
		\For {$j=1,\ldots,N$}
			\State Estimate $\gamma_j$ for each group/individual $j$ using equation~\eqref{estimate}, and compute the localized residuals $\hat{e}_{jt}$ as in the test procedure (see~\eqref{eq:sigmaej2}).
			\State Estimate the variance $\hat{v}_j^2=Tb_j\sum_{t=1}^{T}\left(w_{jt,b}^{+}-w_{jt,b}^-\right)^2\hat{\sigma}_{e,j}^2$, where $\hat{\sigma}_{e,j}^2$ is defined in~\eqref{eq:sigmaej2}.
		\EndFor
		\State Form the estimated correlation matrix $\widetilde{R}=(\widetilde{R}_{ij})_{1\leq i,j\leq N}$ with
		\begin{equation*}
		\widetilde{R}_{ij}=\frac{\sum_{t=1}^{T}\left(w_{it,b}^{+}-w_{it,b}^{-}\right)\left(w_{jt,b}^{+}-w_{jt,b}^{-}\right)\hat{e}_{it}\hat{e}_{jt}}{\hat{v}_{i}\hat{v}_j},
		\end{equation*}
		so that $\widetilde{R}_{jj}=1$.
		\State Regularize $\widetilde{R}$ by the Ledoit--Wolf shrinkage estimator \citep{ledoit2004well} toward the identity, $\hat{R}=(1-\hat{\rho})\widetilde{R}+\hat{\rho}\,I_N$, where $\hat{\rho}\in[0,1]$ is the data-driven shrinkage intensity.
		\State Calculate the feasible test statistic $\hat{\mathcal{I}}:=\max_{1\leq j\leq N}(Tb_j)^{1/2}|\hat{\gamma}_j|/\hat{v}_j$.
		\State For $b=1,\ldots,B$, draw $Z_b\sim N(0,\hat{R})$ independently and compute $|Z_b|_\infty=\max_{1\leq j\leq N}|Z_{b,j}|$. Let $\hat{q}_\alpha^{\,\mathrm{dep}}$ be the empirical $(1-\alpha)$ quantile of $\{|Z_b|_\infty\}_{b=1}^B$.
		\State Reject the null hypothesis if $\hat{\mathcal{I}}>\hat{q}_\alpha^{\,\mathrm{dep}}$.
	\end{algorithmic}
\end{algorithm}

\section{Additional Simulation Results}
\label{sec:simulation_appendix}

\setcounter{table}{0}
\renewcommand{\thetable}{A\arabic{table}}

\begin{table}[tbp]
\centering
\begin{tabular}{rrr|rrr|rrr}
\hline
\multicolumn{3}{r}{} & \multicolumn{3}{c}{Size} & \multicolumn{3}{c}{Power}
\\ 
& $N$ & $T$ & $\alpha=0.1$ & $\alpha=0.05$ & $\alpha=0.01$ & $\alpha=0.1$ & $%
\alpha=0.05$ & $\alpha=0.01$ \\ \hline
\multirow{6}{*}{DGP1} & 10 & 200 & 0.114 & 0.070 & 0.010 & 0.315 & 0.226 & 
0.092 \\ 
& 10 & 400 & 0.114 & 0.065 & 0.014 & 0.374 & 0.267 & 0.118 \\ 
& 10 & 800 & 0.119 & 0.056 & 0.010 & 0.416 & 0.286 & 0.113 \\ 
& 100 & 200 & 0.104 & 0.058 & 0.018 & 0.845 & 0.741 & 0.519 \\ 
& 100 & 400 & 0.118 & 0.062 & 0.014 & 0.897 & 0.824 & 0.603 \\ 
& 100 & 800 & 0.120 & 0.062 & 0.013 & 0.933 & 0.882 & 0.709 \\ \hline
\multirow{6}{*}{DGP2} & 10 & 200 & 0.115 & 0.067 & 0.011 & 0.440 & 0.329 & 
0.160 \\ 
& 10 & 400 & 0.111 & 0.059 & 0.013 & 0.483 & 0.382 & 0.188 \\ 
& 10 & 800 & 0.107 & 0.059 & 0.010 & 0.563 & 0.445 & 0.272 \\ 
& 100 & 200 & 0.111 & 0.052 & 0.012 & 0.954 & 0.922 & 0.816 \\ 
& 100 & 400 & 0.104 & 0.051 & 0.013 & 0.984 & 0.960 & 0.898 \\ 
& 100 & 800 & 0.115 & 0.055 & 0.009 & 0.995 & 0.992 & 0.951 \\ \hline
\multirow{6}{*}{DGP3} & 10 & 200 & 0.115 & 0.070 & 0.012 & 0.315 & 0.216 & 
0.081 \\ 
& 10 & 400 & 0.123 & 0.074 & 0.012 & 0.353 & 0.242 & 0.094 \\ 
& 10 & 800 & 0.113 & 0.059 & 0.015 & 0.386 & 0.269 & 0.122 \\ 
& 100 & 200 & 0.125 & 0.054 & 0.011 & 0.801 & 0.712 & 0.498 \\ 
& 100 & 400 & 0.123 & 0.060 & 0.009 & 0.854 & 0.787 & 0.572 \\ 
& 100 & 800 & 0.128 & 0.058 & 0.009 & 0.923 & 0.873 & 0.701 \\ \hline
\end{tabular}%
\caption{Empirical size and power results for the uniform test of
heterogeneity for DGP1--3.}
\label{table:sim2}
\end{table}

\begin{itemize}
\item[DGP4:] \emph{Strong cross-sectional dependence}. The setting is
similar to that in DGP3, but now the factor term in the running variable and
error dominate the idiosyncratic part via a non-zero mean and a relatively
larger variance, 
\begin{align*}
\epsilon _{jt}& =\left( \lambda _{\epsilon ,j}+2\right) f_{\epsilon
,t}+u_{\epsilon ,jt}/8, \\
X_{jt}& =\frac{1}{4}\left( (\lambda _{x,j}+2)f_{x,t}+u_{x,jt}/8\right) .
\end{align*}

\item[DGP5:] \emph{Semi-strong cross-sectional dependence}. The setting is
similar to that in DGP4, with some more weight put on the idiosyncratic part
of the error term and the running variable, 
\begin{align*}
\epsilon _{jt}& =\left( \lambda _{\epsilon ,j}+2\right) f_{\epsilon
,t}+u_{\epsilon ,jt}/4 \\
X_{jt}& =\frac{1}{4}\left( (\lambda _{x,j}+2)f_{x,t}+u_{x,jt}/4\right) .
\end{align*}

\item[DGP6:] \emph{Semi-strong cross-sectional dependence and time effect}.
The setting is similar to that in DGP5, with an additional time effect for
the error term, 
\begin{equation*}
\epsilon _{jt}=\left( \lambda _{\epsilon ,j}+2\right) f_{\epsilon ,t}+\nu
_{t}+u_{\epsilon ,jt}/4,
\end{equation*}%
where $\nu _{t}\sim i.i.d.\ N(0,0.25)$.
\end{itemize}

\begin{table}[tbp]
\centering
\begin{tabular}{rrr|rrr|rrr}
\hline
\multicolumn{3}{r}{} & \multicolumn{3}{c}{Size} & \multicolumn{3}{c}{Power}
\\ 
& $N$ & $T$ & $\alpha=0.1$ & $\alpha=0.05$ & $\alpha=0.01$ & $\alpha=0.1$ & $%
\alpha=0.05$ & $\alpha=0.01$ \\ \hline
\multirow{6}{*}{DGP4} & 10 & 200 & 0.104 & 0.051 & 0.012 & 0.311 & 0.246 & 
0.157 \\ 
& 10 & 400 & 0.100 & 0.045 & 0.008 & 0.334 & 0.273 & 0.193 \\ 
& 10 & 800 & 0.098 & 0.048 & 0.014 & 0.380 & 0.314 & 0.209 \\ 
& 100 & 200 & 0.084 & 0.042 & 0.006 & 0.755 & 0.711 & 0.625 \\ 
& 100 & 400 & 0.079 & 0.039 & 0.009 & 0.798 & 0.757 & 0.671 \\ 
& 100 & 800 & 0.090 & 0.047 & 0.014 & 0.828 & 0.792 & 0.728 \\ \hline
\multirow{6}{*}{DGP5} & 10 & 200 & 0.108 & 0.058 & 0.010 & 0.320 & 0.240 & 
0.141 \\ 
& 10 & 400 & 0.114 & 0.052 & 0.018 & 0.368 & 0.289 & 0.199 \\ 
& 10 & 800 & 0.100 & 0.051 & 0.011 & 0.397 & 0.312 & 0.199 \\ 
& 100 & 200 & 0.107 & 0.053 & 0.012 & 0.753 & 0.702 & 0.581 \\ 
& 100 & 400 & 0.106 & 0.056 & 0.009 & 0.816 & 0.780 & 0.685 \\ 
& 100 & 800 & 0.102 & 0.058 & 0.009 & 0.858 & 0.807 & 0.713 \\ \hline
\multirow{6}{*}{DGP6} & 10 & 200 & 0.100 & 0.056 & 0.017 & 0.216 & 0.137 & 
0.053 \\ 
& 10 & 400 & 0.119 & 0.059 & 0.015 & 0.259 & 0.174 & 0.068 \\ 
& 10 & 800 & 0.108 & 0.060 & 0.009 & 0.270 & 0.185 & 0.076 \\ 
& 100 & 200 & 0.122 & 0.071 & 0.013 & 0.477 & 0.391 & 0.229 \\ 
& 100 & 400 & 0.110 & 0.056 & 0.010 & 0.564 & 0.468 & 0.283 \\ 
& 100 & 800 & 0.117 & 0.049 & 0.007 & 0.647 & 0.546 & 0.376 \\ \hline
\end{tabular}%
\caption{Empirical results on the size and power of the uniform test for the existence of jump effects for DGP4--6.}
\label{table:sim3}
\end{table}

Apart from the size and power properties of our uniform testing procedure, we also check the estimation accuracy of $\hat{c}_j$ ( estimator of the unknown threshold location $c_{0j}$). For this purpose, we set all $\gamma _{j}=T^{-2/5}(\mathrm{log}N)^{1/2}5B_{j}$, where $B_{j}\sim i.i.d.\ U[2,10]$. We consider a grid of possible threshold locations, i.e. $[-0.3,-0.29,\ldots,0.3]$. The results in Table \ref{table:sim5} reveal that the maximum and averaged error over $j$ can be reduced effectively with increasing $T$.

\begin{table}[H]
\centering
\begin{tabular}{rrr|rr}
\hline
& $N$ & $T$ & $1/N\sum_j|\hat{c}_j-c_{0j}|$ & \textbf{$\max_j|\hat{c}_j-c_{0j}|$} \\[1mm] \hline
\multirow{6}{*}{DGP1} & 10 & 200 & 0.0241 & 0.1399 \\ 
& 10 & 400 & 0.0140 & 0.0979 \\ 
& 10 & 800 & 0.0087 & 0.0668 \\
& 10 & 1600 & 0.0063 & 0.0528 \\ 
& 50 & 200 & 0.0156 & 0.2260 \\ 
& 50 & 400 & 0.0069 & 0.1520 \\ 
& 50 & 800 & 0.0035 & 0.1008 \\
& 50 & 1600 & 0.0021 & 0.0679 \\ \hline
\multirow{8}{*}{DGP2} & 10 & 200 & 0.0085 & 0.0635 \\ 
& 10 & 400 & 0.0058 & 0.0486 \\ 
& 10 & 800 & 0.0038 & 0.0332 \\
& 10 & 1600 & 0.0029 & 0.0272 \\
& 50 & 200 & 0.0048 & 0.1177 \\ 
& 50 & 400 & 0.0024 & 0.0811 \\ 
& 50 & 800 & 0.0015 & 0.0564 \\
& 50 & 1600 & 0.0010 & 0.0381 \\ 
\hline
\end{tabular}%
\caption{Estimation accuracy of the $\hat{c}_j$'s for DGP1--2.}
\label{table:sim5}
\end{table}

In the following, we compare our uniform testing procedure to that of a pooled parametric threshold model. To be precise, as a comparison we assume the following linear panel threshold model \citep{hansen2000sample},
\begin{align*}
    Y_{jt}=\beta_0+\beta_1X_{jt}+\mathbf{1}_{\{X_{jt}\geq c_{0}\}}\left(\beta_{0}^{+}+\beta_{1}^{+}X_{jt}\right)+e_{jt}.
\end{align*}
The parameter of interest is $\gamma:=\beta_{0}^{+}+\beta_{1}^{+}c_0$ and we test the null hypothesis $H_0:\gamma=0$. We consider sup-Wald type statistics discussed by \cite{andrews1993tests}.
For this comparison, we consider the following linear data generating process.
\begin{enumerate}
\item[DGP7:] \emph{Linear model}. The smooth part of the conditional mean function is now linear.
\begin{align*}
Y_{jt}=0.2X_{jt}+\gamma_j\mathbf{1}_{\{X_{jt}\geq c_{0j}\}}+\sigma(X_{jt},U_{jt})\epsilon_{jt}.
\end{align*}
The generation of $X_{jt}$, $U_{jt}$, and $\sigma(\cdot,\cdot)$ is identical to DGP3. A proportion between $0$ and $1$ of the coefficients $\gamma_j$ are drawn from a standard normal distribution, the remaining coefficients are set to zero.
\end{enumerate}

The comparison results are presented in Table \ref{table:sim6}. The results confirm that our uniform testing procedure has similar size properties as a parametric test but has much larger power under the alternative. This is true for the sparse case with a proportion of only $10\%$ non-zero signals, as well as for the relatively dense case with a proportion of $20\%$. The reasons for this are signal-dilution on the one hand and signal cancellation on the other hand. Regarding the former, our uniform testing procedure is robust to settings with sparse signals, in which a pooled test suffers from dilution of signals. And regarding the latter, our uniform testing procedure does not have the problem of signal cancellation in case the signals for different $j$'s have opposite signs and cancel each other out.

\begin{table}[tbp]
\centering
\begin{tabular}{lrr|rrr|rrr}
\hline
\multicolumn{3}{r}{} & \multicolumn{3}{c}{Uniform test} & \multicolumn{3}{c}{Pooled parametric}
\\ 
Proportion & $N$ & $T$ & $\alpha=0.1$ & $\alpha=0.05$ & $\alpha=0.01$ & $\alpha=0.1$ & $%
\alpha=0.05$ & $\alpha=0.01$ \\ \hline
\multirow{4}{*}{0\%} & 50 & 400 & 0.085 & 0.036 & 0.007 & 0.126 & 0.080 & 0.017 \\
& 50 & 800 & 0.117 & 0.054 & 0.013 & 0.106 & 0.058 & 0.011 \\ 
& 100 & 400 & 0.123 & 0.055 & 0.014 & 0.116 & 0.067 & 0.020 \\
 & 100 & 800 & 0.106 & 0.060 & 0.011 & 0.113 & 0.055 & 0.019 \\   \hline
\multirow{6}{*}{10\%} & 50 & 400 & 0.451 & 0.368 & 0.249 & 0.251 & 0.158 & 0.066 \\
& 50 & 800 & 0.743 & 0.682 & 0.576 & 0.300 & 0.215 & 0.115 \\ 
& 100 & 400 & 0.608 & 0.526 & 0.380 & 0.216 & 0.128 & 0.051 \\
& 100 & 800 & 0.894 & 0.858 & 0.753 & 0.307 & 0.231 & 0.120 \\   \hline
\multirow{6}{*}{20\%} & 50 & 400 & 0.660 & 0.583 & 0.430 & 0.325 & 0.225 & 0.122 \\
& 50 & 800 & 0.896 & 0.865 & 0.780 & 0.392 & 0.316 & 0.210 \\ 
& 100 & 400 & 0.842 & 0.792 & 0.661 & 0.298 & 0.225 & 0.110 \\
& 100 & 800 & 0.983 & 0.967 & 0.938 & 0.407 & 0.327 & 0.211 \\   \hline
\end{tabular}%
\caption{Empirical size and power results for the uniform test of the existence of
jump effects under unknown threshold locations for DGP7. Comparison with a sup-Wald test based on a linear panel threshold model.}
\label{table:sim6}
\end{table}

We obtain similar results in the comparison with a pooled nonparametric test in Table \ref{table:sim7}. The results are based on DGP3 and the non-zero coefficients are generated from a standard normal distribution. For the comparison, we pool the observations along the individual dimension. Again, the uniform testing procedure is more powerful than the pooled testing procedure in all settings with existence of some non-zero coefficients.

\begin{table}[tbp]
\centering
\begin{tabular}{lrr|rrr|rrr}
\hline
\multicolumn{3}{r}{} & \multicolumn{3}{c}{Uniform test} & \multicolumn{3}{c}{Pooled nonparametric}
\\ 
Proportion & $N$ & $T$ & $\alpha=0.1$ & $\alpha=0.05$ & $\alpha=0.01$ & $\alpha=0.1$ & $%
\alpha=0.05$ & $\alpha=0.01$ \\ \hline
\multirow{4}{*}{0\%} & 50 & 400 & 0.120 & 0.061 & 0.012 & 0.099 & 0.054 & 0.009 \\
& 50 & 800 & 0.117 & 0.061 & 0.011 & 0.083 & 0.044 & 0.004 \\
& 100 & 400 & 0.122 & 0.067 & 0.019 & 0.094 & 0.050 & 0.011 \\
& 100 & 800 & 0.117 & 0.053 & 0.018 & 0.106 & 0.055 & 0.014 \\   \hline
\multirow{6}{*}{10\%} & 50 & 400 & 0.460 & 0.376 & 0.235 & 0.124 & 0.072 & 0.013 \\
& 50 & 800 & 0.693 & 0.633 & 0.522 & 0.144 & 0.076 & 0.029 \\
& 100 & 400 & 0.651 & 0.569 & 0.414 & 0.140 & 0.082 & 0.021 \\
& 100 & 800 & 0.862 & 0.823 & 0.729 & 0.157 & 0.095 & 0.028 \\   \hline
\multirow{6}{*}{20\%} & 50 & 400 & 0.653 & 0.575 & 0.441 & 0.148 & 0.086 & 0.026 \\
& 50 & 800 & 0.915 & 0.885 & 0.790 & 0.206 & 0.144 & 0.061 \\
& 100 & 400 & 0.807 & 0.737 & 0.591 & 0.144 & 0.097 & 0.027 \\
& 100 & 800 & 0.983 & 0.971 & 0.932 & 0.215  & 0.124 & 0.055 \\   \hline
\end{tabular}%
\caption{Empirical size and power results for the uniform test of the existence of
jump effects under unknown threshold locations for DGP3. Comparison with pooled nonparametric test.}
\label{table:sim7}
\end{table}

We further examine the size and power properties of the test of existence of a threshold effect in the first derivative of the conditional expectations. It is a well known fact in nonparametric regression that estimating derivatives is a harder problem than estimating the regression function itself. For this reason, we restrict our simulations to the known threshold case and we only consider fixed alternatives. Under the null, we set $f_j(x,u)=5x$, whereas we consider $f_j(x,u)=5|x|$ under the alternative hypothesis. The share of individuals' DGPs generated under the alternative is $20\%$ for both $N=10$ and $N=100$. {For this setting, we oversmooth the bandwidth.} The results for DGP1--DGP3 are presented in Table \ref{table:sim8}. The size properties are reasonably close to the nominal size. Similarly, the power approaches one with increasing sample size and dimension.

\begin{table}[tbp]
\centering
\begin{tabular}{rrr|rrr|rrr}
\hline
\multicolumn{3}{r}{} & \multicolumn{3}{c}{Size} & \multicolumn{3}{c}{Power}
\\ 
& $N$ & $T$ & $\alpha=0.1$ & $\alpha=0.05$ & $\alpha=0.01$ & $\alpha=0.1$ & $%
\alpha=0.05$ & $\alpha=0.01$ \\ \hline
\multirow{4}{*}{DGP1} 
& 10 & 400 & 0.102 & 0.052 & 0.007 & 0.923 & 0.831 & 0.527 \\ 
& 10 & 800 & 0.113 & 0.046 & 0.011 & 0.998 & 0.995 & 0.941 \\ 
& 100 & 400 & 0.120 & 0.058 & 0.008 & 0.999 & 0.998 & 0.909 \\ 
& 100 & 800 & 0.115 & 0.060 & 0.010 & 1.000 & 1.000 & 1.000 \\    \hline
\multirow{4}{*}{DGP2} 
& 10 & 400 & 0.109 & 0.050 & 0.014 & 0.718 & 0.590 & 0.327 \\ 
& 10 & 800 & 0.096 & 0.048 & 0.009 & 0.956 & 0.915 & 0.716 \\ 
& 100 & 400 & 0.100 & 0.040 & 0.009 & 0.978 & 0.935 & 0.710 \\ 
& 100 & 800 & 0.138 & 0.054 & 0.012 & 1.000 & 1.000 & 0.997 \\  \hline
\multirow{4}{*}{DGP3} 
& 10 & 400 & 0.108 & 0.064 & 0.014 & 0.710 & 0.564 & 0.303 \\ 
& 10 & 800 & 0.128 & 0.067 & 0.013 & 0.946 & 0.903 & 0.703 \\ 
& 100 & 400 & 0.130 & 0.054 & 0.010 & 0.988 & 0.939 & 0.693 \\ 
& 100 & 800 & 0.126 & 0.057 & 0.016 & 1.000 & 1.000 & 0.999 \\ \hline
\end{tabular}%
\caption{Empirical size and power results for the uniform test of the existence of
derivative jump effects under known threshold locations for DGP1--3.}
\label{table:sim8}
\end{table}

In the following, we study the finite sample performance of our testing procedure under consideration of the higher order terms in the covariance matrix. Instead of calculating critical values analytically as in the previous simulation settings, we simulate $N$-dimensional random variables $Z_b\sim N(0,\hat{R})$, and calculate $|Z_b|_{\infty}$ for $b=1,\ldots,B$, where $\hat{R}$ is the shrinkage version of the estimator of \citet{ledoit2004well} of the following estimate of the correlation matrix
\begin{align*}
    \widetilde{R}_{ij}=\frac{\sum_{t=1}^{T}\left(w_{it,b}^{+}-w_{it,b}^{-}\right)\left(w_{jt,b}^{+}-w_{jt,b}^{-}\right)\hat{e}_{it}\hat{e}_{jt}}{\hat{v}_{i}\hat{v}_j}.
\end{align*}
We consider the following DGP:
\begin{enumerate}
\item[DGP8:] \emph{Higher-order term DGP}.
The setting is
similar to that in DGP3, but now the factor loadings for both the generation $X_{jt}$ and $\epsilon_{jt}$ are identical for all $j$,
\begin{align*}
\epsilon _{jt}& = 2f_{\epsilon
,t}+u_{\epsilon ,jt}/10, \\
X_{jt}& =f_{x,t}/2+u_{x,jt}/40.
\end{align*}
\end{enumerate}
We compare the size and power properties to our testing procedure without estimated correlation matrix, and to the Benjamini--Hochberg approach for multiple testing. The simulation results in Table~\ref{table:sim9} show that both comparison procedures are indeed too conservative, in particular in settings with $N=100$. In contrast, our correlation-adjusted testing procedures have excellent size properties. Under the alternative, we observe that accounting for the higher order terms can indeed lead to a substantial increase in power compared to the procedures that do not take dependence into account.

\begin{table}[tbp]
\centering
\begin{tabular}{rrr|rrr|rrr}
\hline
\multicolumn{3}{r}{} & \multicolumn{3}{c}{Size} & \multicolumn{3}{c}{Power}
\\ 
& $N$ & $T$ & $\alpha=0.1$ & $\alpha=0.05$ & $\alpha=0.01$ & $\alpha=0.1$ & $%
\alpha=0.05$ & $\alpha=0.01$ \\ \hline
\multirow{6}{*}{IND} & 10 & 200 & 0.087 & 0.048 & 0.011 & 0.170 & 0.104 & 0.039 \\ 
& 10 & 400 & 0.087 & 0.047 & 0.010 & 0.178 & 0.102 & 0.023 \\ 
& 10 & 800 & 0.081 & 0.043 & 0.005 & 0.196 & 0.102 & 0.047 \\ 
& 100 & 200 & 0.066 & 0.037 & 0.011 & 0.223 & 0.144 & 0.060 \\ 
& 100 & 400 & 0.054 & 0.024 & 0.008 & 0.294 & 0.222 & 0.095 \\ 
& 100 & 800 & 0.054 & 0.031 & 0.011 & 0.326 & 0.241 & 0.118 \\  \hline
\multirow{6}{*}{COR} & 10 & 200 & 0.118 & 0.053 & 0.013 & 0.215 & 0.131 & 0.042 \\ 
& 10 & 400 & 0.119 & 0.058 & 0.010 & 0.220 & 0.122 & 0.032 \\ 
& 10 & 800 & 0.106 & 0.055 & 0.006 & 0.200 & 0.108 & 0.048 \\ 
& 100 & 200 & 0.100 & 0.050 & 0.006 & 0.239 & 0.152 & 0.068 \\ 
& 100 & 400 & 0.101 & 0.043 & 0.009 & 0.311 & 0.236 & 0.100 \\ 
& 100 & 800 & 0.115 & 0.059 & 0.012 & 0.485 & 0.342 & 0.015 \\  \hline
\multirow{6}{*}{BH} & 10 & 200 & 0.085 & 0.050 & 0.014 & 0.180 & 0.112 & 0.042 \\ 
& 10 & 400 & 0.091 & 0.052 & 0.010 & 0.185 & 0.107 & 0.024 \\ 
& 10 & 800 & 0.090 & 0.049 & 0.005 & 0.241 & 0.134 & 0.049 \\ 
& 100 & 200 & 0.077 & 0.046 & 0.011 & 0.329 & 0.228 & 0.078 \\ 
& 100 & 400 & 0.067 & 0.030 & 0.008 & 0.436 & 0.303 & 0.118 \\ 
& 100 & 800 & 0.065 & 0.037 & 0.012 & 0.349 & 0.253 & 0.124 \\  \hline
\end{tabular}%
\caption{Empirical size and power results of the uniform test for the existence of jump effects for DGP8. The table reports the results based on i.i.d.\ Gaussian critical values (IND), correlation-adjusted critical values (COR), and the Benjamini-Hochberg procedure (BH).}
\label{table:sim9}
\end{table}

As a final check, we show that our results are robust to different specifications for $h_j$. While the previous results all relate to DGP's with additive formulations, we now consider a DGP that exhibits interaction effects, $h_j(x,u)=\cos(x)(1+2\sin^2(u))$. The results in Table~\ref{table:sim10} show that our testing procedure is indeed robust to different specifications.

\begin{table}[tbp]
\centering
\begin{tabular}{rrr|rrr|rrr}
\hline
\multicolumn{3}{r}{} & \multicolumn{3}{c}{Size} & \multicolumn{3}{c}{Power}
\\ 
& $N$ & $T$ & $\alpha=0.1$ & $\alpha=0.05$ & $\alpha=0.01$ & $\alpha=0.1$ & $%
\alpha=0.05$ & $\alpha=0.01$ \\ \hline
\multirow{6}{*}{DGP1} & 10 & 200 & 0.113 & 0.059 & 0.009 & 0.397 & 0.301 & 0.134 \\ 
& 10 & 400 & 0.107 & 0.053 & 0.010 & 0.425 & 0.324 & 0.164 \\ 
& 10 & 800 & 0.114 & 0.060 & 0.012 & 0.485 & 0.376 & 0.202 \\ 
& 100 & 200 & 0.101 & 0.050 & 0.014 & 0.881 & 0.805 & 0.611 \\ 
& 100 & 400 & 0.133 & 0.068 & 0.011 & 0.940 & 0.893 & 0.719 \\ 
& 100 & 800 & 0.103 & 0.052 & 0.018 & 0.952 & 0.909 & 0.792 \\  \hline
\multirow{6}{*}{DGP2} & 10 & 200 & 0.111 & 0.054 & 0.008 & 0.523 & 0.407 & 0.246  \\ 
& 10 & 400 & 0.099 & 0.035 & 0.004 & 0.580 & 0.496 & 0.322 \\ 
& 10 & 800 & 0.097 & 0.048 & 0.008 & 0.638 & 0.549 & 0.390 \\ 
& 100 & 200 & 0.111 & 0.054 & 0.010 & 0.980 & 0.957 & 0.891 \\ 
& 100 & 400 & 0.119 & 0.059 & 0.007 & 0.993 & 0.988 & 0.955 \\ 
& 100 & 800 & 0.111 & 0.064 & 0.017 & 0.996 & 0.993 & 0.971 \\  \hline
\multirow{6}{*}{DGP3} & 10 & 200 & 0.116 & 0.056 & 0.013 & 0.380 & 0.257 & 0.115 \\ 
& 10 & 400 & 0.126 & 0.059 & 0.010 & 0.402 & 0.280 & 0.138 \\ 
& 10 & 800 & 0.116 & 0.050 & 0.016 & 0.468 & 0.359 & 0.181 \\ 
& 100 & 200 & 0.143 & 0.079 & 0.016 & 0.856 & 0.790 & 0.579 \\ 
& 100 & 400 & 0.116 & 0.057 & 0.015 & 0.915 & 0.855 & 0.707 \\ 
& 100 & 800 & 0.114 & 0.063 & 0.016 & 0.947 & 0.910 & 0.797 \\  \hline
\end{tabular}%
\caption{Empirical size and power results of the uniform test for the existence of jump effects for DGP1--3 with $h_j(x,u)=\cos(x)(1+2\sin^2(u))$.}
\label{table:sim10}
\end{table}

\section{Empirical Application}
\label{sec:application}

We apply our testing procedure to two empirical applications. First, we study possible jumps in the news impact curve for constituents of the S\&P500. As a second application, we study the heterogeneity across time and states of the incumbency effect in U.S. House elections.

\subsection{Application I: Jumps in the News Impact Curve}
\label{application1}

As a motivating example for our model, we explore an application involving stock lagged returns and volatility using threshold regression. Threshold regression, dating back to \cite{chan1985multiple} and more recently studied in works such as \cite{duffy2023stationarity}, is mainly formulated within a linear parametric setting. Our approach naturally extends this framework to the nonparametric setting. Specifically, in our first application, we revisit the news impact curve introduced by \cite{engle1993measuring}. In particular, we want to identify possible jumps in the relationship between news events in the form of past returns on the volatility of stocks. The application serves as an illustration of our uniform testing procedure when the threshold $c_{0j}$ is unknown. Originally proposed in the context of autoregressive volatility models (e.g., ARCH), \cite{engle1993measuring} and \cite{linton2005estimating} consider partially nonparametric specifications for the news impact curve which are able to capture possible asymmetries. A limitation of these specifications is the restriction to continuous functions. However, \cite{zakoian1994threshold} and \cite{fornari1997sign} advocate for (parametric) GARCH specifications that allow for discontinuities in the way past returns impact volatility. We therefore want to investigate this issue by testing for the presence of jumps. 

In our model setup, the dependent variable $Y_{jt}$ is the Garman-Klass volatility of stock $j$ at time $t$, and $X_{jt}$ is the lagged stock return.
We want to test the null hypothesis $H_0^{(1)}:\gamma_i=0$ for all $i\in[N]$ and all $c_{0j}$, versus the alternative $H_a^{(1)}:\gamma_i\neq0$ for some $i$ and some $c_{0j}$.  Our testing procedure is robust to the presence of unobserved risk factors, $U_{jt}$, which are more than likely to occur in our application. Our data includes the daily stock returns and volatility data of the S\&P 500 constituents from January 2010 to May 2024, which we obtained from Kaggle \footnote{https://www.kaggle.com/datasets/andrewmvd/sp-500-stocks.}. Therefore we have a fairly high-dimensional setting with $N=500$ and $T_j$ is the number of days of stock $j$ being a member of the stock index. See Figure \ref{figure:return_vola} for a visualization of the time series of returns and volatility for the stock of AT\&T. As candidates for the threshold location, we consider $c_{i}\in\{-0.01,-0.005,0,0.005,0.01\}$.

\setcounter{figure}{0}
\counterwithin{figure}{section}
\begin{figure}[tbp]
\centering
\includegraphics[width=.49\linewidth]{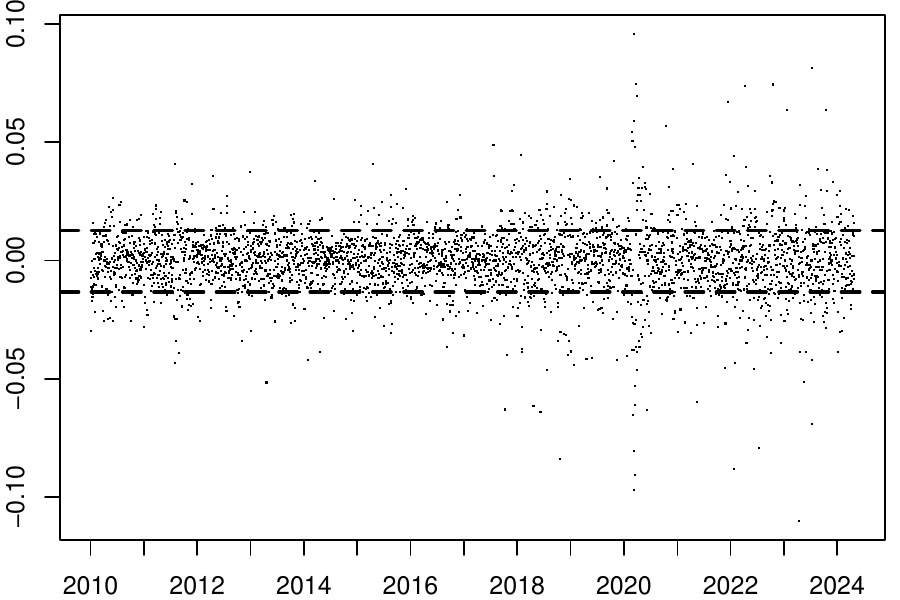}\hfill
\includegraphics[width=.49\linewidth]{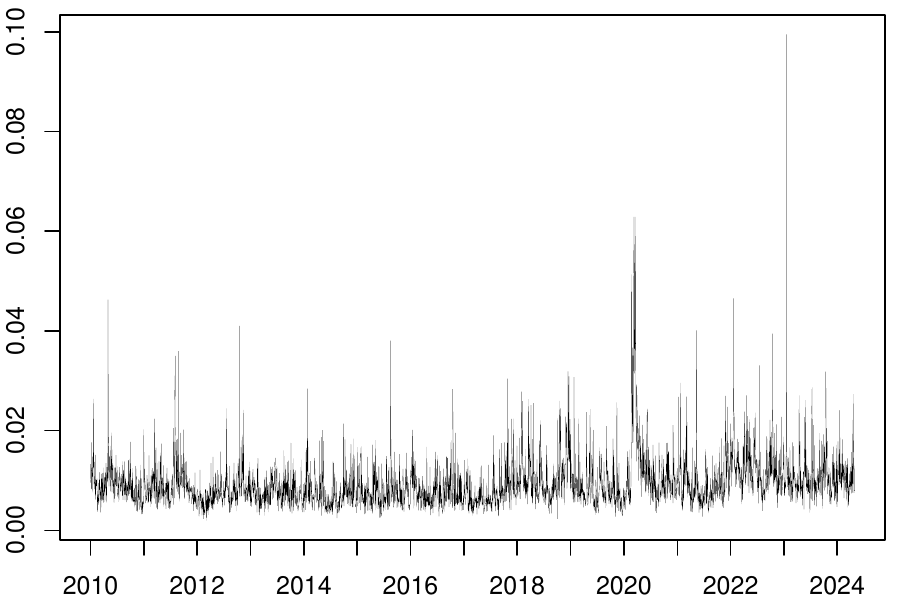}
\caption{Time series for daily stock returns (left panel) and daily volatility (right panel) for AT\&T. The dashed lines in the left panel indicate the $10\%$ and $90\%$ quantiles of the return distribution.}
\label{figure:return_vola}
\end{figure}

\begin{figure}[tbp]
\centering
\includegraphics[width=.49\linewidth]{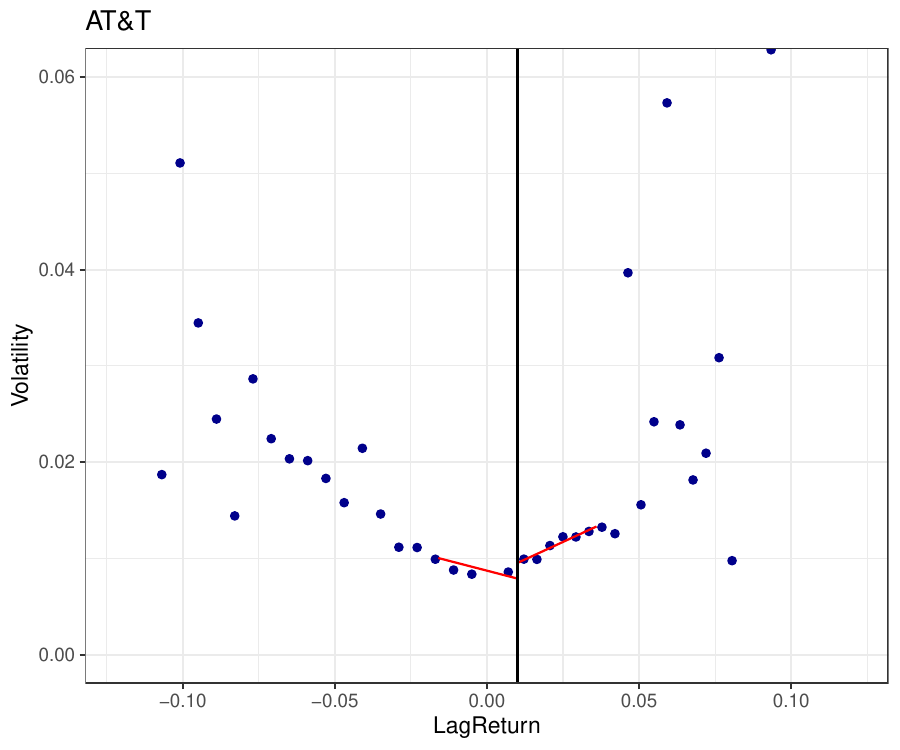}\hfill
\includegraphics[width=.49\linewidth]{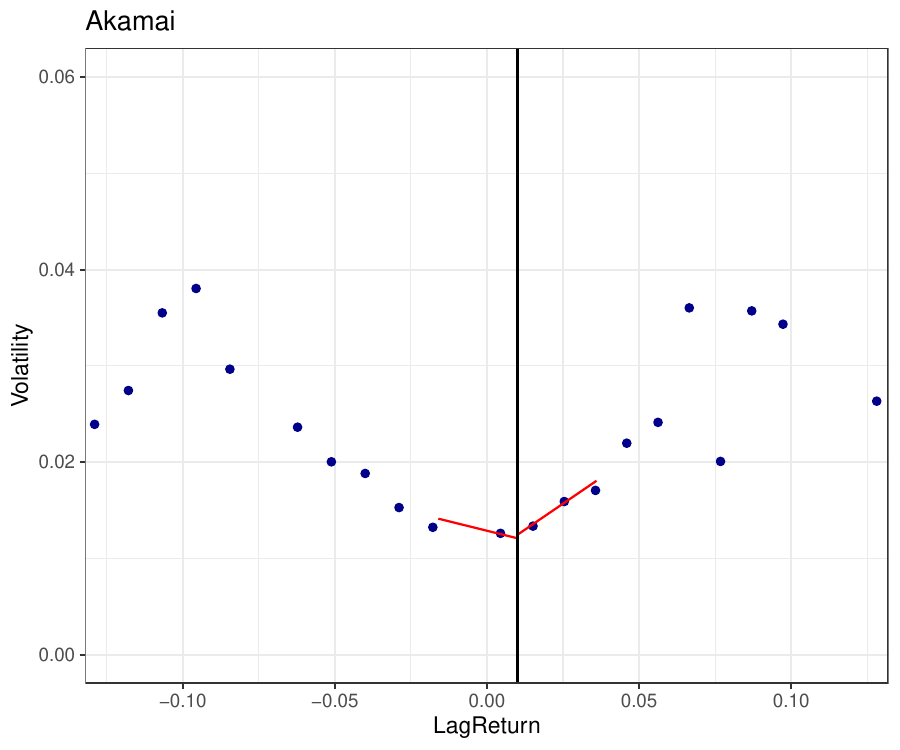}
\caption{Local linear fit of a significant threshold effect for AT\&T (left panel) and an insignificant effect for Akamai (right panel) at $1\%$ significance level}.
\label{figure:comparison_application}
\end{figure}

\begin{table}[tbp]
\centering%
\begin{tabular}{l|rrrr}
\hline
Symbol & $\hat{c}_{0j}$ & $100\cdot\hat{\gamma}_{j}(\hat{c}_{0j})$ & $100\cdot(T_{j}b_{j})^{-1/2}\hat{v}_j(\hat{c}_{0j})$ & $%
(T_jb_j)^{1/2}\hat{\gamma}_j(\hat{c}_{0j})/\hat{v}_j(\hat{c}_{0j})$  \\ \hline
ALL & --0.010 & --0.240 & 0.043 & --5.535 \\ 
  AXP & --0.010 & --0.223 & 0.048 & --4.702 \\ 
  CTAS & --0.010 & --0.189 & 0.041 & --4.632 \\ 
  ORLY & --0.010 & --0.211 & 0.044 & --4.750 \\ 
  SJM & --0.010 & --0.191 & 0.040 & --4.742 \\ 
  NXPI & --0.005 & --0.529 & 0.078 & --6.793 \\ 
  HON & 0.005 & 0.136 & 0.028 & 4.763 \\ 
  T & 0.010 & 0.162 & 0.033 & 4.857 \\ 
  MCK & 0.010 & 0.211 & 0.042 & 5.000 \\ 
  TRV & 0.010 & 0.156 & 0.031 & 4.978 \\ 
  WM & 0.010 & 0.181 & 0.034 & 5.290 \\ 
\hline
$\hat{\mathcal{I}}$ &  &  & & 6.793  \\ 
\hline
\end{tabular}%
\caption{Uniform test for the existence of jump effects. This table reports the selected threshold locations, the estimated
coefficients, the standard errors, and the corresponding individual test
statistic. Only the significant results at the $1\%$ level are reported. The last line reports the test statistic of the uniform
testing procedure, with critical values $4.610$ at $1\%$ level.}
\label{table:stocks}
\end{table}

The main findings are summarized in Table \ref{table:stocks}.
We find significant results for $11$ out of the $500$ stocks we consider. For stocks with significant threshold effects, the locations of the estimated thresholds vary, and none of the significant threshold effects occur precisely at zero. This contradicts the rationale of the sign-switching ARCH model of \cite{fornari1997sign}, which models a jump effect based on the sign of the lagged returns, and thus expects a significant threshold effect at zero. For $c_{i}=-0.01$, we get significant negative effects at the $1\%$ level for five stocks, including Allstate Corporation and American Express. On the other hand, we obtain significant positive effects for the $c_{i}=0.01$ threshold for four stocks, e.g.  AT\&T and McKesson. These results reaffirm the findings of \cite{chen2011news}, who find that both
bad and good news have an increasing effect on volatility. Additionally, we get another  significant negative coefficient at the $-0.005$ threshold location and another positively significant coefficient at $c_{i}=0.005$. Overall, we get evidence for the presence of jump effects in the news impact curve. Figure \ref{figure:comparison_application} shows a visualization of the significant positive threshold effect for AT\&T at $c_i=0.01$, and as a comparison we show the fit for Akamai for which we did not find a significant effect at any threshold location.

\subsection{Application II: Party Incumbency Effects on U.S. House Elections}\label{application2}

In the second application, we are interested in estimating the advantage a
candidate has in the U.S. House elections when the seat is currently
occupied by the party of the candidate\footnote{We use the data provided by the MIT Election Data and Science Lab. The data include the
vote shares of all candidates in the elections for the US House of
Representatives from 1976--2020. Due to redistricting at the start of each
decade, we have to exclude years ending with a `0' or `2'.}. This is called the party incumbency
effect. We follow the research design of \cite{lee2008randomized}. See %
\cite{caughey2011elections} for an analysis of more recent elections. The
dependent variable $Y_{jt}$ is the Democratic two-party vote share in year $%
j $ in district $t$. The covariate $X_{jt}$ is the difference in the
two-party vote share in the previous election. Since the winner of the
election is determined by the 0 threshold, our model setup is appropriate.
We want to test the one-sided null hypothesis $H_{0}^{(1)}:\gamma _{i}>0$ for all $%
i\in \lbrack N]$, versus the alternative hypothesis $H_{a}^{(1)} $: $\gamma
_{i}\leq 0$ for some $i$. Unlike \cite{lee2008randomized} who pools the
data together to estimate a single incumbency effect, we conduct a separate
analysis for different election years and also a separate analysis for
different states by interchanging the roles of $j$ and $t$ and allowing for
possible heterogeneity of incumbency effects along either the $j$ or $t$
dimension. In addition, it is easy to see our theories extend to the
unbalanced data in which case we can use $T_{j}$ to denote the number of
observations associated with individual/group $j$.

First, we are interested in the election year-specific effects. The election
year-specific results are reported in Table \ref{table:vote_new}. Our
analysis covers 13 election years, so $N=13$ and $T_{j}$ is equal to the
number of districts included in the sample of the respective election
year/group. For each $j,$ the effective number of observations is given by
the number of non-zero weights in the local linear estimation, which depends
on the bandwidth parameters $b_{j}$'s which are chosen as in the simulation
section. Table \ref{table:vote_new} reports the results for the effect
estimate ($\hat{\gamma}_{j}$)$,$ the standard
error ($(T_{j}b_{j})^{-1/2}\hat{v}_{j}$)$,$ the individual test statistic ($(T_{j}b_{j})^{1/2}%
\hat{\gamma}_{j}/\hat{v}_{j}$)$,$ the number of observations $T_{j}$ (Obs)
and the number of effective observations (EObs) as well. We find five years
with significant effects at the $1\%$ level and one year which is significant at the 
$5\%$ level by using the simulated critical value for $\mathcal{\hat{I}}$.
Since we are interested in the maximum of the test statistics in our uniform
testing procedure, the null hypothesis of no effect ($H_{0}^{\left( 1\right)
}$) can be rejected at the $1\%$ level. We can observe that the estimated
incumbency effect is stronger in the period from 1978 to 1998, while in the
most recent elections the effect is insignificant. However, the
sign of the estimated coefficients is positive for all election years. We
also run the test for the homogeneity of incumbency effects ($H_{0}^{\left(
2\right) }$) over the 13 election years, see Table \ref%
{table:vote_new_hetero}. We rely on the median ($\hat{\gamma}_{0.5}$) as a
more robust estimator of the average effect across election years. The test statistic is given by $\hat{\mathcal{%
Q}}=3.011$, which implies that we can reject the null hypothesis of
homogeneous effects at a 5\% level.

Now, we are interested in the state-specific effects. We restrict our
attention to states with a sufficient number of elections held in the
sample. We choose a threshold of 50 elections. This gives us $N=29$ states
in our analysis, and $T_{j}$ now denotes the number of elections included
for the $j$th state. We argue that this setting constitutes a fairly high-dimensional setting since the effective number of observations is smaller
than $N$ for a fairly large proportion of the states. The results are
reported in Table \ref{table:vote_new_states}. Consistent with the previous
results, almost all of the estimated coefficients are positive. By using the
critical values for the simultaneous tests, we find significant effects for
six states, for Missouri at $1\%$ level, and for
California, Iowa, Virginia, Alabama and South Carolina at $5\% $. Again, we
can reject the null hypothesis of no jump effect for our uniform testing
procedure at a $1\%$ level. The significant states tend to belong to those
with the largest number of observations. However, for the other two largest
states, New York and Texas, the estimated effect is rather small and far
from being significant. The incumbency effect seems to be stronger in some
states compared to others. See Figure \ref{figure:application_election} for a visualization of the estimated threshold effects for the states of California and Pennsylvania. We also run the test for the homogeneity of
incumbency effects ($H_{0}^{\left( 2\right) }$) over the 29 states. The
results are displayed in Table \ref{table:vote_new_states_bw_hetero} and the
test statistic is given by $\hat{\mathcal{Q}}=3.096$, which implies that we
can reject the null hypothesis of homogeneous effects at the $5\%$ level. We
find heterogeneity in the effects across election periods and states.

\begin{table}[H]
\centering%
\begin{tabular}{l|rrrrr}
\hline
Year & $\hat{\gamma }_{j}$ & $(T_{j}b_{j})^{-1/2}\hat{v}_{j}$ & $%
(T_{j}b_{j})^{1/2}\hat{\gamma }_{j}/\hat{v}_{j}$ & Obs & EObs \\ \hline
  1978 & 0.174 & 0.043 & \cellcolor{blue!50}4.103 & 412 & 127 \\ 
  1984 & 0.163 & 0.030 & \cellcolor{blue!50}5.515 & 413 & 154 \\ 
  1986 & 0.098 & 0.041 & \cellcolor{blue!10}2.418 & 408 & 86 \\ 
  1988 & 0.161 & 0.042 & \cellcolor{blue!50}3.830 & 395 & 79 \\ 
  1994 & 0.117 & 0.035 & \cellcolor{blue!50}3.325 & 387 & 102 \\ 
  1996 & 0.037 & 0.034 & 1.087 & 394 & 112 \\ 
  1998 & 0.112 & 0.035 & \cellcolor{blue!50}3.167 & 369 & 132 \\ 
  2004 & 0.045 & 0.066 & 0.676 & 359 & 65 \\ 
  2006 & 0.063 & 0.052 & 1.222 & 403 & 102 \\ 
  2008 & 0.063 & 0.029 & 2.179 & 405 & 142 \\ 
  2014 & 0.075 & 0.026 & \cellcolor{blue!25}2.912 & 380 & 103 \\ 
  2016 & 0.036 & 0.040 & 0.902 & 387 & 42 \\ 
  2018 & 0.037 & 0.032 & 1.171 & 403 & 37 \\ \hline
$\hat{\mathcal{I}}$ &  &  & \cellcolor{blue!50}5.515 &  &  \\ \hline
\end{tabular}%
\caption{Uniform test for the existence of jump effects. This table reports the
estimated coefficients, the standard errors, the corresponding individual
test statistic, the number of observations (Obs) and the effective number of
observations (EObs). The last line reports the test statistic of the uniform
testing procedure, where the critical values at $10\%$, $5\%$ and $1\%$
levels are $2.406$, $2.657$ and $3.166$, respectively.}
\label{table:vote_new}
\end{table}

\begin{table}[H]
\centering
\begin{tabular}{l|rrrrr}
\hline
Year & $\hat{\gamma}_j-\hat{\gamma}_{0.5}$ & $(T_jb_j)^{-1/2}\hat{\tilde{v}}%
_j$ & $(T_jb_j)^{1/2}(\hat{\gamma}_j-\hat{\gamma}_{0.5})/\hat{\tilde{v}}_j$ & 
Obs & EObs \\ \hline
  1978 & 0.100 & 0.041 & \cellcolor{blue!10}2.451 & 412 & 127 \\ 
  1984 & 0.089 & 0.029 & \cellcolor{blue!25}3.011 & 413 & 154 \\ 
  1986 & 0.024 & 0.039 & 0.606 & 408 & 86 \\ 
  1988 & 0.086 & 0.040 & 2.140 & 395 & 79 \\ 
  1994 & 0.042 & 0.034 & 1.229 & 387 & 102 \\ 
  1996 & --0.038 & 0.033 & --1.146 & 394 & 112 \\ 
  1998 & 0.037 & 0.034 & 1.078 & 369 & 132 \\ 
  2004 & --0.030 & 0.062 & --0.481 & 359 & 65 \\ 
  2006 & --0.011 & 0.049 & --0.234 & 403 & 102 \\ 
  2008 & --0.012 & 0.029 & --0.414 & 405 & 142 \\ 
  2014 & 0.000 & 0.026 & 0.000 & 380 & 103 \\ 
  2016 & --0.038 & 0.039 & --0.986 & 387 & 42 \\ 
  2018 & --0.038 & 0.031 & --1.198 & 403 & 37 \\ \hline
$\hat{\mathcal{Q}}$ &  &  & \cellcolor{blue!25}3.011
&  &  \\ \hline
\end{tabular}%
\caption{Uniform test for heterogeneity. This table reports the estimated
coefficients, the standard errors, the corresponding individual test
statistic, the number of observations and the effective number of
observations. The last line reports the test statistic of the uniform
testing procedure. Critical values at $10\%$, $5\%$ and $1\%$ level are $%
2.406$, $2.657$ and $3.166$.}
\label{table:vote_new_hetero}
\end{table}

\begin{table}[tbp]
\centering%
\begin{tabular}{ll|rrrrrr}
\hline
State & State\_ic & $\hat{\gamma }_{j}$ & $(T_{j}b_{j})^{-1/2}\hat{v}_j$ & $%
(T_jb_j)^{1/2}\hat{\gamma}_j/\hat{v}_j$ & Obs & EObs &  \\ \hline
  CONNECTICUT & 1 & 0.142 & 0.048 & \cellcolor{blue!10}2.949 & 72 & 41 \\ 
  MASSACHUSETTS & 3 & 0.251 & 0.184 & 1.362 & 129 & 22 \\ 
  NEW JERSEY & 12 & 0.092 & 0.047 & 1.953 & 171 & 55 \\ 
  NEW YORK & 13 & 0.053 & 0.026 & 2.032 & 387 & 153 \\ 
  PENNSYLVANIA & 14 & 0.096 & 0.036 & \cellcolor{blue!10}2.699 & 237 & 100 \\ 
  ILLINOIS & 21 & 0.114 & 0.043 & 2.658 & 250 & 85 \\ 
  INDIANA & 22 & 0.080 & 0.031 & 2.546 & 125 & 62 \\ 
  MICHIGAN & 23 & 0.067 & 0.032 & 2.086 & 206 & 79 \\ 
  OHIO & 24 & 0.133 & 0.047 & \cellcolor{blue!10}2.815 & 230 & 69 \\ 
  WISCONSIN & 25 & 0.128 & 0.087 & 1.472 & 97 & 31 \\ 
  IOWA & 31 & 0.125 & 0.038 & \cellcolor{blue!25}3.247 & 64 & 50 \\ 
  MISSOURI & 34 & 0.258 & 0.050 & \cellcolor{blue!50}5.117 & 115 & 37 \\ 
  VIRGINIA & 40 & 0.223 & 0.069 & \cellcolor{blue!25}3.209 & 102 & 54 \\ 
  ALABAMA & 41 & 0.212 & 0.068 & \cellcolor{blue!25}3.124 & 74 & 21 \\ 
  FLORIDA & 43 & 0.039 & 0.066 & 0.593 & 221 & 85 \\ 
  GEORGIA & 44 & 0.101 & 0.075 & 1.340 & 126 & 40 \\ 
  LOUISIANA & 45 & --0.403 & 0.386 & --1.044 & 80 & 16 \\ 
  MISSISSIPPI & 46 & 0.101 & 0.121 & 0.836 & 51 & 13 \\ 
  NORTH CAROLINA & 47 & 0.027 & 0.039 & 0.684 & 147 & 77 \\ 
  SOUTH CAROLINA & 48 & 0.352 & 0.106 & \cellcolor{blue!25}3.324 & 65 & 25 \\ 
  TEXAS & 49 & 0.016 & 0.060 & 0.271 & 344 & 117 \\ 
  KENTUCKY & 51 & 0.067 & 0.086 & 0.786 & 70 & 33 \\ 
  MARYLAND & 52 & 0.126 & 0.096 & 1.314 & 105 & 29 \\ 
  TENNESSEE & 54 & 0.049 & 0.128 & 0.381 & 101 & 20 \\ 
  ARIZONA & 61 & 0.131 & 0.071 & 1.836 & 75 & 34 \\ 
  COLORADO & 62 & 0.070 & 0.053 & 1.322 & 78 & 44 \\ 
  CALIFORNIA & 71 & 0.088 & 0.027 & \cellcolor{blue!25}3.252 & 639 & 189 \\ 
  OREGON & 72 & 0.046 & 0.042 & 1.086 & 64 & 33 \\ 
  WASHINGTON & 73 & 0.036 & 0.037 & 0.981 & 114 & 55 \\ \hline
$\hat{\mathcal{I}}$ &  &  &  & \cellcolor{blue!50}5.117 &  &  &  \\ 
\hline
\end{tabular}%
\caption{Uniform test for the existence of jump effects. This table reports the
estimated coefficients, the standard errors, the corresponding individual
test statistic, the number of observations and the effective number of
observations. The last line reports the test statistic of the uniform
testing procedure, where the critical values at $10\%$, $5\%$ and $1\%$
levels are $2.685$, $2.917$ and $3.392,$ respectively. Due to the low
effective sample size, the bandwidth is chosen uniformly the same over all states.}
\label{table:vote_new_states}
\end{table}

\begin{table}[tbp]
\centering%
\begin{tabular}{ll|rrrrrrr}
\hline
State & State\_ic & $\hat{\gamma}_j-\hat{\gamma}_{0.5}$ & $%
(T_{j}b_{j})^{-1/2}\hat{\tilde{v}}_{j}$ & $(T_{j}b_{j})^{1/2}(\hat{\gamma}_j-%
\hat{\gamma}_{0.5})/\hat{\tilde{v}}_{j}$ & Obs & EObs &  &  \\ \hline
CONNECTICUT & 1 & 0.046 & 0.050 & 0.913 & 72 & 41 \\ 
  MASSACHUSETTS & 3 & 0.155 & 0.179 & 0.866 & 129 & 22 \\ 
  NEW JERSEY & 12 & --0.004 & 0.049 & --0.077 & 171 & 55 \\ 
  NEW YORK & 13 & --0.043 & 0.032 & --1.347 & 387 & 153 \\ 
  PENNSYLVANIA & 14 & 0.000 & 0.039 & 0.000 & 237 & 100 \\ 
  ILLINOIS & 21 & 0.017 & 0.045 & 0.384 & 250 & 85 \\ 
  INDIANA & 22 & --0.016 & 0.036 & --0.453 & 125 & 62 \\ 
  MICHIGAN & 23 & --0.029 & 0.036 & --0.789 & 206 & 79 \\ 
  OHIO & 24 & 0.037 & 0.049 & 0.744 & 230 & 69 \\ 
  WISCONSIN & 25 & 0.031 & 0.086 & 0.366 & 97 & 31 \\ 
  IOWA & 31 & 0.028 & 0.042 & 0.684 & 64 & 50 \\ 
  MISSOURI & 34 & 0.161 & 0.052 & \cellcolor{blue!25}3.096 & 115 & 37 \\ 
  VIRGINIA & 40 & 0.127 & 0.070 & 1.818 & 102 & 54 \\ 
  ALABAMA & 41 & 0.116 & 0.068 & 1.698 & 74 & 21 \\ 
  FLORIDA & 43 & --0.057 & 0.066 & --0.86 & 221 & 85 \\ 
  GEORGIA & 44 & 0.005  & 0.075 & 0.064 & 126 & 40 \\ 
  LOUISIANA & 45 & --0.499 & 0.373 & --1.339 & 80 & 16 \\ 
  MISSISSIPPI & 46 & 0.005 & 0.118 & 0.040 & 51 & 13 \\ 
  NORTH CAROLINA & 47 & --0.069 & 0.043 & --1.627 & 147 & 77 \\ 
  SOUTH CAROLINA & 48 & 0.256 & 0.104 & 2.463 & 65 & 25 \\ 
  TEXAS & 49 & --0.080 & 0.061 & --1.301 & 344 & 117 \\ 
  KENTUCKY & 51 & --0.029 & 0.085 & --0.342 & 70 & 33 \\ 
  MARYLAND & 52 & 0.029 & 0.094 & 0.313 & 105 & 29 \\ 
  TENNESSEE & 54 & --0.048 & 0.125 & --0.381 & 101 & 20 \\ 
  ARIZONA & 61 & 0.034 & 0.071 & 0.484 & 75 & 34 \\ 
  COLORADO & 62 & --0.026 & 0.055 & --0.475 & 78 & 44 \\ 
  CALIFORNIA & 71 & --0.009 & 0.032 & --0.269 & 639 & 189 \\ 
  OREGON & 72 & --0.050 & 0.045 & --1.114 & 64 & 33 \\ 
  WASHINGTON & 73 & --0.060 & 0.040 & --1.496 & 114 & 55 \\ \hline
$\hat{\mathcal{Q}}$ &  &  &  & \cellcolor{blue!25}3.096 &  &  &  &  \\ 
\hline
\end{tabular}%
\caption{Uniform test for heterogeneity. This table reports the estimated
coefficients, the standard errors, the corresponding individual test
statistic, the number of observations and the effective number of
observations. The last line reports the test statistic of the uniform
testing procedure, where the critical values at $10\%$, $5\%$ and $1\%$
levels are $2.685$, $2.917$ and $3.392,$ respectively. Due to the low
effective sample size, the bandwidth is chosen uniformly over all states.}
\label{table:vote_new_states_bw_hetero}
\end{table}

\begin{figure}[tbp]
\centering
\includegraphics[width=.49\linewidth]{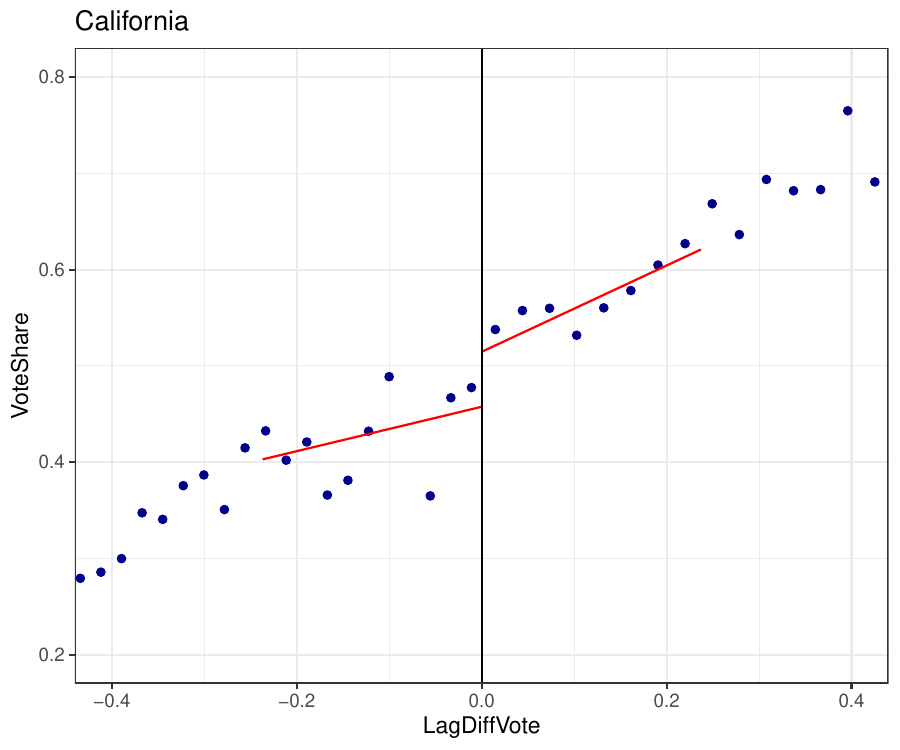}\hfill
\includegraphics[width=.49\linewidth]{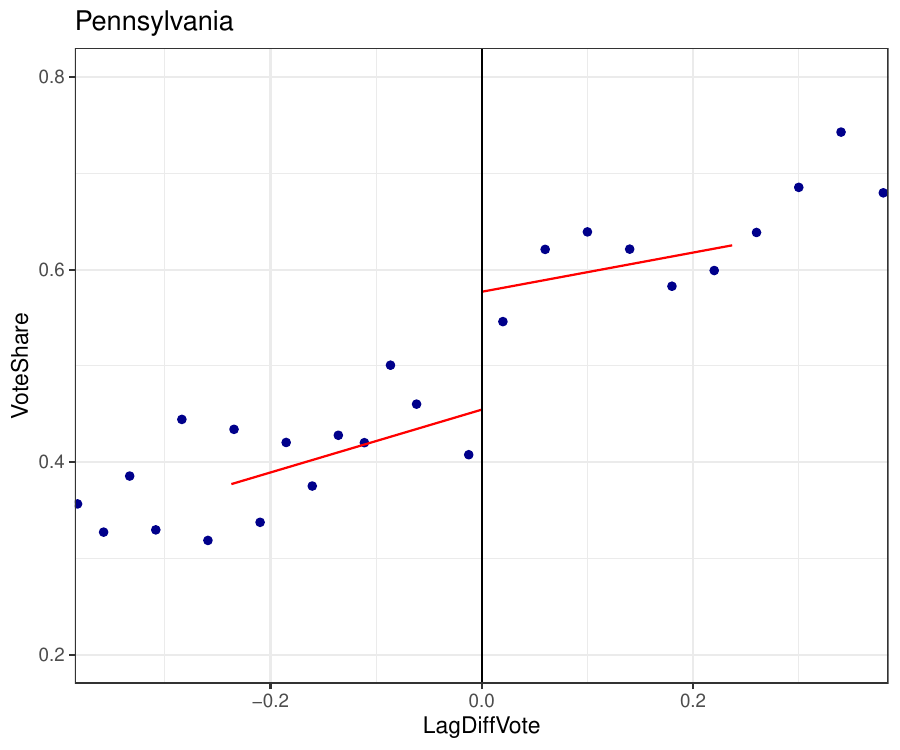}
\caption{Local linear fit of a significant threshold effect for California (left panel) and an insignificant effect for Pennsylvania (right panel).}
\label{figure:application_election}
\end{figure}

\section{Additional Examples and Remarks}
\label{addtionalcovariate}

\noindent\textbf{Example of Assumption \ref{asmp:dependence}}. 
We consider a special case of $X_{jt}$ in which it takes the form
\begin{align*}
X_{jt}=g_{j1}(\iota_{t})+g_{j2}(\iota_{t-1},\iota_{t-2},\ldots)
=g_{j1}(\iota_{t})+g_{j2}(\F_{t-1}),
\end{align*}
where $g_{j1},g_{j2}$ are measurable functions. Assume that $g_1(\iota_{t})$ has density function $f_j(x)$, then 
\begin{align}
\label{eq:example1}
g_{j,t}(x|\F_{t-1})
=f_j(x-g_{j2}(\F_{t-1})).
\end{align}
If the density function $f_j$ has uniformly bounded derivative and $g_{j2}(x_1,x_2,\ldots)$ is uniformly Lipschitz continuous over $j\in[N]$, that is 
\begin{align}
\label{eq:example12}
\max_{j,(x_i)_{i\neq k},u,v}|g_{j2}(x_1,\ldots,x_{k-1},u,x_{k+1},\ldots)
-g_{j2}(x_1,\ldots,x_{k-1},v,x_{k+1},\ldots)|
\leq a_k|u-v|,
\end{align}
then Assumption \ref{asmp:dependence} (i) holds if $\sum_{k\geq m}a_k\leq m^{-\alpha}.$
\\

\noindent\textbf{Example of Assumption \ref{asmp:bddxj1j2density}.}
Suppose $\iota_t = (\iota_{1t}, \ldots, \iota_{Nt})^\top \in \RR^N$, where the components $\iota_{jt}$ are i.i.d.\ across both $j$ and $t$, with common density function $f_\iota$. Consider the following factor time series:
\begin{align*}
X_{j_1t}=\lambda_{j_1} \eta_{t-1}+\iota_{j_1t}\quad
\textrm{and}\quad
X_{j_2t}=\lambda_{j_2} \eta_{t-    1}+\iota_{j_2t},
\end{align*}
where $\eta_{t-1}$ is measurable with respect to $\mathcal{F}_{t-1}$. Then, the conditional joint density of $(X_{j_1t}, X_{j_2t})$ given $\mathcal{F}_{t-1}$ is
\begin{align*}
g_{j_{1},j_{2},t}(x_{1},x_{2}|\mathcal{F}_{t-1})  
=f_\iota(x_1-\lambda_{j_1} \eta_{t-1})f_\iota(x_2-\lambda_{j_2} \eta_{t-1}).
\end{align*}
If the density function $f_\iota$ is bounded, then Assumption \ref{asmp:bddxj1j2density} is satisfied. Furthermore, if 
\begin{align*}
\eta_t = \sum_{k \geq 0} a_k \iota_{1(t-k)},    
\end{align*}
and $|a_k| \leq c k^{-\alpha - 1}$ for some constant $c > 0$,
then \eqref{eq:example1}, \eqref{eq:example12} hold, thus Assumption~\ref{asmp:dependence} (i) is satisfied.

\begin{remark}
    Let
$U_{jt}'=H_j(X_{jt},\tilde \iota_t'),$
where \(\tilde \iota_t'\) is an independent copy of \(\tilde \iota_t\), independent of
\((\mathcal F_T,\tilde{\mathcal F}_T)\). Conditional on \(X_{jt}\), \(U_{jt}'\) has the same distribution as \(U_{jt}\) and is independent of \(U_{jt}\). Hence
$\tilde h_j(X_{jt})
=
\mathbb E\{h_j(X_{jt},U_{jt}')\mid X_{jt}\}.
$
Therefore,
\[
h_j(X_{jt},U_{jt})-\tilde h_j(X_{jt})
=
\mathbb E\left[
h_j(X_{jt},U_{jt})-h_j(X_{jt},U_{jt}')
\mid X_{jt},U_{jt}
\right].
\]
By Assumption \ref{boundedness},
\[
\left|
h_j(X_{jt},U_{jt})-\tilde h_j(X_{jt})
\right|
\leq
L\,
\mathbb E\left[
|U_{jt}-U_{jt}'|_2
\mid X_{jt},U_{jt}
\right].
\]
Hence
\[
\mathbb E\left[
\left|
h_j(X_{jt},U_{jt})-\tilde h_j(X_{jt})
\right|^q
\mid \mathcal F_T
\right]
\leq
C L^q
\mathbb E\left[
|U_{jt}|_2^q+|U_{jt}'|_2^q
\mid \mathcal F_T
\right]
<\infty .
\]
The same argument applied to \(\tau_j(X_{jt},U_{jt})-\tilde \tau_j(X_{jt})\) yields
\[
\mathbb E\left[
\left|
\tau_j(X_{jt},U_{jt})-\tilde \tau_j(X_{jt})
\right|^q
\mid \mathcal F_T
\right]
<\infty .
\]
It follows that
\[
\mathbb E\left(
|\varepsilon_j(X_{jt},U_{jt})|^q
\mid \mathcal F_T
\right)
<\infty .
\]

\end{remark}

\begin{remark}
(Additive time and individual effects)

Our model can account for both individual fixed effects $\alpha _{j}$ and time fixed effects $\nu _{t}$. Specifically, our working model (\ref{mainmodel}%
) becomes 
\begin{equation}
Y_{jt}=\alpha_j+\nu_t+\tilde{h}_{j}(X_{jt})+\tilde{\tau}_{j}(X_{jt})\mathbf{1}_{\{X_{jt}\geq c_{0j}\}}+e_{jt}.
\label{model3}
\end{equation}
If {$\mathbb{E}(\nu_{t}|X_{jt})=d(X_{jt})$ for some function }$d(\cdot)$ under a covariance-stationarity condition along the $t$-dimension, then by defining $\tilde{e}_{jt}=e_{jt}+[\nu _{t}-d(X_{jt})]$ and $\tilde{\tilde h}_j(X_{jt})=\alpha _{j}+d(X_{jt})+\tilde h_{j}(X_{jt}),$
the above model can be reformulated as follows:
\begin{equation}
Y_{jt}=\tilde{\tilde h}_j(X_{jt})+\tilde\tau_j(X_{jt})\mathbf{1}_{\{X_{jt}\geq c_{0j}\}}+\tilde{e}_{jt}.
\label{model}
\end{equation}
In the absence of covariance-stationarity, the estimation would become
much more involved and we have to leave it for future research.
\end{remark}

\begin{remark}(Adding more covariates to fit the threshold model)

Suppose that we want to include more covariates $Z_{jt} \in \mathbb{R}^d$ and $\theta \in \mathbb{R}^d$, we can impose a partial linear structure. Then the working model in (\ref{mainmodel}) can be of the following form:
\begin{eqnarray}
&&Y_{jt}=\tilde h_{j}(X_{jt})+\tilde\tau(X_{jt})\mathbf{1}_{\{X_{jt}\geq c_{0j}\}}+Z_{jt}^{\top}\theta+e_{jt}.
\end{eqnarray}
Our methodology can be extended to the model above by adapting the approach used in \cite{li2022simultaneous}.
\end{remark}

\end{document}

%% file: 20260713_theorem.tex
\section{Proofs of the Main Results in the Paper \label{SecA}}
\subsection{Proof of Theorem \ref{thm:ga}}
In this section, we first outline the main idea for the proof of Theorem \ref%
{thm:ga} and present the main lemma for the Gaussian approximation. Then we
prove Theorem \ref{thm:ga}, Proposition \ref{thm:hatv}, and Corollary \ref{thm:power} in turn.

\subsubsection{Outline of the Proof of Theorem \protect\ref{thm:ga} \label%
{SecA.1}}

For exposition purposes, we outline the main steps in the proof of Theorem \ref%
{thm:ga}. The intuition of proving Theorem \ref{thm:homotesting} is similar. 
Define the statistic $I_{\epsilon }$ after canceling the smooth component in
the main text, 
\begin{equation}
I_{\epsilon }=\max_{1\leq j\leq N}(Tb_{j})^{1/2}\Big|%
\sum_{t=1}^{T}(w_{jt,b}^{+}-w_{jt,b}^{-})[\varepsilon
_{j}(X_{jt},U_{jt})+\sigma_{j}(X_{jt},U_{jt})\epsilon _{jt}]+\gamma
_{j}\Big|/v_{j}.  \label{eq:iepsilon}
\end{equation}%
By the continuity of $f_{j}\left( \cdot \right) $ in Assumption \ref%
{asmp:smooth} and Lemma \ref{lem:propwitb} $(i)$ below, we have 
\begin{equation}
|\mathcal{I}-I_{\epsilon }|\leq \max_{1\leq j\leq N}(Tb_{j})^{1/2}\Big|%
\sum_{t=1}^{T}(w_{jt,b}^{+}-w_{jt,b}^{-})f_{j}(X_{jt})\Big|/v_{j}=O(T^{1/2}%
\bar{b}^{5/2}).
\end{equation}%
Thus we only need to work with $I_{\epsilon }.$ We prove by an $m$-lag
truncation to facilitate us to construct an $m$-dependent sequence. Specifically, for some 
$m>0,$ consider a truncated version of the error term: 
\begin{equation}
\epsilon _{t,m}=(\epsilon _{1t,m},\epsilon _{2t,m},\ldots ,\epsilon
_{Nt,m})^{\top }=\sum_{k=0}^{m-1}A_{k}\eta _{t-k}.  \label{eq:defofmdep}
\end{equation}%
Let $I_{\epsilon ,m}$ be the $m$-dependent approximation of $I_{\epsilon },$
where $I_{\epsilon ,m}$ is $I_{\epsilon }$ with $\epsilon _{t}$ therein
replaced by $\epsilon _{t,m},$ namely, 
\begin{equation}
I_{\epsilon ,m}=\max_{1\leq j\leq N}(Tb_{j})^{1/2}\Big|\sum_{t=1}^{T}w_{jt,b}%
\big[\varepsilon
_{j}(X_{jt},U_{jt})+\sigma _{j}(X_{jt},U_{jt})\epsilon _{jt,m}\big]+\gamma _{j}\Big|/v_{j},  \label{eq:iem}
\end{equation}
where $w_{jt,b}:=(w_{jt,b}^{+}-w_{jt,b}^{-})\mathbf{1}_{\mathcal{A}_{T}}$
where $\mathcal{A}_{T}$ is defined in Lemma \ref{lem:propwitb} $(ii)$ below.
For large $m,$ due to the decay of temporal dependence assumption, we expect 
\begin{equation*}
I_{\epsilon }\approx I_{\epsilon ,m}.
\end{equation*}%
Let $Z_{j,m}$'s be normally distributed random variables that are
independent across $j$.
Denote the $m$-dependent Gaussian counterpart as 
\begin{equation*}
\tilde{I}_{z,m}=\max_{1\leq j\leq N}\Big|Z_{j,m}+\frac{(Tb_{j})^{1/2}}{v_{j}}%
\gamma _{j}\Big|.
\end{equation*}%
Similarly, let $I_{z}$ (resp. $I_{z,m}$) be $I_{\epsilon }$ (resp. $%
I_{\epsilon ,m}$) with $\eta _{t}$ therein replaced by $z_{t}$, where $%
\left\{ z_{t}\right\} _{t\in \mathbb{Z}}$ are i.i.d. Gaussian vectors with
zero mean and identity covariance matrix. Since $I_{\epsilon ,m}$ can be rewritten as the maximum of a sum of independent vectors conditional on $X_{jt}, U_{jt}$, we can apply the GA theorem from \cite{MR3693963} to show the GA for the conditional part. For the remaining part, we need to demonstrate GA again, as detailed in Lemma \ref{lem:gapprox} below, then we have 
\begin{equation*}
I_{\epsilon ,m}\overset{\mathbb{P}}{\approx }I_{z,m}.
\end{equation*}
We complete the proof by showing that the distributions of $I_{z,m}$ and $|Z+%
\underline{d}|_{\infty }$ are close, and the continuity of the maximum of a
non-centered Gaussian vector. Then the conclusion follows.

\subsubsection{A Lemma on Gaussian Approximation}

This subsection presents a Gaussian approximation lemma that utilizes a nonstandard technique compared to the existing literature. The lemma specifically focuses on the $m$-dependent part $I_{\epsilon,m}$ in \eqref{eq:iem}.
The difference between $I_{\epsilon,m}$ and $I_{\epsilon}$ is effectively controlled in Lemmas \ref{lem:mdepend} and \ref{lem:mdepend_supexp} below. Recall that $\mathcal{F}_{t}=(e_{t},e_{t-1},\ldots )$ and $
\tilde{\mathcal{F}}_{t}=(\tilde{e}_{t},\tilde{e}_{t-1},\ldots ).$

The non-triviality of the proof of the theorems lies in the fact that the
interplay between the two terms in \eqref{eq:iem}:
\begin{align*}
\frac{(Tb_{j})^{1/2}}{v_{j}}
\sum_{t=1}^{T}w_{jt,b}\sigma_{j}(X_{jt},U_{jt})\epsilon _{jt,m}  \quad\textrm{and}\quad
\frac{(Tb_{j})^{1/2}}{v_{j}}\sum_{t=1}^{T}w_{jt,b}\varepsilon
_{j}(X_{jt},U_{jt}).
\end{align*} 
The second term arises from the potential presence of the latent variable $U_{jt}$. If $U_{jt}$ is absent, the second term will disappear, allowing time-dependent Gaussian approximation to be applied to the first term.
This second term causes the derivation steps to differ from those in the existing literature. Noting that the second term becomes determined when we condition on {$\{X_{jt},U_{jt}\}_{1\leq t\leq T, 1\leq j\leq N}$},
we will first prove a GA for the first term under this conditioning. Then we shall replace the first term by its Gaussian counterparts 
$Z_{j,1}=(Tb_{j})^{1/2}v_{j}^{-1}\sum_{t=1}^{T}\tilde{\xi}_{jt,1}$, where $\tilde{\xi}_{jt,1}$ are i.i.d. 
normal
and independent from $\mathcal{%
F}_{t}$ and $\tilde{\mathcal{F}}_{t}$.
Then we can show the GA for the summation of this counterpart and the second term.

The following lemma is the key lemma for Gaussian approximation, whose proof
can be found in Section \ref{SecB.1}. 




\begin{lemma}
\label{lem:gapprox} 
Let $T/2\geq m\rightarrow \infty .$

$(i)$ Suppose that the conditions in Theorem \ref{thm:ga} $(i)$ hold. Then we
have 
\begin{align*}
& \sup_{u\in \mathbb{R}}\Big|\mathbb{P}(I_{\epsilon ,m}\leq u)-\mathbb{P}%
\Big(\max_{1\leq j\leq N}\big|Z_{j}+(Tb_{j})^{1/2}\gamma _{j}/v_{j}\big|\leq
u\Big)\Big| \\
\lesssim & (\underline{b}T)^{-1/6}\mathrm{log}^{7/6}(NT)+(T^{2/q}/(%
\underline{b}T))^{1/3}\mathrm{log}(NT)+(\bar{b}+m^{-\beta })^{1/3}\mathrm{log%
}^{2/3}(N)+T^{-(\alpha p)\wedge (p/2-1)},
\end{align*}%
where the constant in $\lesssim $ is independent of $N,T.$

$(ii)$ Suppose that the conditions in Theorem \ref{thm:ga} $(ii)$ hold. Then we
have 
\begin{align*}
& \sup_{u\in \mathbb{R}}\Big|\mathbb{P}(I_{\epsilon ,m}\leq u)-\mathbb{P}%
\Big(\max_{1\leq j\leq N}\big|Z_{j}+(Tb_{j})^{1/2}\gamma _{j}/v_{j}\big|\leq
u\Big)\Big| \\
\leq & (\underline{b}T)^{-1/6}\mathrm{log}(NT)^{7/6}+(\bar{b}+m^{-\beta
})^{1/3}\mathrm{log}^{2/3}(N)+T^{-(\alpha p)\wedge (p/2-1)},
\end{align*}%
where the constant in $\lesssim $ is independent of $N,T.$
\end{lemma}

\subsubsection{Main Steps of Proof of Theorem \protect\ref{thm:ga} \label{SecA.2}}

To prove Theorem \ref{thm:ga}, we need the following seven lemmas. See
Section \ref{SecB.2} for the proofs of Lemmas \ref{lem:propwitb}, \ref
{lem:mdepend}, and \ref{lem:mdepend_supexp}.
It should be noted that, for simplicity in the exposition of $W_{jt,b}$, we define an event $\mathcal{A}_T$ in Lemma \ref{lem:propwitb}, which occurs with probability $1-b_{T}$, where $b_{T} \to 0$.

\begin{lemma}
(Properties for weights) \label{lem:propwitb} Under Assumptions \ref
{asmp:moment}-{\ref{asmp:bddxj1j2density}} and \ref{boundedness}-\ref{kernel}, the followings hold:

(i) $\sum_{t=1}^{T}w_{jt,b}^{+}=\sum_{t=1}^{T}w_{jt,b}^{-}=1\quad $and$\quad
\sum_{t=1}^{T}(X_{jt}-c_{j0})w_{jt,b}^{+}=%
\sum_{t=1}^{T}(X_{jt}-c_{j0})w_{jt,b}^{-}=0.$

(ii) For $l=0,1,2,$ on some set $\mathcal{A}_{T}$ with 
\begin{equation}
\mathbb{P}(\mathcal{A}_{T})\geq 1-O(T^{-(\alpha p)\wedge (p/2-1)})
\end{equation}%
we have 
\begin{equation}
\max_{1\leq j\leq N}\big|S_{jl,b}^{+}/(Tb_{j}^{l+1})-K_{l}^{+}g_{j}(c_{j0})%
\big|\lesssim \sqrt{\mathrm{log}(TN)/(T\underline{b})}+\bar{b},
\label{eq:Sjlplus}
\end{equation}%
where the constants in $O(\cdot )$ and $\lesssim $ are independent of $N,T,%
\underline{b},$ and $\bar{b}.$ And thus on set $\mathcal{A}_{T}$, 
\begin{align}
& \max_{1\leq j\leq N}\Big|w_{jt,b}^{+}-\frac{K((X_{jt}-c_{j0})/b_{j})}{%
g_{j}(c_{j0})Tb_{j}}\frac{K_{2}^{+}-K_{1}^{+}(X_{jt}-c_{j0})/b_{j}}{%
K_{2}^{+}K_{0}^{+}-K_{1}^{+2}}\mathbf{1}_{\{X_{jt}\geq c_{j0}\}}\Big|  \notag \\
\lesssim & \sqrt{\mathrm{log}(TN)/(\underline{b}T)^{3}}+1/T,
\label{eq:witb}
\end{align}%
and moreover, 
\begin{equation}
\sum_{t=1}^{T}|w_{jt,b}^{+}|=O(1)\quad \text{and}\quad
\sum_{t=1}^{T}|w_{jt,b}^{-}|=O(1)  \label{eq:bddforsumw}
\end{equation}%
uniformly over $j$.
\end{lemma}

The following is a high dimensional extension of the concentration
inequality in \cite{freedman1975tail}.

\begin{lemma}
(Freedman's inequality) \label{lem:freedman} Let $\mathcal{A}$ be an index
set with $|\mathcal{A}|<\infty $. For each $a\in \mathcal{A}$, let $\{\xi
_{a,i}\}_{i=1}^{n}$ be a martingale difference sequence with respect to the
filtration $\{\mathcal{F}_{i}\}_{i=1}^{n}.$ Let $M_{a}=\sum_{i=1}^{n}\xi
_{a,i}$ and $V_{a}=\sum_{i=1}^{n}\mathbb{E}[\xi _{a,i}^{2}|\mathcal{F}%
_{i-1}] $. Then for all $z,u,v>0$ 
\begin{equation*}
\mathbb{P}\Big(\max_{a\in \mathcal{A}}|M_{a}|\geq z\Big)\leq \sum_{i=1}^{n}%
\mathbb{P}\Big(\max_{a\in \mathcal{A}}|\xi _{a,i}|\geq u\Big)+2\mathbb{P}%
\Big(\max_{a\in \mathcal{A}}V_{a}\geq v\Big)+2|\mathcal{A}%
|e^{-z^{2}/(2zu+2v)}.
\end{equation*}
\end{lemma}

\begin{lemma}[\cite{burkholder1988sharp}, \cite{MR2472010}]
\label{burkholder} Let $q>1$, $q^{\prime }=\min\{q,2\}$. Let $%
M_T=\sum_{t=1}^T \xi_t,$ where $\xi_t\in \mathcal{L}^q$ are martingale
differences. Then 
\begin{align*}
\|M_T\|_q^{q^{\prime }} \leq K_q^{q^{\prime
}}\sum_{t=1}^T\|\xi_t\|_q^{q^{\prime }}, \mbox{ where } K_q = \max\big(%
(q-1)^{-1}, \sqrt{q-1}\big).
\end{align*}
\end{lemma}

\begin{lemma}[\cite{MR2083397}]
\label{lem:anticon} Let $X=(X_{1},X_{2},\ldots ,X_{v})^{\top }$ be a
centered Gaussian vector in $\mathbb{R}^{v}.$ Assume $\mathbb{E}%
(X_{i}^{2})\geq b$ for some $b>0$ and all $1\leq i\leq v.$ Then for any $e>0$
and $d\in \mathbb{R}^{v},$ 
\begin{equation}
\sup_{x\in \mathbb{R}}\mathbb{P}\Big(\big||X+d|_{\infty }-x\big|%
\leq e\Big)\leq ce\sqrt{\mathrm{log}(v)},
\end{equation}%
where $c$ is some constant depending only on $b$.
\end{lemma}

The following lemma is Lemma 3 in \cite{chen2022inference}.

\begin{lemma}[Comparison]
\label{lem:comparison} Let $X=(X_{1},X_{2},\ldots ,X_{v})^{\top }$ and $%
Y=(Y_{1},Y_{2},\ldots ,Y_{v})^{\top }$ be two centered Gaussian vectors in $%
\mathbb{R}^{v}$ and let $d=(d_{1},d_{2},\ldots ,d_{v})^{\top }\in \mathbb{R}%
^{v}.$ We denote $\Delta =\max_{1\leq i,j\leq v}|\sigma _{i,j}^{X}-\sigma
_{i,j}^{Y}|,$ where we define $\sigma _{i,j}^{X}=\mathbb{E}(X_{i}X_{j})$
(resp. $\sigma _{i,j}^{Y}=\mathbb{E}(Y_{i}Y_{j})$). Assume that $Y_{i}$s
have the same variance $\sigma ^{2}>0.$ Then we have 
\begin{equation}
\sup_{x\in \mathbb{R}}\Big|\mathbb{P}\big(|X+d|_{\infty }\leq x\big)-\mathbb{%
P}\big(|Y+d|_{\infty }\leq x\big)\Big|\lesssim \Delta ^{1/3}\mathrm{log}%
(v)^{2/3},
\end{equation}%
where the constant involved in $\lesssim $ only depends on $\sigma .$
\end{lemma}

\begin{lemma}[$m$-dependent approximation for polynomial case]
\label{lem:mdepend} Assume that the conditions in Theorem \ref{thm:ga} $(i)$
hold. For some $1\ll m\leq T/2,$ we have 
\begin{equation*}
\mathbb{P}\big(|I_{\epsilon}-I_{\epsilon ,m}|\geq z\big)\lesssim N\mathrm{%
log}^{q}(NT)(T\underline{b})^{1-q/2}m^{-q\beta }z^{-q}+Ne^{-z^{2}m^{2\beta }}+%
\mathbb{P}(\mathcal{A}_{T}^{c}).
\end{equation*}%
where $\mathcal{A}_{T}$ is defined in Lemma \ref{lem:propwitb}.
\end{lemma}

\begin{lemma}[$m$-dependent approximation for sub-exponential case]
\label{lem:mdepend_supexp} Assume that the conditions in Theorem \ref{thm:ga} 
$(ii)$ hold. For some $1\ll m\leq T/2,$ we have 
\begin{equation*}
\mathbb{P}\big(|I_{\epsilon }-I_{\epsilon ,m}|\geq z\big)\lesssim N\mathrm{%
exp}\Big\{-c\frac{z^{2}}{m^{-\beta }(\underline{b}T)^{-1/2}z+m^{-2\beta }}%
\Big\}+\mathbb{P}(\mathcal{A}_{T}^{c}),
\end{equation*}%
where $c>0$ is some constant and $\mathcal{A}_{T}$ is defined in Lemma \ref%
{lem:propwitb}.
\end{lemma}

\noindent \textbf{Proof of Theorem \ref{thm:ga}. }

$(i)$ Recall the definition of $I_{\epsilon }$ in \eqref{eq:iepsilon} and $%
I_{\epsilon ,m}$ is the $m$-dependent part of $I_{\epsilon }$. Let $m=T/2.$
For any $l>0,$ we have 
\begin{equation*}
\mathbb{P}\big(\mathcal{I}\leq u\big)\leq \mathbb{P}\big(|\mathcal{I}%
-I_{\epsilon ,m}|\geq l\big)+\mathbb{P}\big(I_{\epsilon ,m}\leq u+l\big),
\end{equation*}%
and similarly 
\begin{equation*}
\mathbb{P}\big(I_{z}\leq u+2l\big)\geq -\mathbb{P}\big(|I_{z}-I_{z,m}|\geq l%
\big)+\mathbb{P}\big(I_{z,m}\leq u+l\big).
\end{equation*}%
Combining the above inequalities leads to 
\begin{align}
& \sup_{u\in \mathbb{R}}\Big[\mathbb{P}\big(\mathcal{I}\leq u)-\mathbb{P}%
(I_{z}\leq u\big)\Big]  \notag \\
\leq & \mathbb{P}\big(|\mathcal{I}-I_{\epsilon ,m}|\geq l\big)+\mathbb{P}%
\big(|I_{z}-I_{z,m}|\geq l\big)+\sup_{u\in \mathbb{R}}\big|\mathbb{P}%
(I_{\epsilon ,m}\leq u)-\mathbb{P}(I_{z,m}\leq u)\big|  \notag \\
& +\sup_{u\in \mathbb{R}}\big|\mathbb{P}\big(I_{z}\leq u+2l\big)-\mathbb{P}%
\big(I_{z}\leq u\big)\big|  \notag \\
=&:\mathrm{I}_{1}+\mathrm{I}_{2}+\mathrm{I}_{3}+\mathrm{I}_{4}.
\label{eq:thmmaindecomp}
\end{align}%
For part $\mathrm{I}_{1}$, we have $|\mathcal{I}-I_{\epsilon ,m}|\leq |%
\mathcal{I}-I_{\epsilon }|+|I_{\epsilon }-I_{\epsilon ,m}|.$ By Assumption
\ref{asmp:smooth} and Lemma \ref{lem:propwitb}, 
\begin{align*}
& \max_{1\leq j\leq N}\Big|%
\sum_{t=1}^{T}w_{jt,b}^{+}f_{j}(X_{jt})-f_{j}(c_{j0})\Big| \\
=& \max_{1\leq j\leq N}\Big|\sum_{t=1}^{T}w_{jt,b}^{+}\big[%
f_{j}(c_{j0})+\partial _{+}f_{j}(c_{j0})(X_{jt}-c_{j0})+O((X_{jt}-c_{j0})^{2})%
\big]-f_{j}(c_{j0})\Big| \\
=& O(\bar{b}^{2}),
\end{align*}%
where the constant in $O(\cdot )$ here and all the followings are
independent of $N,T$. The same bound can be derived with $%
w_{jt,b}^{+}$ therein replaced by $w_{jt,b}^{-}$, and thus 
\begin{equation}
|\mathcal{I}-I_{\epsilon }|\leq \max_{1\leq j\leq N}(Tb_{j})^{1/2}\Big|%
\sum_{t=1}^{T}(w_{jt,b}^{+}-w_{jt,b}^{-})f_{j}(X_{jt})\Big|%
/v_{j}=O(T^{1/2}\bar{b}^{5/2}).  \label{eq:bddiiep}
\end{equation}%
We define $l'=N^{1/q}T^{-\beta }\sqrt{\mathrm{log}(NT)}$. Then by Lemma %
\ref{lem:mdepend} with $m=T/2$, we have 
\begin{equation}
\mathbb{P}\big(|I_{\epsilon }-I_{\epsilon ,m}|\geq l^{\prime }\big)=O\big\{%
\mathrm{log}(NT)^{q/2}(T\underline{b})^{-q/2+1}+T^{-(\alpha p)\wedge
(p/2-1)}\big\}.  \label{eq:t832}
\end{equation}%
Hence for $l=l'+cT^{ 1/2}\bar{b}^{5/2}$, where $c>0$ is some constant large
enough, we have 
\begin{equation*}
\mathrm{I}_{1}=O\big\{\mathrm{log}(NT)^{q/2}(T\underline{b})^{-q/2+1}+T^{-(\alpha p)\wedge (p/2-1)}\big\}.
\end{equation*}%
Similar argument leads to the same bound for part $\mathrm{I}_{2}$.

For part $\mathrm{I}_{3}$, by Lemma \ref{lem:gapprox} $(i)$ we have 
\begin{equation*}
\mathrm{I}_{3}\lesssim (\underline{b}T)^{-1/6}\mathrm{log}%
^{7/6}(NT)+(T^{2/q}/(\underline{b}T))^{1/3}\mathrm{log}(NT)+(\bar{b}+T^{-\beta })^{1/3}\mathrm{log%
}^{2/3}(N)+T^{-(\alpha
p)\wedge (p/2-1)}.
\end{equation*}

For part $\mathrm{I}_{4}$, 
we use Lemma \ref{lem:anticon} to obtain 
\begin{equation*}
\mathrm{I}_{4}\leq \sup_{u\in \mathbb{R}}\mathbb{P}\big(|I_{z}-u\big|\leq 2l%
\big)\lesssim l\mathrm{log}(N)^{1/2}.
\end{equation*}%
We complete the proof by combining the $\mathrm{I}_{1}$-$\mathrm{I}_{4}$
parts and a similar argument for the other side of the inequality.

$(ii)$ As in part $(i)$, we have the decomposition as in \eqref{eq:thmmaindecomp}%
. By \eqref{eq:bddiiep} and Lemma \ref{lem:mdepend_supexp} with $m=T/2$, we
have $\mathrm{I}_{1}=O((NT)^{-q})$ with $l=O(T^{-\beta }\mathrm{log}%
(NT)^{1/2}+T^{1/2}b^{5/2}).$ The same bound can be derived for part $\mathrm{%
I}_{2}$. By Lemma \ref{lem:gapprox} $(ii)$, we have 
\begin{equation*}
\mathrm{I}_{3}\lesssim (bT)^{-1/6}\mathrm{log}^{7/6}(NT)+T^{-(\alpha
p)\wedge (p/2-1)}.
\end{equation*}%
For part $\mathrm{I}_{4}$, we use the same argument as in part $(i)$ to obtain
\begin{equation*}
\mathrm{I}_{4}\lesssim l\mathrm{log}(N)^{1/2}\lesssim T^{-\beta }\mathrm{log}
(NT)+\mathrm{log}^{1/2}(N)T^{1/2}\bar{b}^{5/2}.
\end{equation*}
The desired result follows by combining the $\mathrm{I}_{1}$-$\mathrm{I}_{4}$
and a similar argument for the other side of the inequality.


\noindent \textbf{Proof of Remark \ref{rmk:rmkvj1j2}.}
The proof follows from the same argument as that of Theorem \ref{thm:ga}. 
The only difference is that, without Assumption \ref{asmp:bddxj1j2density}, we no longer use the localization argument to show that the off-diagonal covariance terms are asymptotically negligible.

More specifically, in the proof of Lemma \ref{lem:gapprox}, we keep the full covariance structure
\[
\Sigma_{j_1j_2}
=
\frac{v_{j_1j_2}}{v_{j_1}v_{j_2}},
\qquad 1\leq j_1,j_2\leq N,
\]
instead of setting \(v_{j_1j_2}=0\) for \(j_1\neq j_2\). Then, by directly applying
\eqref{eq:bdd4iemandzj} with this covariance matrix, we obtain
\[
\sup_{u\in\mathbb R}
\left|
\mathbb P(\mathcal I\leq u)
-
\mathbb P\left(
\left|\Sigma^{1/2}Z+\underline d\right|_\infty\leq u
\right)
\right|
\lesssim
\mathcal R_{NT},
\]
where \(Z\sim N(0,I_N)\). This proves the desired claim.
\hfill\(\square\)

\subsection{Proof of Theorem \protect\ref{thm:homotesting} \label{SecA.3}}
\label{proofhomo}
Denote 
\begin{equation}
a_{jt,b}=w_{jt,b}\sigma _{j}(X_{jt},U_{jt})  \label{eq:ajtb}
\end{equation}%
and let 
\begin{align}
\mathcal{Q}_{\epsilon }=\max_{1\leq j\leq N}\Bigg|\sum_{t=1}^{T}& \Big(%
a_{jt,b}\epsilon _{jt}-\sum_{l=1}^{N}a_{lt,b}\epsilon
_{lt}/N+w_{jt,b}\varepsilon _{j}(X_{jt},U_{jt})  \notag \\
& -\sum_{l=1}^{N}w_{lt,b}\varepsilon _{l}(X_{lt},U_{lt})/N\Big)+(\gamma _{j}-%
\bar{\gamma})\Bigg|/\tilde{v}_j.  \label{eq:Qepsilon}
\end{align}%
Denote $\Q_z$ be $\Q_\epsilon$ with $\epsilon_{jt}$ and $U_{jt}$ replaced by Gaussian distributed random variable with the same covariance structure and $\Q_{\epsilon, m}$ be $\Q_\epsilon$ with $\epsilon_{jt}$ therein replaced by $\epsilon_{jt,m}.$
To prove Theorem \ref{thm:homotesting}, we add two lemmas.

\begin{lemma}[$m$-dependent approximation]
\label{lem:mdependQ_homo} Let $0\ll m\leq T/2.$

(i) Under the conditions in Theorem \ref{thm:homotesting}(i) we have 
\begin{equation*}
\mathbb{P}\big(|\mathcal{Q}_{\epsilon }-\mathcal{Q}_{\epsilon ,m}|\geq z\big)%
\lesssim N\mathrm{log}(NT)^{q}(T\underline{b})^{1-q/2}m^{-q\beta
}z^{-q}+Ne^{-z^{2}m^{2\beta }}+\mathbb{P}(\mathcal{A}_{T}^{c});
\end{equation*}%
(ii) Under the conditions in Theorem \ref{thm:homotesting}(ii), we have 
\begin{equation*}
\mathbb{P}\big(|\mathcal{Q}_{\epsilon }-\mathcal{Q}_{\epsilon ,m}|\geq z\big)%
\lesssim N\mathrm{exp}\Big\{-c\frac{z^{2}}{m^{-\beta
}(\underline{b}T)^{-1/2}z+m^{-2\beta }}\Big\}+\mathbb{P}(\mathcal{A}_{T}^{c}).
\end{equation*}
\end{lemma}

\begin{lemma}
\label{lem:gapprox_homo} Let $0<m\leq T/2.$

(i) Assume the conditions in Theorem \ref{thm:homotesting} $(i)$ hold. Then we
have 
\begin{align*}
& \sup_{u\in \mathbb{R}}|\mathbb{P}(\mathcal{Q}_{\epsilon ,m}\leq u)-\mathbb{%
P}(\mathcal{Q}_{z,m}\leq u)| \\
\lesssim & (\underline{b}T)^{-1/6}\mathrm{log}^{7/6}(NT)+(T^{2/q}/(\underline{b}T))^{1/3}\mathrm{log}%
(NT)+T^{-(\alpha p)\wedge (p/2-1)}.
\end{align*}

(ii) Assume the conditions in Theorem \ref{thm:homotesting} $(ii)$ hold. Then
we have 
\begin{equation*}
\sup_{u\in \mathbb{R}}|\mathbb{P}(\mathcal{Q}_{\epsilon ,m}\leq u)-\mathbb{P}%
(\mathcal{Q}_{z,m}\leq u)|\lesssim (\underline{b}T)^{-1/6}\mathrm{log}%
^{7/6}(NT)+T^{-(\alpha p)\wedge (p/2-1)}.
\end{equation*}%
Here the constants in $\lesssim $ are independent of $\left( N,T,b\right) .$
\end{lemma}

\noindent \textbf{Proof of Theorem \ref{thm:homotesting}}

$(i)$ Let $m=T/2.$ For any $l>0,$ as in \eqref{eq:thmmaindecomp}, we have 
\begin{align}
& \sup_{u\in \mathbb{R}}\Big|\mathbb{P}\big((T\underline{b})^{1/2}\mathcal{Q}\leq u)-%
\mathbb{P}((T\underline{b})^{1/2}\mathcal{Q}_{z}\leq u\big)\Big|  \notag \\
\leq & \mathbb{P}\Big((T\underline{b})^{1/2}|\mathcal{Q}-\mathcal{Q}_{\epsilon ,m}|\geq l%
\Big)+\mathbb{P}\Big((T\underline{b})^{1/2}|\mathcal{Q}_{z}-\mathcal{Q}_{z,m}|\geq l\Big)%
+\sup_{u\in \mathbb{R}}\Big|\mathbb{P}(\mathcal{Q}_{\epsilon ,m}\leq u)-%
\mathbb{P}(\mathcal{Q}_{z,m}\leq u)\Big|  \notag \\
& +\sup_{u\in \mathbb{R}}\mathbb{P}\big(|(T\underline{b})^{1/2}\mathcal{Q}_{z}-u|\leq 2l%
\big)=:\mathrm{I}_{1}+\mathrm{I}_{2}+\mathrm{I}_{3}+\mathrm{I}_{4}.
\label{eq:thmmaindecomp_homo}
\end{align}%
For part $\mathrm{I}_{1}$, by \eqref{eq:bddiiep}, we have 
\begin{equation*}
|\mathcal{Q}-\mathcal{Q}_{\epsilon ,m}|\leq |\mathcal{Q}-\mathcal{Q}%
_{\epsilon }|+|\mathcal{Q}_{\epsilon }-\mathcal{Q}_{\epsilon
,m}|=O(\bar{b}^{2})+|\mathcal{Q}_{\epsilon }-\mathcal{Q}_{\epsilon ,m}|.
\end{equation*}%
We define $l^{\prime}=N^{1/q}T^{-\beta }\sqrt{\mathrm{log}(NT)}$. Then by Lemma %
\ref{lem:mdependQ_homo} with $m=T/2$, we have 
\begin{equation}
\mathbb{P}\big((T\underline{b})^{1/2}|\mathcal{Q}_{\epsilon }-\mathcal{Q}_{\epsilon
,m}|\geq l^{\prime }\big)=O\big(\mathrm{log}%
(NT)^{q/2}(T\underline b)^{-q/2+1}+T^{-(\alpha p)\wedge (p/2-1)}\big).
\label{eq:t832_homo}
\end{equation}%
Hence for $l=l^{\prime}+cT^{1/2}\bar{b}^{5/2},$ some constant $c>0$ large enough, we have 
$$
\mathrm{I}_{1}=O(\mathrm{log}(NT)^{q/2}(T\underline b)^{-q/2+1}+T^{-(\alpha
p)\wedge (p/2-1)}).$$ Similar argument leads to the same bound for $\mathrm{I}%
_{2}$ part. For part $\mathrm{I}_{3}$, by Lemma \ref{lem:gapprox_homo} we
have 
\begin{equation*}
\mathrm{I}_{3}\lesssim (bT)^{-1/6}\mathrm{log}^{7/6}(NT)+(T^{2/q}/(bT))^{1/3}%
\mathrm{log}(NT)+T^{-(\alpha p)\wedge (p/2-1)}.
\end{equation*}

For part $\mathrm{I}_{4}$, denote $E=(E_{j_{1}j_{2}})_{1\leq j_{1},j_{2}\leq
N}$ as the conditional covariance matrix for $(b_jT)^{1/2}%
\sum_{t=1}^{T}(a_{jt,b}\epsilon _{jt}$ $-\sum_{l=1}^{N}a_{lt,b}\epsilon
_{lt}/N)$ given $\mathcal{F}_{T}.$ The diagonal entity takes the form 
\begin{align*}
E_{jj}& =b_jT\sum_{h=1-T}^{T-1}\sum_{t=1\vee (1-h)}^{T\wedge (T-h)}\mathbb{E}%
\Big[\big(a_{jt,b}\epsilon _{jt}-\sum_{l_{1}=1}^{N}a_{l_{1}t,b}\epsilon
_{l_{1}t}/N\big)\big(a_{j(t+h),b}\epsilon
_{j(t+h)}-\sum_{l_{2}=1}^{N}a_{l_{2}(t+h),b}\epsilon _{l_{2}(t+h)}/N\big)%
\Big] \\
& \cdot (1+O(b_j)).
\end{align*}%
Similarly to the proof of Lemma \ref{lem:gapprox}, with probability
greater than $1-O(T^{-(\alpha p)\wedge (p/2-1)})$, for some constant $c'>0$, we have 
\begin{equation*}
\min_{1\leq j\leq N}E_{jj}\geq c^{\prime }.
\end{equation*}
Hence by Lemma \ref{lem:anticon}, we have 
\begin{equation}
\sup_{u\in \mathbb{R}}\mathbb{P}\Big((T\underline{b})^{1/2}|\mathcal{Q}_{z}-u|<2l\Big)\lesssim l
\mathrm{log}(N)^{1/2}.  \label{eq:i4part2df_homo}
\end{equation}%
Hence
\begin{equation*}
\mathrm{I}_{4}=O\Big(\big(T^{1/2}\bar{b}^{5/2}+N^{1/q}T^{-\beta }\sqrt{\mathrm{log}(NT)}\big)
\mathrm{log}(N)^{1/2}\Big).
\end{equation*}%
By Lemma \ref{lem:comparison}, we have 
\begin{equation}
\sup_{u\in \mathbb{R}}\Big|\mathbb{P}\big((\underline{b}T)^{1/2}\mathcal{Q}_{z}\leq u%
\big)-\mathbb{P}\big(|Z+\tilde{\underline{d}}|_{\infty }\leq u\big)\Big|\lesssim
(\underline{b}+T^{-\beta})^{1/3}\mathrm{log}(N)^{2/3}.
\label{eq:i4part1df_homo}
\end{equation}
Desired result follows by combining above with $\mathrm{I}_{1}$-$\mathrm{I}_{4}$
parts.
${\tiny \blacksquare }$

\subsection{Proof of Proposition \protect\ref{thm:hatv} and Corollary \protect\ref{thm:power}}
\label{proofunkown}

\noindent \textbf{Proof of Proposition \ref{thm:hatv}. }Let $\bar{Y}_{jt}$
(resp. $\bar{f}_{jt}$ and $\bar{e}_{jt}$) be the local linear estimator of $%
Y_{jt}$ (resp. $f_{j}(X_{jt})$ and $e_{jt}$) at location $X_{jt}.$ Consider 
\begin{equation*}
Y_{jt}-\bar{Y}_{jt}=f_{j}(X_{jt})-\bar{f}_{jt}+e_{jt}-\bar{e}_{jt}
\end{equation*}%
Due to the continuity of $f_{j}$, we have $f_{j}(X_{jt})-\bar{f}%
_{jt}=O(b_{j})$. Random variables $e_{jt}$ are
weakly temporal dependent, by Theorem 2 in \cite{wu2016performance}, we have 
\begin{equation*}
\mathbb{P}\Big(\Big|\sum_{t=1}^{T}\mathbf{1}_{|X_{jt}-c_{0,j}|\leq b_{j}}\big(%
e_{jt}^{2}-\mathbb{E}(e_{jt}^{2})\big)\Big|>x\Big)\lesssim
b_{j}Tx^{-q/2}+e^{-x^{2}/(b_{j}T)}.
\end{equation*}%
Then by taking $x=c\mathrm{log}^{1/2}(N)$ for some constant $c>0$ large enough, we
have 
\begin{align*}
& \mathbb{P}\Big(\max_{1\leq j\leq N}(Tb_{j})^{-1/2}\Big|\sum_{t=1}^{T}%
\mathbf{1}_{|X_{jt}-c_{0,j}|\leq b_{j}}\big(e_{jt}^{2}-\mathbb{E}%
(e_{jt}^{2})\big)\Big|>x\Big) \\
\lesssim & N(\underline{b}T)^{1-q/4}\mathrm{log}^{-q/4}(N).
\end{align*}%
Hence $\max_{1\leq j\leq N}|\widehat{\sigma }_{e,j}^{2}-\sigma _{e,j}^{2}|=O_{\mathbb{P}}(\bar{b}+(\underline{b}T)^{-1/2}\mathrm{log}^{1/2}(N))$. For $%
v_{j}^{2}$, note that 
\begin{align*}
|v_{j}^{2}-\hat{v}_{j}^{2}|=&
(Tb_{j})\sum_{t=1}^{T}(w_{jt,b}^{+}-w_{jt,b}^{-})^{2}\big|\sigma _{e,j}^{2}-\widehat{\sigma }_{e,j}^{2}\big| \\
\leq & (Tb_{j})\sum_{t=1}^{T}(w_{jt,b}^{+}-w_{jt,b}^{-})^{2}\max_{1\leq j\leq N}|%
\widehat{\sigma }_{e,j}^{2}-\sigma _{e,j}^{2}|.
\end{align*}%
Since $\sum_{t=1}^{T}(w_{jt,b}^{+}-w_{jt,b}^{-})^{2}\lesssim
(T\underline{b})^{-1}$, we complete the proof. For the case in Assumption \ref{asmp:moment} (ii), the moment condition is even stronger, so the conclusion natually holds as well.
${\tiny \blacksquare }$%
\bigskip 

\noindent \textbf{Proof of Corollary \ref{thm:power}. } $(i)$ The
result follows from Theorem \ref{thm:ga} and Proposition \ref{thm:hatv}. To see this,
notice that $\hat{v}_{j}^{2}\overset{p}{\rightarrow }v_{j}^{2}>0$ under $%
H_{a}^{\left( 1\right) }$ where the convergence in probability holds
uniformly in $j\in \left[ N\right] .$ Then 
\begin{eqnarray*}
\mathcal{\hat{I}} &=&\max_{j\in \left[ N\right] }\left( Tb_{j}\right)
^{1/2}\left\vert \hat{\gamma}_{j}\right\vert /\hat{v}_{j} \\
&\leq &\max_{j\in \left[ N\right] }\left( Tb_{j}\right) ^{1/2}\left\vert 
\hat{\gamma}_{j}-\gamma _{j}\right\vert /\hat{v}_{j}+\max_{j\in \left[ N%
\right] }\left( Tb_{j}\right) ^{1/2}\left\vert \gamma _{j}\right\vert /\hat{v%
}_{j} \\
&\lesssim &\max_{j\in \left[ N\right] }\left( Tb_{j}\right) ^{1/2}\left\vert 
\hat{\gamma}_{j}-\gamma _{j}\right\vert /v_{j}+\max_{j\in \left[ N\right]
}\left( Tb_{j}\right) ^{1/2}\left\vert \gamma _{j}\right\vert :=\text{I}_{1}+%
\text{I}_{2}.
\end{eqnarray*}%
As in the proof of Theorem \ref{thm:ga}, I$_{1}$ can be approximated by $%
\max_{j\in \left[ N\right] }\left\vert Z_{j}\right\vert =O_{\mathbb{P}}\left( (%
\mathrm{log}N)^{1/2}\right) .$ Under the stated conditions, I$_{2}\gg (%
\mathrm{log}N)^{1/2}.$ It follows that $\lim_{\left( N,T\right) \rightarrow
\infty }\mathbb{P(}\mathcal{\hat{I}}>q_{\alpha })=1$ by noticing that $%
q_{\alpha}\asymp (\mathrm{log}N)^{1/2}.$

$(ii)$ The result follows from Theorem \ref{thm:homotesting} and Proposition \ref%
{thm:hatv} and arguments as used in part $(i)$. ${\tiny \blacksquare }$
\subsection{Proof of Theorem \ref{thm:unknown}}
\begin{proof} 
Denote
\begin{equation}
\mathcal{I}_{\epsilon}^{C}=\max_{1\leq i\leq K}\max_{1\leq j\leq N}\frac{(Tb_{j})^{1/2}}{v_j(c_i)}\Big|\sum_{t=1}^{T}w_{jt,b}(c_{i})\big[\varepsilon
_{j}(X_{jt},U_{jt})+\sigma _{j}(X_{jt},U_{jt})\epsilon _{jt}\big]
+
\gamma_j(c_i)\Big|.
\end{equation}
Let $\mathcal{I}_{\epsilon,m}^{C}$ be $\mathcal{I}_{\epsilon}^{C}$ with $\epsilon_{jt}$ therein replaced by $\epsilon_{jt,m}.$
Let $m=T/2.$ For any $l>0,$ similar to \eqref{eq:thmmaindecomp}, we have 
\begin{align}
& \sup_{u\in \mathbb{R}}\Big|\mathbb{P}\big(\mathcal{I}^C\leq u)-%
\mathbb{P}(|Z^C+\underline{d}^C|\leq u\big)\Big|  \notag \\
\leq & \mathbb{P}\Big(|\mathcal{I}^C-\mathcal{I}^C_{\epsilon ,m}|\geq l%
\Big)
+\sup_{u\in \mathbb{R}}\Big|\mathbb{P}(\mathcal{I}_{\epsilon ,m}^C\leq u)-%
\mathbb{P}(|Z^C+\underline{d}^C|_\infty\leq u)\Big|  \notag \\
& +\sup_{u\in \mathbb{R}}\mathbb{P}\Big(\big||Z^C+\underline{d}^C|_\infty-u\big|\leq l
\Big)=:\mathrm{I}_{1}+\mathrm{I}_{2}+\mathrm{I}_{3}.
\end{align}
For part $\mathrm{I}_{1}$, by \eqref{eq:bddiiep}, we have 
\begin{equation*}
|\mathcal{I}^C-\mathcal{I}^C_{\epsilon ,m}|\leq |\mathcal{I}^C-\mathcal{I}^C
_{\epsilon }|+|\mathcal{I}^C_{\epsilon }-\mathcal{I}^C_{\epsilon
,m}|=O(\bar{b}^{2})+|\mathcal{I}^C_{\epsilon }-\mathcal{I}^C_{\epsilon ,m}|.
\end{equation*}
We define $l^{\prime}=(NK)^{1/q}T^{-\beta }\sqrt{\mathrm{log}(NKT)}$. Then by a similar argument as in Lemma 
\ref{lem:mdependQ_homo} with $m=T/2$, we have 
\begin{equation}
\mathbb{P}\big(|\mathcal{I}^C_{\epsilon }-\mathcal{I}^C_{\epsilon
,m}|\geq l^{\prime }\big)=O\big(\mathrm{log}%
(NKT)^{q/2}(T\underline b)^{-q/2+1}+T^{-(\alpha p)\wedge (p/2-1)}\big).
\end{equation}%
Hence for $l=l^{\prime}+cT^{1/2}\bar{b}^{5/2},$ some constant $c>0$ large enough, we have 
$$
\mathrm{I}_{1}=O(\mathrm{log}(NKT)^{q/2}(T\underline b)^{-q/2+1}+T^{-(\alpha
p)\wedge (p/2-1)}).$$ 
For part $\mathrm{I}_2$, similar to Lemma \ref{lem:gapprox_homo} and \eqref{eq:i4part2df_homo}, we
have 
\begin{align*}
\mathrm{I}_{2}
&\lesssim (\underline{b}T)^{-1/6}\mathrm{log}^{7/6}(NKT)+(T^{2/q}/(\underline{b}T))^{1/3}%
\mathrm{log}(NKT)\\
&+\big(T^{1/2}\bar{b}^{5/2}+(NK)^{1/q}T^{-\beta }\sqrt{\mathrm{log}(NKT)}\big)
\mathrm{log}(NK)^{1/2}+T^{-(\alpha p)\wedge (p/2-1)}.
\end{align*}
For part $\I_3,$ by Lemma \ref{lem:anticon}, we have 
\begin{equation}
\sup_{u\in \mathbb{R}}\mathbb{P}\Big(\big||Z^C+\underline{d}^C|_\infty-u\big|<2l\Big)\lesssim l
\mathrm{log}(NK)^{1/2}. 
\end{equation}%
Desired result follows by combining $\mathrm{I}_{1}$-$\mathrm{I}_{3}$
parts.
\end{proof}

\subsection{Proof of Theorem  \ref{consistency}}
For any $c_{0j}-b_j<c\leq c_{0j}$, we can decompose the difference into three parts 
\begin{align*}
&\hat\gamma_j(c)-\hat\gamma_j(c_{0j})\\
=&\sum_{t=1}^{T}\big(w_{jt,b}(c)-w_{jt,b}(c_{0j})\big)Y_{jt}\\
=&\sum_{t=1}^{T}\big(w_{jt,b}(c)-w_{jt,b}(c_{0j})\big)f_{j}(X_{jt})+\sum_{t=1}^{T}\big(w_{jt,b}(c)-w_{jt,b}(c_{0j})\big)\gamma _{j}\mathbf{1}_{\{X_{jt}\geq c_{0j}\}}\\
&+\sum_{t=1}^{T}\big(w_{jt,b}(c)-w_{jt,b}(c_{0j})\big)e_{jt}
=:\I_{1j}+\I_{2j}+\I_{3j}.
\end{align*}
For part $\I_{1j}$, due to the smoothness of the function $f_j(\cdot)$, 
\begin{align*}
\sum_{t=1}^{T}w_{jt,b}(c)f_{j}(X_{jt})\lesssim b_j . 
\end{align*}
For part $\I_{2j},$ since $\sum_{t=1}^{T}w_{jt,b}^+(c)=\sum_{t=1}^{T}w_{jt,b}^-(c)=1,$ we have
\begin{align*}
&\sum_{t=1}^{T}\big(w_{jt,b}(c_{0j})-w_{jt,b}(c)\big)\gamma _{j}\mathbf{1}_{\{X_{jt}\geq c_{0j}\}}\\
=&\gamma_j\sum_t \Big[w_{jt,b}(c_{0j})- w_{jt,b}(c-c_{0j}+ c_{0j})\Big]\1_{\{X_{jt}\geq c_{0j}\}}\\
\lesssim &\gamma_j\sum_t \Big[\frac{1}{Tb_j}\frac{|c-c_{0j}|}{b_j}\Big]\1_{\{|X_{jt} -c_{0j}|<b_j\}}\\
\asymp &\gamma_j|c-c_{0j}|/b_j.
\end{align*}
For part $\I_{3j},$ consider the observations in region
\begin{align*}
\A_1:=\Big\{t: X_{jt}\in [c_{0j}-b_j, c)\Big\}.    
\end{align*}
In this region, both $w_{jt,b}(c)$ and $w_{jt,b}(c_{0j})$ are non-zero. By Lemma \ref{lem:propwitb}, the variance of $e_{jt}$ in this region can be bounded as
\begin{align*}
\var\Big(\sum_{t\in\A_1}(w_{jt,b}(c)-w_{jt,b}(c_{0j})) e_{jt}\Big)\lesssim 
(Tb_j)^{-2}(1/g_j(c_{0j})-1/g_j(c))^2Tb_jg_j(c_{0j})
\lesssim (Tb_j)^{-1}|c-c_{0j}|^2b_j^{-2}.
\end{align*}
Consider another kind of region
\begin{align*}
\A_2:=\Big\{t: X_{jt}\in [c-b_j, c_{0j}-b_j)\Big\}.  
\end{align*}
In region $\A_2,$ different from $\A_1,$ only 
$w_{jt,b}(c)$ is non-zero. The variance of noise in this region follows
\begin{align*}
\var\Big(\sum_{t\in\A_2}(w_{jt,b}(c)-w_{jt,b}(c_{0j})) e_{jt}\Big)
\lesssim 
(Tb_j)^{-2}g_j(c_{0j})^{-2}|c-c_{0j}|Tg_j(c_{0j})
\lesssim (Tb_j)^{-2}|c-c_{0j}|T.
\end{align*}
The rest regions can be similarly bounded. A similar GA can be constructed as in the proof of Theorem \ref{thm:ga}, which can be used to handle the maximum over $j$. Hence we have
\begin{align*}
\max_{1\leq j\leq N}|\I_{3j}|=O_\PP\Big((Tb_j)^{-1/2}\log^{1/2}(N)|c-c_{0j}|/b_j+(Tb_j)^{-1}\log^{1/2}(N)|c-c_{0j}|^{1/2}T^{1/2} \Big).    
\end{align*}
Since $\hat\gamma_j(\hat c)$ shall be the largest, for $c=\hat c,$
\begin{align*}
\gamma_j|c-c_{0j}|/b_j\lesssim (Tb_j)^{-1/2}\log^{1/2}(N)|c-c_{0j}|/b_j+(Tb_j)^{-1}\log^{1/2}(N)|c-c_{0j}|^{1/2}T^{1/2},    \end{align*} 
which can only be achieved if the following bound uniformly over $j$, $$|c-c_{0j}|=O_{\PP}( \log(N)/(\gamma_j^2T)).$$

\subsection{Proofs for the Unknown Threshold Case }

\begin{proof}(Proof of Proposition \ref{thm:hatvap})
Let $\bar f_{jt}^*$ and $\bar e_{jt}^*$ be the local linear estimator of $f_{j}(X_{jt})$ and 
$e_{jt}$ respectively with bandwidth $b^*$, where $b^* \ll \underline{b}$.  
Recall that  $\hat e_{jt} = e_{jt}-\bar e_{jt}^*+ f_{j}(X_{jt})- \bar f_{jt}^*+d_{jt}^*$, where $d_{jt}^*=0$ if $|X_{jt}-c_{0j}|>b^*$ and is $\sum_{t=1}^T w_{jt}^*\gamma_j$ for $w_{jt}^*$ being the local linear weight.
For $c_i$ satisfying $\min_{1\leq i\leq K}|c_i -c_{0j}|\leq b^*, $ we have
\begin{align*}
\hat{\sigma }_{e,j}^{2}(c_i) 
=&\frac{\sum_{t=1}^{T}%
\mathbf{1}_{\{|X_{jt}-c_{i}|\leq b_j,|X_{jt}-c_{0j}|\leq b^*\}}\phi_{\acute{a}}(\hat{e}_{jt}^{2})}{\sum_{t=1}^{T}%
\mathbf{1}_{\{|X_{jt}-c_{i}|\leq b_{j}\}}}
+\frac{\sum_{t=1}^{T}%
\mathbf{1}_{\{|X_{jt}-c_{0j}|>b^*, |X_{jt}-c_{i}|\leq b_j\}}\phi_{\acute{a}}(\hat{e}_{jt}^{2})}{\sum_{t=1}^{T}%
\mathbf{1}_{\{|X_{jt}-c_{i}|\leq b_{j}\}}}\\
=&:\I_{1ij}+\I_{2ij}.
\end{align*}
By a similar argument as in Lemma \ref{lem:propwitb}, we have with probability greater than $1-O(T^{-(\alpha p)\wedge(p/2-1)}),$
\begin{align}
\label{eq:concenforb2thm}
&\max_{1\leq i\leq K, 1\leq j\leq N}\Big|\sum_{t=1}^{T}
\big(\mathbf{1}_{\{|X_{jt}-c_{i}|\leq b_{j}\}} -\EE \mathbf{1}_{\{|X_{jt}-c_{i}|\leq b_{j}\}}\big) \Big|
\lesssim \sqrt{\log(TNK)(T\underline b)}+(T\bar b)\bar b.
\end{align}
Same argument can be applied on $\sum_{t=1}^{T}%
\mathbf{1}_{\{|X_{jt}-c_{i}|\leq b_j,|X_{jt}-c_{0j}|\leq b^*\}}$. 
For part $\I_{1ij},$ since $|\phi_{\acute{a}}|\leq \acute{a}$, we have
\begin{align*}
\max_{1\leq i\leq K, 1\leq j\leq N}|\I_{1ij}|\leq  \max_{1\leq i\leq K, 1\leq j\leq N}\frac{\sum_{t=1}^{T}%
\mathbf{1}_{\{|X_{jt}-c_{i}|\leq b_j,|X_{jt}-c_{0j}|\leq b^*\}}\acute{a}}{\sum_{t=1}^{T}%
\mathbf{1}_{\{|X_{jt}-c_{i}|\leq b_{j}\}}} 
=O_\PP( \acute{a} b^*/\bar b )
\end{align*}
For $\I_{2ij}$ part, $d_{jt}^*=0$. Note that by Markov's inequality
\begin{align*}
 \max_{1\leq i\leq K, 1\leq j\leq N}\big|\EE(\phi_{\acute{a}}({e}_{jt}^{2}))-\EE({e}_{jt}^{2})\big|   
 \leq \acute{a}^{-p/2}.
\end{align*}
Moreover by an argument similar to \eqref{eq:concenforb2thm}, we have
\begin{align*}
&\max_{1\leq i\leq K, 1\leq j\leq N}\bigg|\frac{\sum_{t=1}^{T}%
\mathbf{1}_{\{|X_{jt}-c_{0j}|>b^*, |X_{jt}-c_{i}|\leq b_j\}}\big[\phi_{\acute{a}}({e}_{jt}^{2})-\EE(\phi_{\acute{a}}({e}_{jt}^{2}))\big]}{\sum_{t=1}^{T}%
\mathbf{1}_{\{|X_{jt}-c_{i}|\leq b_{j}\}}}
\bigg|
=O_{\PP}\big(\sqrt{\log(NTK)/(T\underline b)}\big).
\end{align*}
Hence $\max_{1\leq i\leq K, 1\leq j\leq N}|\I_{2ij}|=O_{\PP}(\sqrt{\log(NTK)/(T\underline b)}+\bar b+\acute{a}b^*/\bar b+\acute{a}^{-p/2}).$
\end{proof}

\subsection{Theorems and Proofs on Derivatives}
\label{derivatives}
In this subsection, we present derivative estimators for the comparison with the slope coefficients in linear models.
Consider the test statistic 
\begin{equation}
\mathcal{I}^{[1]}=\max_{1\leq j\leq N}(Tb_{j}^3)^{1/2}|\hat{\gamma}_{j}^{[1]}/v_{j}^{[1]}|,
\label{I1_derivative}
\end{equation}%
where 
\begin{equation}
v_{j}^{[1]2}=v_{j}^{[1]2}(c_{0j})=(Tb_{j}^3)%
\sum_{t=1}^{T}(w_{jt,b}^{+[1]}-w_{jt,b}^{-[1]})^{2}\mathrm{Var}\big(e_{jt}|X_{jt}\big).
\label{eq:def_sigma_derivative}
\end{equation}
Define $d_{j}^{[1]}=(Tb_j^3)^{1/2}\gamma^{[1]} _{j}/v_{j}^{[1]}$ and $\underline{d}^{[1]}
=(d_{1}^{[1]},d_{2}^{[1]},\ldots ,d_{N}^{[1]})^{\top }$.
\begin{theorem}
\label{thm:ga_derivative}
Let Assumptions \ref{asmp:moment}-\ref{kernel} hold. Then we have
\begin{equation}
\sup_{u\in \mathbb{R}}\big|\mathbb{P}(\mathcal{I}^{[1]}\leq u)-\mathbb{P}(|Z+%
\underline{d}^{[1]}|_{\infty }\leq u)\big|\lesssim \mathcal{R}_{NT}.
\end{equation}%
\end{theorem}

Denote $\mathcal{I}^{C[1]}$, $Z^{C[1]}$ and $\underline{d}^{C[1]}$ as the derivative counterparts of $\mathcal{I}^{C[1]}$,  $Z^{C}$ and $\underline{d}^{C}$.
Let $$d_{jK}^{[1]}=(Tb_j^3)^{1/2}(\gamma _{j}^{[1]}(c_{1})/v_{j}^{[1]}(c_{1}),\ldots ,\gamma
_{j}^{[1]}(c_{K})/v_{j}^{[1]}(c_{K}))^{\top },$$
where 
$\underline{d}%
^{C[1]}=(d_{1K}^{[1]\top },d_{2K}^{[1]\top },\ldots ,d_{NK}^{[1]\top })^{\top }.$
Let $\Sigma^{C[1]}$  be the variance covariance matrix of $Z^{C[1]}$.
For each $j\in \lbrack N]$ and $1\leq i_{1},i_{2}\leq K$,
the covariance between the 
$i_{1}$- and $i_{2}$-th elements of $Z_{j}^{C[1]}$  {{is
given by{%
\begin{equation*}
\cov(Z_{j,i_1}^{C[1]}, Z_{j,i_2}^{C[1]})=(v_{j}(c_{i_{1}}^{[1]})v_{j}^{[1]}(c_{i_{2}}))^{-1}(Tb_{j}^3)%
\sum_{t=1}^{T}w^{[1]}_{jt,b}(c_{i_{1}})w^{[1]}_{jt,b}(c_{i_{2}})\mathrm{Var}\big(%
e_{jt}|X_{jt}\big).
\end{equation*}%
}}}${\normalsize {{Z_{j_1}^{C[1]}}}}$ and $Z_{j_2}^{C[1]}$ are
independent for $ j_1\neq j_2$,  so that the matrix $\Sigma^{C[1]}$ is block diagonal.

\begin{equation}
\mathcal{I}^{C[1]}=\max_{1\leq i\leq K}\max_{1\leq j\leq N}(Tb_{j}^3)^{1/2}|\hat{%
\gamma}_{j}^{[1]}(c_{i})/v_{j}^{[1]}(c_{i})|.
\end{equation}

\begin{theorem}\label{thm:unknownde}
{\normalsize (Gaussian Approximation) Suppose that Assumptions \ref
{asmp:moment}-\ref{kernel} hold. Suppose that }$\Delta_{c,\max }\rightarrow 0.$ Then 
\begin{equation}
\sup_{u\in \mathbb{R}}\big|\mathbb{P}(\mathcal{I}^{C[1]}\leq u)-\mathbb{P}%
(|Z^{C}+\underline{d}^{C[1]}|_{\infty }\leq u)\big|\lesssim \mathcal{R}_{(NK)T}.
\label{eq:ga11}
\end{equation}
\end{theorem}
Define
$\hat{v}_j^{[1]2}=Tb_j^3\sum_{t=1}^{T}\left(w_{jt,b}^{+[1]}-w_{jt,b}^{-[1]}\right)^2\hat{\sigma}_{e,j}^2$ and 
$\hat{\tilde{v}}_j^{[1]}=\left(\hat{v}_j^{[1]2}\left(1-\frac{1}{N}\right)^2+\frac{1}{N^2}\sum_{i\neq j}^N\hat{v}_i^{[1]2}\right)^{1/2}$.
\begin{proposition}[Consistency of the variance estimator]
{\normalsize \label{thm:hatvde} Suppose that Assumptions {\ref{asmp:moment}}-\ref{kernel} hold and $N(\underline{b}T)^{-q/4}\rightarrow 0.$ Then we have $
\max_{j\in \lbrack N]}|\hat{v}_{j}^{[1]2}-v_{j}^{[1]2}|$ $=O_{\mathbb{P}}((%
\underline{b}T)^{-1/2}\mathrm{log}(N)).$}
\end{proposition}
Denote $\mathcal{\hat{I}}^{[1]}$ and $\mathcal{\hat{Q}}^{[1]}$ as $\mathcal{{I}}^{[1]}$ and $\mathcal{{Q}}^{[1]}$ replaced by estimated variance $\hat{v}_j^{[1]2}$ and $\hat{\tilde{v}}_j^{[1]}$.
\begin{corollary}[Power]
\label{thm:powerde}Suppose that Assumptions \ref%
{asmp:moment}-\ref{kernel} hold. Let $\bar{\gamma}^{[1]}=%
\sum_{j=1}^{N}\gamma _{j}^{[1]}/N.$ 

{\normalsize $(i)$ If $\mathcal{R}_{NT}\to 0$ in Theorem \ref{thm:ga} and $\max_{j\in \left[ N\right] }(Tb_{j}^3)^{1/2}|\gamma
_{j}^{[1]}|\gg (\mathrm{log}N)^{1/2},$ then we have $\lim_{\left( N,T\right) \rightarrow
\infty }\mathbb{P}(\mathcal{\hat{I}}^{[1]}>q_{\alpha})=1;$ }

{\normalsize $(ii)$ If $\mathcal{R}_{NT}\to 0$ in Theorem \ref{thm:homotesting} and $\max_{j\in \left[ N\right] }(Tb_{j}^3)^{1/2}|%
\gamma _{j}^{[1]}-\bar{\gamma}^{[1]}|\gg (\mathrm{log}N)^{1/2},$ then $\lim_{\left(
N,T\right) \rightarrow \infty }\mathbb{P}(\mathcal{\hat{Q}}^{[1]}>q_{\alpha})=1. $ }
\end{corollary}
Regarding the unknown-threshold case for derivatives, the algorithms and theoretical results can be developed in a similar manner, and we omit the repetitive details here. The convergence rate of the derivative estimator can be further improved by using local quadratic estimation, see, for example, \cite{fan2018local}.

To prove the above results we introduce a few technical lemmas.
\begin{lemma}
(Properties for weights in derivatives) \label{lem:propwitb[1]} Under Assumptions \ref
{asmp:moment}-{\ref{asmp:bddxj1j2density}} and \ref{boundedness}-\ref{kernel}, the followings hold:

(i) $\sum_{t=1}^{T}w_{jt,b}^{+[1]}=\sum_{t=1}^{T}w_{jt,b}^{-[1]}=0\ $and$\
 \sum_{t=1}^{T}(X_{jt}-c_{j0})w_{jt,b}^{+[1]}=%
\sum_{t=1}^{T}(X_{jt}-c_{j0})w_{jt,b}^{-[1]}=1.$

(ii) For $l=0,1,2,$
 on set $\mathcal{A}_{T}$, with $\mathcal{A}_T$ defined in Lemma \ref{lem:propwitb},
\begin{align}
& \max_{1\leq j\leq N}\Big|w_{jt,b}^{+[1]}-\frac{K((X_{jt}-c_{j0})/b_{j})}{%
g_{j}(c_{j0})Tb^2_{j}}\frac{K_1^{+}-K_0^{+}(X_{jt}-c_{j0})/b_{j}}{%
-K_{2}^{+}K_{0}^{+}+K_{1}^{+2}}\mathbf{1}_{\{X_{jt}\geq c_{j0}\}}\Big|  \notag \\
\lesssim & \sqrt{\mathrm{log}(TN)/(T^3\underline{b}^5)}+1/(T\underline b),
\end{align}%
and moreover, 
\begin{equation}
\sum_{t=1}^{T}|w_{jt,b}^{+}|=O(1/\underline b)\quad \text{and}\quad
\sum_{t=1}^{T}|w_{jt,b}^{-}|=O(1/\underline b)  \label{eq:bddforsumw1}
\end{equation}%
uniformly over $j$.
\end{lemma}

\noindent \textbf{Proof of Lemma \ref{lem:propwitb[1]}. }
Part (i) can be calculated through elementary calculation. 
Part (ii) can be deduced by a similar argument as in Lemma \ref{lem:propwitb}.
${\tiny \blacksquare }$

\begin{lemma}[$m$-dependent approximation for derivative case]
\label{lem:mdepend_derivative} 
For some $1\ll m\leq T/2,$ 
\begin{itemize}
    \item[(i)] assume that the conditions in Theorem \ref{thm:ga_derivative} $(i)$
hold, we have 
\begin{equation*}
\mathbb{P}\big(|I_{\epsilon}^{[1]}-I_{\epsilon ,m}^{[1]}|\geq z\big)\lesssim N\mathrm{%
log}^{q}(NT)(T\underline{b})^{1-q/2}m^{-q\beta }z^{-q}+Ne^{-z^{2}m^{2\beta }}+%
\mathbb{P}(\mathcal{A}_{T}^{c}).
\end{equation*}%
    \item[(ii)] assume that the conditions in Theorem \ref{thm:ga_derivative} $(ii)$
hold, we have 
\begin{equation*}
\mathbb{P}\big(|I_{\epsilon }^{[1]}-I_{\epsilon ,m}^{[1]}|\geq z\big)\lesssim N\mathrm{%
exp}\Big\{-c\frac{z^{2}}{m^{-\beta }(\underline{b}T)^{-1/2}z+m^{-2\beta }}%
\Big\}+\mathbb{P}(\mathcal{A}_{T}^{c}),
\end{equation*}%
\end{itemize}

where $\mathcal{A}_{T}$ is defined in Lemma \ref{lem:propwitb}.
\end{lemma}

\noindent \textbf{Proof of Lemma \ref{lem:mdepend_derivative}. }
We only show case $(i)$, for the case $(ii)$ the result can be similary derived. Let
\begin{align}
r_{j}^{[1]}(X_{jt},U_{jt})
=\sigma_j(X_{jt},U_{jt})/v_j^{[1]}
\end{align}
and
\begin{equation}
\xi _{jk}^{[1]}=(Tb_{j}^3)^{1/2}\sum_{t=1\vee
(k+m)}^{T}w_{jt,b}^{[1]}r_{j}^{[1]}(X_{jt},U_{jt})A_{t-k,j,\cdot}^\top\eta _{k}.
\end{equation}
Similar to \eqref{eq:bddiepandepm}, we have
\begin{align*}
|I_{\epsilon }^{[1]}-I_{\epsilon ,m}^{[1]}|\mathbf{1}_{\mathcal{A}_{T}}\leq  \max_{1\leq
j\leq N}\Big|\sum_{k\leq T-m}\xi _{jk}^{[1]}\Big|.
\end{align*}
Then by \eqref{eq:funlmd}, we have 
\begin{align}
\sigma^{[1]2}& =\max_{1\leq j\leq N}\sum_{k\leq T-m}%
\mathbb{E}(\xi _{jk}^{[1]2}|\mathcal{F}_{T},\tilde{\mathcal{F}}_{T})  \notag \\
& \lesssim \max_{1\leq j\leq N}Tb_{j}^3\sum_{k\leq T-m}\Big(\sum_{t=1\vee
(k+m)}^{T}|w_{jt,b}^{[1]}r_j^{[1]}(X_{jt},U_{jt})A_{t-k,j,\cdot }|_{2}\Big)^{2}\nonumber\\
&\lesssim m^{-2\beta}. 
\end{align}
By Lemma \ref{lem:freedman}, for any $z>0,$ 
\begin{align}
\label{eq:funlmd_derivative}
& \mathbb{P}\Big(|I_{\epsilon }^{[1]}-I_{\epsilon ,m}^{[1]}|\mathbf{1}_{\mathcal{A}%
_{T}}\geq z\big|\mathcal{F}_{T},\tilde{\mathcal{F}}_{T}\Big)  \notag \\
\lesssim & \sum_{k\leq T-m}\sum_{1\leq
j\leq N}\mathbb{P}\Big(|\xi _{jk}^{[1]}|\geq u|\mathcal{F}_{T},\tilde{\mathcal{F}}_{T}\Big)+2N%
\mathrm{exp}\Big(-\frac{z^{2}}{zu+m^{-2\beta}}\Big).
\end{align}
By Markov's inequality, we have 
\begin{align}
\sum_{k\leq T-m}\mathbb{P}\Big(|\xi _{jk}^{[1]}|\geq u\Big)& \leq \sum_{k\leq T-m}u^{-q}\EE\big(\EE(|\xi_{jk}^{[1]}|^q|\F_T,\tilde\F_T)\big) \notag \\
&\lesssim u^{-q} \sum_{k\leq T-m}(Tb_{j}^3)^{q/2}
\EE\Big|\sum_{t=1\vee
(k+m)}^{T}w_{jt,b}^{[1]}r_{j}^{[1]}(X_{jt},U_{jt})A_{t-k,j,\cdot}^\top\Big|_2^q\notag\\
& \lesssim (T\underline{b})^{1-q/2}m^{-q\beta }u^{-q}.  \label{eq:bdforeta_derivative}
\end{align}%
We complete the proof by implementing 
\eqref{eq:bdforeta_derivative} into \eqref{eq:funlmd_derivative} with $u=z/\mathrm{log}(NT)$. 
${\tiny \blacksquare }$

\begin{lemma}
\label{lem:gapprox_derivarive}
Let $T/2\geq m\rightarrow \infty .$

$(i)$ Suppose that the conditions in Theorem \ref{thm:ga_derivative} $(i)$ hold. Then we
have 
\begin{align*}
& \sup_{u\in \mathbb{R}}\Big|\mathbb{P}(I_{\epsilon ,m}^{[1]}\leq u)-\mathbb{P}%
\Big(\max_{1\leq j\leq N}\big|Z_{j}+(Tb_{j}^3)^{1/2}\gamma _{j}^{[1]}/v_{j}^{[1]}\big|\leq
u\Big)\Big| \\
\lesssim & (\underline{b}T)^{-1/6}\mathrm{log}^{7/6}(NT)+(T^{2/q}/(%
\underline{b}T))^{1/3}\mathrm{log}(NT)+(\bar{b}+m^{-\beta })^{1/3}\mathrm{log%
}^{2/3}(N)+T^{-(\alpha p)\wedge (p/2-1)},
\end{align*}%
where the constant in $\lesssim $ is independent of $N,T.$

$(ii)$ Suppose that the conditions in Theorem \ref{thm:ga_derivative} $(ii)$ hold. Then we
have 
\begin{align*}
& \sup_{u\in \mathbb{R}}\Big|\mathbb{P}(I_{\epsilon ,m}^{[1]}\leq u)-\mathbb{P}%
\Big(\max_{1\leq j\leq N}\big|Z_{j}+(Tb_{j}^3)^{1/2}\gamma^{[1]} _{j}/v_{j}^{[1]}\big|\leq
u\Big)\Big| \\
\leq & (\underline{b}T)^{-1/6}\mathrm{log}(NT)^{7/6}+(\bar{b}+m^{-\beta
})^{1/3}\mathrm{log}^{2/3}(N)+T^{-(\alpha p)\wedge (p/2-1)},
\end{align*}%
where the constant in $\lesssim $ is independent of $N,T.$
\end{lemma}

\noindent \textbf{Proof of Lemma \ref{lem:gapprox_derivarive}. }
Note that
\begin{align}
I_{\epsilon ,m}^{[1]}\mathbf{1}_{\mathcal{A}_{T}}& =\max_{1\leq j\leq N}\frac{%
(Tb_{j}^3)^{1/2}}{v_{j}^{[1]}}\Big|\sum_{t=1}^{T}w_{jt,b}^{[1]}\big[\sigma
_{j}(X_{jt},U_{jt})\epsilon _{jt,m}+\varepsilon _{j}(X_{jt},U_{jt})+f_j(X_{jt})\big]\Big|  \notag \\
& =\max_{1\leq j\leq N}\frac{(Tb_{j}^3)^{1/2}}{v_{j}^{[1]}}\Big|\sum_{k=2-m}^{T}\xi
_{jk,1}^{[1]}+\sum_{t=1}^{T}\xi _{jt,2}^{[1]}+\gamma _{j}^{[1]}+O(\bar b)\Big|,
\label{eq:iepsilonmdecomp_derivative}
\end{align}
where $\xi
_{jk,1}^{[1]}$ and $\xi
_{jk,2}^{[1]}$ are the same as $\xi_{jk,1}$ and $\xi_{jk,2}$ in \eqref{eq:xijk12def} respectively. By Lemma \ref{lem:propwitb[1]},  $b_j w_{jt,b}^{[1]}$ has similar performance as $w_{jt,b}$, hence similar to \eqref{eq:bdd4iemandzj}, we have
\begin{align}
& \sup_{u\in \mathbb{R}}\bigg|\mathbb{P}\big( I_{\epsilon ,m}^{[1]}\mathbf{1}_{%
\mathcal{A}_{T}}\leq u\big)-\mathbb{P}\Big(\max_{1\leq j\leq N}\big|%
Z_{j}+(Tb_{j}^3)^{1/2}\gamma _{j}^{[1]}/v_{j}^{[1]}\big|\leq u\Big)\bigg|  \notag \\
\lesssim & (\underline{b}T)^{-1/6}\mathrm{log}^{7/6}(NT)+(T^{2/q}/(%
\underline{b}T))^{1/3}\mathrm{log}(NT)+(\bar{b}+m^{-\beta })^{1/3}\mathrm{log%
}^{2/3}(N).  \label{eq:bdd4iemandzj_derivative}
\end{align}

${\tiny \blacksquare }$

Then we show the proofs of the main theorems in this subsections.\\
\noindent \textbf{Proof of Theorem \ref{thm:ga_derivative}. }

$(i)$ Consider the notation
\begin{equation}
I_{\epsilon }^{[1]}=\max_{1\leq j\leq N}(Tb_{j}^3)^{1/2}\Big|%
\sum_{t=1}^{T}(w_{jt,b}^{+[1]}-w_{jt,b}^{-[1]})\big(\varepsilon
_{j}(X_{jt},U_{jt})+\sigma_{j}(X_{jt},U_{jt})\epsilon _{jt}\big)+\gamma
_{j}^{[1]}\Big|/v_{j}^{[1]}. 
\end{equation}%
Let $I_{\epsilon, m}^{[1]}$ be $I_{\epsilon }^{[1]}$ with $\epsilon_{jt}$ replaced by $\epsilon_{jt,m}.$ Similarly, let $I_{z}^{[1]}$ (resp. $I_{z,m}^{[1]}$) be $I_{\epsilon }$ (resp. $%
I_{\epsilon ,m}$) with $\eta_{t}$ therein replaced by $z_{t}$, where $%
\left\{ z_{t}\right\} _{t\in \mathbb{Z}}$ are i.i.d. Gaussian vectors with
zero mean and identity covariance matrix. 

Similar to the proof of Theorem \ref{thm:ga}, for $m=T/2$, by \eqref{eq:thmmaindecomp} we have
\begin{align}
& \sup_{u\in \mathbb{R}}\Big[\mathbb{P}\big(\mathcal{I}^{[1]}\leq u)-\mathbb{P}%
( I_{z}^{[1]}\leq u\big)\Big]  \notag \\
\leq & \mathbb{P}\big(|\mathcal{I}^{[1]}-I_{\epsilon ,m}^{[1]}|\geq l\big)+\mathbb{P}%
\big(| I_{z}^{[1]}-I_{z,m}^{[1]}|\geq l\big)+\sup_{u\in \mathbb{R}}\big|\mathbb{P}%
(I_{\epsilon ,m}^{[1]}\leq u)-\mathbb{P}(I_{z,m}^{[1]}\leq u)\big|  \notag \\
& +\sup_{u\in \mathbb{R}}\big|\mathbb{P}\big( I_{z}^{[1]}\leq u+2l\big)-\mathbb{P}%
\big(I_{z}^{[1]}\leq u\big)\big|  \notag \\
=&:\mathrm{I}_{1}+\mathrm{I}_{2}+\mathrm{I}_{3}+\mathrm{I}_{4}.
\end{align}%
For part $\mathrm{I}_{1}$, we have $|\mathcal{I}^{[1]}-I_{\epsilon ,m}^{[1]}|\leq |%
\mathcal{I}^{[1]}-I_{\epsilon }^{[1]}|+|I_{\epsilon }^{[1]}-I_{\epsilon ,m}^{[1]}|.$ By Assumption
\ref{asmp:smooth} and Lemma \ref{lem:propwitb[1]}, we have
\begin{align*}
& \max_{1\leq j\leq N}\Big|%
\sum_{t=1}^{T}w_{jt,b}^{+[1]}f_{j}(X_{jt})-\partial _{+}f_{j}(c_{j0})\Big| \\
=& \max_{1\leq j\leq N}\Big|\sum_{t=1}^{T}w_{jt,b}^{+[1]}\big[%
f_{j}(c_{j0})+\partial _{+}f_{j}(c_{j0})(X_{jt}-c_{j0})+O((X_{jt}-c_{j0})^{2})%
\big]-\partial _{+}f_{j}(c_{j0})\Big| \\
=& O(\bar{b}).
\end{align*}
Thus 
\begin{equation}
|\mathcal{I}^{[1]}-I_{\epsilon }^{[1]}|
\leq \max_{1\leq j\leq N}(Tb_{j}^3)^{1/2}\Big|%
\sum_{t=1}^{T}(w_{jt,b}^{+[1]}-w_{jt,b}^{-[1]})f_{j}(X_{jt})\Big|%
/v_{j}^{[1]}=O(T^{1/2}\bar{b}^{5/2}). 
\end{equation}%
We define $l'=N^{1/q}T^{-\beta }\sqrt{\mathrm{log}(NT)}$. Then by Lemma \ref{lem:mdepend_derivative} with $m=T/2$, we have 
\begin{equation}
\mathbb{P}\big(|I_{\epsilon }^{[1]}-I_{\epsilon ,m}^{[1]}|\geq l^{\prime }\big)=O\big\{%
\mathrm{log}(NT)^{q/2}(T\underline{b})^{-q/2+1}+T^{-(\alpha p)\wedge
(p/2-1)}\big\}.
\end{equation}%
Hence for $l=l'+cT^{ 1/2}\bar{b}^{5/2}$, where $c>0$ is some constant large
enough, we have 
\begin{equation*}
\mathrm{I}_{1}=O\big\{\mathrm{log}(NT)^{q/2}(T\underline{b})^{-q/2+1}+T^{-(\alpha p)\wedge (p/2-1)}\big\}.
\end{equation*}%
Similar argument leads to the same bound for part $\mathrm{I}_{2}$.

For part $\mathrm{I}_{3}$, by Lemma \ref{lem:gapprox_derivarive} we have 
\begin{equation*}
\mathrm{I}_{3}\lesssim (\underline{b}T)^{-1/6}\mathrm{log}%
^{7/6}(NT)+(T^{2/q}/(\underline{b}T))^{1/3}\mathrm{log}(NT)+(\bar{b}+T^{-\beta })^{1/3}\mathrm{log%
}^{2/3}(N)+T^{-(\alpha
p)\wedge (p/2-1)}.
\end{equation*}

For part $\mathrm{I}_{4}$, 
we use Lemma \ref{lem:anticon} to obtain 
\begin{equation*}
\mathrm{I}_{4}\leq \sup_{u\in \mathbb{R}}\mathbb{P}\big(| I_{z}^{[1]}-u\big|\leq 2l%
\big)\lesssim l\mathrm{log}(N)^{1/2}.
\end{equation*}%
We complete the proof by combining the $\mathrm{I}_{1}$-$\mathrm{I}_{4}$
parts and a similar argument for the other side of the inequality.

Similar justification can be made for part $(ii)$.
${\tiny \blacksquare }$

\noindent \textbf{Proof of Theorem \ref{thm:unknownde}, Proposition \ref{thm:hatvde}, and Corollary \ref{thm:powerde} }
Theorem \ref{thm:unknownde}, Proposition \ref{thm:hatvde} and Corollary \ref{thm:powerde} are derived from Theorem \ref{thm:unknown}, Proposition \ref{thm:hatv} and Corollary \ref{thm:power} respectively using arguments similar to those in the proof of Theorem \ref{thm:ga_derivative}.

\subsection{Gaussian Approximation for Joint Inference on Level and Derivative Jumps}
Denote
\begin{align*}
I^{\mathrm{comb}}
=
\max\left\{I,I^{[1]}\right\},    
\end{align*}
where $I$ and $I^{[1]}$ are defined in \eqref{I1} and \eqref{I1_derivative} respectively. For \(1\leq j\leq N\), define
\begin{align}
v_{j}^{\mathrm{comb}}
&=
T b_{j}^2
\sum_{t=1}^{T}
\bigl(w_{jt,b}^{+}-w_{jt,b}^{-}\bigr)
\bigl(w_{jt,b}^{+[1]}-w_{jt,b}^{-[1]}\bigr)
\operatorname{Var}
\bigl(e_{jt}\mid \mathcal F_T\bigr).
\end{align}
For each \(j\in[N]\), let
\[
\rho_j=v_{jj}^{\mathrm{comb}}/(v_jv_j^{[1]}).
\]
Define the \(2N\times 2N\) block diagonal covariance matrix
\begin{align*}
\Sigma^{\mathrm{comb}}
=
\operatorname{diag}
\left(
\Sigma^{\mathrm{comb}}_1,\ldots,\Sigma^{\mathrm{comb}}_N
\right),\quad\mathrm{where}\ 
\Sigma^{\mathrm{comb}}_j
=
\begin{pmatrix}
1 & \rho_j \\
\rho_j & 1
\end{pmatrix}.
\end{align*}
Let
\[
\underline d^{\mathrm{comb}}
=
\left(
d_1,d_1^{[1]},\ldots,d_N,d_N^{[1]}
\right)^\top.
\]
Let \(Z^{\mathrm{comb}}\sim N(0,\Sigma^{\mathrm{comb}})\) be a \(2N\)-dimensional
Gaussian vector. The following result follows by applying the Gaussian
approximation arguments for the level statistic in Theorem \ref{thm:ga} and for
the derivative statistic in Theorem \ref{thm:ga_derivative}.
\begin{corollary}    
\label{cor:ga_comb}
Let Assumptions \ref{asmp:moment}--\ref{kernel} hold. Then
\[
\sup_{u\in\mathbb R}
\left|
\mathbb P\left(\mathcal I^{\mathrm{comb}}\leq u\right)
-
\mathbb P\left(
\left|Z^{\mathrm{comb}}+\underline d^{\mathrm{comb}}\right|_\infty\leq u
\right)
\right|
\lesssim
\mathcal R_{NT}.
\]
\end{corollary}
Note that \(\Sigma^{\mathrm{comb}}\) is block diagonal across the cross-sectional index
\(j\). Thus the Gaussian limits are independent across different individuals, while
the level and derivative statistics for the same individual are allowed to be correlated.

Under the local alternative
  \begin{align*}
      (\gamma_j, \gamma_j^{[1]}) = (Tb)^{-1/2} c_j \cdot
  \left((\log N)^{1/2}, b^{-1}(\log N)^{1/2}\right),
  \end{align*}
   the two tests have comparable
  detection power, and neither component of the stacked statistic
  dominates the other. Under the contiguous alternative
  \begin{align*}
      (\gamma_j, \gamma_j^{[1]}) = \left(a (Tb)^{-1/2}, c (Tb^3)^{-1/2}\right),
  \end{align*}
the level test is asymptotically dominant for fixed $a, c$ as $b \to 0$,
  reflecting the slower convergence rate of the derivative estimator.

\section{Proofs of the Technical Lemmas in Appendix \protect\ref{SecA} \label%
{SecB}}

In this section, we prove the technical lemmas in Section \ref{SecA}.

\subsection{Proof of Lemma\ \protect\ref{lem:gapprox} in Section \protect
\ref{SecA.1}\label{SecB.1}}

\noindent \textbf{Proof of Lemma \ref{lem:gapprox}. }

\textbf{(i).} Recall $\tilde{\mathcal{F}}_{t}$ in \eqref{eq:xjtujtdef}. Define 
\begin{equation}
w_{jt,b}=(w_{jt,b}^{+}-w_{jt,b}^{-})\mathbf{1}_{\mathcal{A}_{T}}\text{ and }%
r_{j}(X_{jt},U_{jt})=\sigma _{j}(X_{jt},U_{jt})/v_{j}.  \label{eq:wjtandrjt}
\end{equation}%
Note that $I_{\epsilon ,m}$ on $\mathcal{A}_{T}$ can be rewritten as 
\begin{align}
I_{\epsilon ,m}\mathbf{1}_{\mathcal{A}_{T}}& =\max_{1\leq j\leq N}\frac{%
(Tb_{j})^{1/2}}{v_{j}}\Big|\sum_{t=1}^{T}w_{jt,b}\big[\sigma
_{j}(X_{jt},U_{jt})\epsilon _{jt,m}+\varepsilon _{j}(X_{jt},U_{jt})\big]%
+\gamma _{j}\Big|  \notag \\
& =\max_{1\leq j\leq N}\frac{(Tb_{j})^{1/2}}{v_{j}}\Big|\sum_{k=2-m}^{T}\xi
_{jk,1}+\sum_{t=1}^{T}\xi _{jt,2}+\gamma _{j}\Big|,
\label{eq:iepsilonmdecomp}
\end{align}%
where 
\begin{equation}
\label{eq:xijk12def}
\xi _{jk,1}=\Big(\sum_{t=1\vee k}^{T\wedge (k+m-1)}w_{jt,b}\sigma
_{j}(X_{jt},U_{jt})A_{t-k,j,\cdot }^{\top }\Big)\eta _{k}\quad \text{and}%
\quad \xi _{jt,2}=w_{jt,b}\varepsilon _{j}(X_{jt},U_{jt}).
\end{equation}

In the following we shall show that for $B_{T}=c\underline{b}^{-1/2}$ with
some constant $c>0$ being large enough:

\begin{enumerate}
\item[(a)] with probability greater than $1-O(T^{-(\alpha p)\wedge (p/2-1)})$%
, for some constant $c^{\prime }>0,$ we have 
\begin{equation*}
\min_{1\leq j\leq N}(Tb_{j})\sum_{k=2-m}^{T}\mathbb{E}\big(\xi _{jk,1}^{2}|%
\mathcal{F}_{T},\tilde{\mathcal{F}}_{T}\big)\geq c^{\prime };
\end{equation*}

\item[(b)] with probability greater than $1-O(T^{-(\alpha p)\wedge (p/2-1)})$%
, for $l=1,2,$ we have 
\begin{equation*}
\max_{1\leq j\leq N}T^{1+l}b_{j}^{(2+l)/2}\sum_{k=2-m}^{T}\mathbb{E}(|\xi
_{jk,1}|^{2+l}|\mathcal{F}_{T},\tilde{\mathcal{F}}_{T})\leq B_{T}^{l}.
\end{equation*}
\end{enumerate}
Since $|w_{jt,b}|\lesssim (Tb_{j})^{-1}$, we have $\max_{k}\mathbb{E}%
(\max_{j}|\xi _{jk,1}|^{q}|\mathcal{F}_{T},\tilde{\mathcal{F}}_{T})\lesssim
(Tb_{j})^{-q},$ and thus 
\begin{equation*}
\mathbb{E}\left( \max_{1\leq j\leq N}(Tb_{j}^{1/2}|\xi _{jk,1}|/B_{T})^{q}%
\big|\mathcal{F}_{T},\tilde{\mathcal{F}}_{T}\right) \leq 2,
\end{equation*}%
for all $k.$ 
Denote 
\[
S_1=
\sum_{k=2-m}^{T}\left(
(Tb_1)^{1/2}v_1^{-1}\xi_{1k,1},
\ldots,
(Tb_N)^{1/2}v_N^{-1}\xi_{Nk,1}
\right)^\top
\]
and define its conditional covariance matrix
\[
\Sigma_{1}
=
\operatorname{Cov}
\left(
S_1\mid \mathcal F_T,\widetilde{\mathcal F}_T
\right).
\]
Let \(G\sim N(0,I_N)\) be independent of
\(\mathcal F_T\) and \(\widetilde{\mathcal F}_T\), and define
\[
\widetilde Z_1
=
\Sigma_{1}^{1/2}G.
\]
Then, conditional on \(\mathcal F_T\) and \(\widetilde{\mathcal F}_T\),
$\widetilde Z_1
=(\widetilde Z_{1,1},\ldots, \widetilde Z_{N,1})^\top$ is a centered Gaussian vector with covariance matrix
\(\Sigma_{1,T}\), and hence has the same conditional covariance structure as
\(S_1\).


For any fixed $j$,
conditional on $\mathcal{F}_{T}$ and $\tilde{\mathcal{F}}_{T},$ $\xi _{jk,1}$%
's are independent for different $k$. Thus combining the above with (a) and
(b), by Proposition 2.1 in \cite{MR3693963}, we have
\begin{align*}
& \sup_{u\in \mathbb{R}}\bigg|\mathbb{P}\big(I_{\epsilon ,m}\mathbf{1}_{%
\mathcal{A}_{T}}\leq u\big)-\mathbb{P}\Big(\max_{1\leq j\leq N}\big|\tilde{Z}%
_{j,1}+(Tb_{j})^{1/2}v_{j}^{-1}\big(\sum_{t=1}^{T}\xi _{jt,2}+\gamma _{j}%
\big)\big|\leq u\Big)\bigg| \\
\lesssim & (\underline{b}T)^{-1/6}\mathrm{log}^{7/6}(NT)+(T^{2/q}/(%
\underline{b}T))^{1/3}\mathrm{log}(NT). 
\end{align*}
Note that the $(j_1,j_2)$th entity of the conditional covariance matrix $\Sigma_{1}$ denoted as $\Sigma_{j_1j_2,1}$ follows,
\begin{align}
\label{eq:dsfec}
& \mathbb{E}(\Sigma_{j_1j_2,1})\nonumber\\
=& T(b_{j_1}b_{j_2})^{1/2}v_{j_1}^{-1}v_{j_2}^{-1}\mathbb{E}\Big(\sum_{t_1=1}^{T}\sum_{t_2=1}^{T}w_{j_1t_1,b}\sigma
_{j_1}(X_{j_1t_1},U_{j_1t_1})\epsilon _{j_1t_1,m}w_{j_2t_2,b}\sigma
_{j_2}(X_{j_2t_2},U_{j_2t_2})\epsilon _{j_2t_2,m}\Big)\nonumber\\
=& \mathbb{E}(v_{j_1j_2,1})+T(b_{j_1}b_{j_2})^{1/2}v_{j_1}^{-1}v_{j_2}^{-1}\sum_{1\leq t_{1}\neq t_{2}\leq T}\mathbb{
E}\Big(w_{j_1t_{1},b}\sigma_{j_1}(X_{j_1t_{1}},U_{j_1t_{1}})w_{j_2t_{2},b}\sigma
_{j_2}(X_{j_2t_{2}},U_{j_2t_{2}})\Big)\mathbb{E}\big(\epsilon
_{jt_{1},m}\epsilon _{jt_{2},m}\big)+O(m^{-\beta })\nonumber \\
=& \mathbb{E}(v_{j_1j_2,1})+O(\bar b+m^{-\beta }),
\end{align}
with 
\begin{equation}
v_{j_1j_2,1}=T(b_{j_1}b_{j_1})^{1/2}v_{j_1}^{-1}v_{j_2}^{-1}\sum_{t=1}^{T}w_{j_1t,b}w_{j_2t,b}\mathrm{Cov}\big(\sigma
_{j_1}(X_{j_1t},U_{j_1t})\epsilon _{j_1t}, \sigma
_{j_2}(X_{j_2t},U_{j_2t})\epsilon _{j_2t}|\mathcal F_T\big).  \label{eq:defvj12}
\end{equation}%
Consider the random variable 
\begin{equation}
Z_{j,1}=(Tb_{j})^{1/2}v_{j}^{-1}\sum_{t=1}^{T}\tilde{\xi}_{jt,1},
\label{eq:zj1}
\end{equation}%
where $\tilde{\xi}_{t,1}=(\tilde{\xi}_{1t,1},\ldots,\tilde{\xi}_{Nt,1})^\top$ are i.i.d. from $
N(0,\tilde\Sigma)$ where
$\tilde\Sigma_{j_1j_2}=v_{j_1j_2,1}v_{j_1}v_{j_2}b_{j_1}^{-1/2}b_{j_2}^{-1/2}T^{-2}$ and independent from $\tilde{\mathcal{F}}_{T}$
and $\epsilon _{jt}$.

By a similar argument as in the proof of part (a) below, with probability
greater than $1-O(T^{-(\alpha p)\wedge (p/2-1)}),$ the maximum difference of 
$\mathrm{Var}(\tilde{Z}_{j,1}|\mathcal{F}_{T},\tilde{\mathcal{F}}_{T})$
(resp. $\mathrm{Var}(Z_{j,1}|\mathcal{F}_{T})$) and $\mathrm{Cov}(\tilde{Z}%
_{j_{1},1},\tilde{Z}_{j2,1}|$ $\mathcal{F}_{T},\tilde{\mathcal{F}}_{T})$
(resp. $\mathrm{Cov}(Z_{j_{1},1},Z_{j2,1}|\mathcal{F}_{T})$) from their
expectation are bounded by $\bar{b}+\sqrt{\mathrm{log}(NT)/(\underline{b}T)}%
. $ Hence by Lemma \ref{lem:comparison}, we further have 
\begin{align*}
& \sup_{u\in \mathbb{R}}\bigg|\mathbb{P}\big(I_{\epsilon ,m}\mathbf{1}_{%
\mathcal{A}_{T}}\leq u\big)-\mathbb{P}\Big(\max_{1\leq j\leq N}\Big|%
Z_{j,1}+(Tb_{j})^{1/2}v_{j}^{-1}\big(\sum_{t=1}^{T}\xi _{jt,2}+\gamma _{j}%
\big)\Big|\leq u\Big)\bigg| \\
\lesssim & (\underline{b}T)^{-1/6}\mathrm{log}^{7/6}(NT)+(T^{2/q}/(%
\underline{b}T))^{1/3}\mathrm{log}(NT)+(\bar{b}+m^{-\beta })^{1/3}\mathrm{log%
}^{2/3}(N).
\end{align*}%
Then by the construction of $Z_{j,1}$ in \eqref{eq:zj1}, we have 
\begin{equation*}
Z_{j,1}+(Tb_{j})^{1/2}v_{j}^{-1}\Big(\sum_{t=1}^{T}\xi _{jt,2}+\gamma _{j}%
\Big)=(Tb_{j})^{1/2}v_{j}^{-1}\Big(\sum_{t=1}^{T}(\tilde{\xi}_{jt,1}+\xi
_{jt,2})+\gamma _{j}\Big).
\end{equation*}%
Given $\mathcal{F}_{T}$, $\tilde{\xi}_{jt,1}+\xi _{jt,2}$ are independent
for different $t.$ Recall the definition of $\Sigma_{j_1j_2}$ in \eqref{eq:def_sigmaj12}, we have 
\begin{equation}
\Sigma_{j_1j_2}=(Tb_{j})v_{j_1}^{-1}v_{j_2}^{-1}\sum_{t=1}^{T}\mathrm{Cov}\big(\tilde{\xi}_{j_1t,1}+\xi
_{j_1t,2}, \tilde{\xi}_{j_2t,1}+\xi
_{j_2t,2}|\mathcal{F}_{T}\big).  \label{eq:2momtlwbd}
\end{equation}%
Recall the definition of $v_{j}^{2}$ in
\eqref{eq:def_sigma}, then $\Sigma_{jj}=1$.
%
%
By Assumption \ref{boundedness}, $\max_{t}\mathbb{%
E}(|\tilde{\xi}_{jt,1}|+|\xi _{jt,2}||\mathcal{F}_{T})=O((b_{j}T)^{-1}).$
With probability greater than $1-O(T^{-(\alpha p)\wedge (p/2-1)})$, for $%
l=1,2,$ 
\begin{align}
& \max_{1\leq j\leq N}T^{1+l}b_{j}^{(2+l)/2}\sum_{t=1}^{T}\mathbb{E}\Big(|(%
\tilde{\xi}_{jt,1}+\xi _{jt,2})/v_{j}|^{2+l}|\mathcal{F}_{T}\Big)  \notag \\
\lesssim & \max_{1\leq j\leq N}Tb_{j}^{(2-l)/2}\sum_{t=1}^{T}\mathbb{E}\Big(%
((\tilde{\xi}_{jt,1}+\xi _{jt,2})/v_{j})^{2}|\mathcal{F}_{T}\Big)\leq
B_{T}^{l}.  \label{eq:2plmomtupbd}
\end{align}
For all $k$, we have 
\begin{equation*}
\mathbb{E}\Big(\max_{1\leq j\leq N}(b_{j}^{1/2}Tv_{j}^{-1}|\tilde{\xi}%
_{jt,1}+\xi _{jt,2}|/B_{T})^{q}\big|\mathcal{F}_{T}\Big)\leq 2.
\end{equation*}%
Combining the above results with \eqref{eq:2momtlwbd} and %
\eqref{eq:2plmomtupbd}, for $Z=\Sigma^{1/2}G,$ by Proposition 2.1 in \cite{MR3693963}, and Lemma \ref%
{lem:comparison}, we have 
\begin{align}
& \sup_{u\in \mathbb{R}}\bigg|\mathbb{P}\big(I_{\epsilon ,m}\mathbf{1}_{%
\mathcal{A}_{T}}\leq u\big)-\mathbb{P}\Big(\max_{1\leq j\leq N}\big|%
Z_{j}+(Tb_{j})^{1/2}v_{j}^{-1}\gamma _{j}\big|\leq u\Big)\bigg|  \notag \\
\lesssim & (\underline{b}T)^{-1/6}\mathrm{log}^{7/6}(NT)+(T^{2/q}/(%
\underline{b}T))^{1/3}\mathrm{log}(NT)+(\bar{b}+m^{-\beta })^{1/3}\mathrm{log%
}^{2/3}(N).  \label{eq:bdd4iemandzj}
\end{align}%
Under Assumption \ref{asmp:bddxj1j2density}, one can set $v_{j_1j_2,1}=0$ for $j_1\neq j_2$, and \eqref{eq:dsfec} still holds. Then $\Sigma=I_N$ becomes the identity matrix. Hence we complete the proof.

\textbf{Proof of part (a).} Since $\mathbb{E}(\eta _{jk}^{2})=1,$ we have 
\begin{align}
\sum_{k=2-m}^{T}\mathbb{E}(\xi _{jk,1}^{2}|\mathcal{F}_{T},\tilde{\mathcal{F}%
}_{T})& \geq \sum_{k=1}^{T-m}\Big|\sum_{t=k}^{k+m-1}w_{jt,b}\sigma
_{j}(X_{jt},U_{jt})A_{t-k,j,\cdot }\Big|_{2}^{2}  \notag \\
& =\sum_{k=1}^{T-m}\Big|\sum_{t=k}^{k+m-1}(M_{jt,1}+D_{jt,1})A_{t-k,j,\cdot }%
\Big|_{2}^{2},  \label{eq:lower4xijk1}
\end{align}%
where 
\begin{equation}
D_{jt,1}=\mathbb{E}\big(w_{jt,b}\sigma _{j}(X_{jt},U_{jt})|\mathcal{F}_{t-1}%
\big)\quad \text{and}\quad M_{jt,1}=w_{jt,b}\sigma
_{j}(X_{jt},U_{jk})-D_{jt,1}.
\end{equation}%
For the simplicity of notation, denote 
\begin{equation}
Y_{jk}=\Big|\sum_{t=k}^{k+m-1}M_{jt,1}A_{t-k,j,\cdot }\Big|_{2}^{2}.
\label{eq:yjk}
\end{equation}%
Since $|D_{jt,1}|\lesssim 1/T,$ \eqref{eq:lower4xijk1} can be further lower
bounded by 
\begin{equation}
\sum_{k=2-m}^{T}\mathbb{E}(\xi _{jk,1}^{2}|\mathcal{F}_{T},\tilde{\mathcal{F}%
}_{T})\geq \sum_{k=1}^{T-m}\Big(Y_{jk}^{1/2}-O(1/T)\Big)^{2}\geq
\sum_{k=1}^{T-m}Y_{jk}-O\Big(\Big(\sum_{k=1}^{T-m}Y_{jk}/T\Big)^{1/2}\Big).
\label{eq:xijk12lbdyjk}
\end{equation}%
We shall bound the difference between $Y_{jk}$ and $\mathbb{E}Y_{jk}$. Note
that $\mathbb{E}_{0}Y_{jk}$ can be rewritten as 
\begin{align*}
\sum_{k=1}^{T-m}\mathbb{E}_{0}Y_{jk}& =\sum_{t=1}^{T}\sum_{k=1\vee
(t-m+1)}^{(T-m)\wedge t}(M_{jt,1}L_{k,t,1}+\mathbb{E}_{0}L_{k,t,2}) \\
& =\sum_{t=1}^{T}\sum_{k=1\vee (t-m+1)}^{(T-m)\wedge t}\big[%
M_{jt,1}L_{k,t,1}+L_{k,t,2}-\mathbb{E}(L_{k,t,2}|\mathcal{F}_{t-1})\big] \\
& +\sum_{t=1}^{T}\sum_{k=1\vee (t-m+1)}^{(T-m)\wedge t}\big[\mathbb{E}%
(L_{k,t,2}|\mathcal{F}_{t-1})-\mathbb{E}(L_{k,t,2})\big]=:\mathrm{I}_{j1}+%
\mathrm{I}_{j2},
\end{align*}%
where 
\begin{equation*}
L_{k,t,1}=2\sum_{s=k}^{t-1}M_{js,1}A_{t-k,j,\cdot }^{\top }A_{s-k,j,\cdot
}\quad \text{and}\quad L_{k,t,2}=M_{jt,1}^{2}|A_{t-k,j,\cdot }|_{2}^{2}.
\end{equation*}%
Let 
\begin{equation*}
\varsigma _{jt}:=\sum_{k=1\vee (t-m+1)}^{(T-m)\wedge t}\big[%
M_{jt,1}L_{k,t,1}+L_{k,t,2}-\mathbb{E}(L_{k,t,2}|\mathcal{F}_{t-1})\big].
\end{equation*}%
Then $\mathrm{I}_{j1}=\sum_{t=1}^{T}\varsigma _{jt}$, the summation of
martingale differences. Since $|M_{jt,1}|=O((Tb_{j})^{-1})$, we have $%
|\varsigma _{jt}|=O((Tb_{j})^{-2})$ and $\mathbb{E}(\varsigma _{jt}^{2}|%
\mathcal{F}_{t-1})=O(b_{j}(Tb_{j})^{-4})$. Thus by Freedman's inequality, 
\begin{equation}
\mathbb{P}\big(b_{j}^{-1/2}(Tb_{j})^{2}|\mathrm{I}_{j1}|\geq z\big)\lesssim 
\mathrm{exp}\Big\{-\frac{z^{2}}{b_{j}^{-1/2}z+T}\Big\}.  \label{eq:ij1apart}
\end{equation}%
Hence with probability greater than $1-T^{-p}$ we have 
\begin{equation}
\max_{1\leq j\leq N}(Tb_{j})|\mathrm{I}_{j1}|\lesssim \sqrt{\mathrm{log}%
(NT)/(\underline{b}T)}.  \label{eq:Ij1bdd}
\end{equation}%
For $\mathrm{I}_{j2}$ part, let $\tilde{\sigma}_{j}^{2}(x)=\mathbb{E}(\sigma
_{j}^{2}(X_{jt},U_{jk})|X_{jt}=x)$. By Assumption \ref{asmp:smooth}, 
\begin{equation}
\mathbb{E}(M_{jt,1}^{2}\mathbf{1}_{X_{jt}\geq c_{j0}}|\mathcal{F}_{t-1})=%
\frac{1}{b_{j}T^{2}}\int_{0}^{1}K^{2}(x)\tilde{\sigma}%
_{j}^{2}(c_{j0})g_{j}(c_{j0})^{-1}\frac{K_{2}^{+}-K_{1}^{+}x}{%
K_{2}^{+}K_{0}^{+}-K_{1}^{+2}}g_{j,t}(c_{j0}|\mathcal{F}_{t-1})dx+O(T^{-2}).
\label{eq:ij2parta}
\end{equation}%
By Theorem 6.2 in \cite{zhang2017gaussian}, we have 
\begin{equation*}
\mathbb{P}\Big(\max_{1\leq j\leq N}(b_{j}T^{2})|\mathrm{I}_{j2}|\geq
z+O(Tb_{j})\Big)\lesssim z^{-p}T^{(p/2-\alpha p)\vee 1}\mathrm{log}%
(N)^{p/2}+e^{-z^{2}/T}.
\end{equation*}%
Hence with probability greater than $1-O(T^{-(\alpha p)\wedge (p/2-1)}),$ we
have 
\begin{equation}
\max_{1\leq j\leq N}(Tb_{j})|\mathrm{I}_{j2}|\lesssim \sqrt{\mathrm{log}%
(NT)/T}+\bar{b}.  \label{eq:Ij2bdd}
\end{equation}%
For the expectation part, since $M_{jt,1}$ are martingale differences, for $%
m\leq T/2,$ 
\begin{equation*}
\frac{Tb_{j}}{v_{j}^{2}}\sum_{k=1}^{T-m}\mathbb{E}(Y_{jk})=\frac{Tb_{j}}{%
v_{j}^{2}}\sum_{k=1}^{T-m}\sum_{t=k}^{k+m-1}\mathbb{E}%
(M_{jt,1}^{2})|A_{t-k,j,\cdot }|_{2}^{2}\asymp c,
\end{equation*}%
where $c>0$ is some constant independent of $j$. We complete the proof by
combining above with \eqref{eq:Ij1bdd} and \eqref{eq:Ij2bdd}. 

\textbf{Proof of part (b).} Recall the definition of $Y_{jk}$ in %
\eqref{eq:yjk}. Since $|Y_{jk}|\lesssim (Tb_{j})^{-2}$, we have 
\begin{align}
\sum_{k=2-m}^{T}\mathbb{E}(|\xi _{jk,1}|^{2+l}|\mathcal{F}_{T},\tilde{%
\mathcal{F}}_{T})& \lesssim \sum_{k=2-m}^{T}\Big|\sum_{t=1\vee k}^{T\wedge
(k+m-1)}w_{jt,b}\sigma _{j}(X_{jt},U_{jt})A_{t-k,j,\cdot }\Big|_{2}^{2+l} 
\notag \\
& \lesssim \sum_{k=2-m}^{T}Y_{jk}^{1+l/2}+\sum_{k=2-m}^{T}\Big(\sum_{t=1\vee
k}^{T\wedge (k+m-1)}|D_{jt,2}||A_{t-k,j,\cdot }|_{2}\Big)^{2+l}  \notag \\
& \lesssim (Tb_{j})^{-l}\sum_{k=2-m}^{T}Y_{jk}+T^{-1-l}.
\label{eq:xijk1plus}
\end{align}%
By a similar argument in $(a)$, with probability greater than $%
1-O(T^{-(\alpha p)\wedge (p/2-1)})$, we have 
\begin{equation}
\Big|\sum_{k=2-m}^{T}Y_{jk}\Big|\lesssim (Tb_{j})^{-1}.  \label{eq:sumkyjk}
\end{equation}%
Then part $(b)$ follows from \eqref{eq:xijk1plus} and \eqref{eq:sumkyjk}.

\textbf{(ii)} Since 
\begin{equation*}
\Big|\sum_{t=1\vee k}^{T\wedge
(k+m-1)}w_{jt,b}r_{j}(X_{jt},U_{jt})A_{t-k,j,\cdot }^{\top }\Big|%
_{2}=O(1/(b_{j}T)),
\end{equation*}%
we have 
\begin{equation*}
\mathbb{E}\Big(\mathrm{exp}\big\{T\sqrt{b_{j}}|\tilde{\xi}_{jt,1}+\xi
_{jt,2}|/B_{T}\big\}\big|\mathcal{F}_{T}\Big)\leq 2,
\end{equation*}%
where $B_{T}=c\underline{b}^{-1/2}$, where $c>0$ is a sufficiently large
constant. We complete the proof in view of Proposition 2.1 in \cite%
{MR3693963}. 
${\tiny \blacksquare }$

\subsection{Proofs of Lemmas in Section \protect\ref{SecA.2}\label{SecB.2}}

\noindent \textbf{Proof of Lemma \ref{lem:propwitb}. } $(i)$\textbf{\ }The
results can be derived by elementary calculus.

$(ii)$ {Define $\mathbb{E}_{0}(X)=X-\mathbb{E}(X)$.} without loss of generality and for
notation simplicity, we assume $c_{j0}=0.$ For random variable $X,$ denote $%
X^{+}=X\mathbf{1}_{\{X\geq 0\}}.$ We shall show that $S_{jl,b}^{+}$ is close
to its expectation. To this end, denote 
\begin{align*}
& M_{jt}=X_{jt}^{+l}K(X_{jt}^{+}/b_{j})-\mathbb{E}%
(X_{jt}^{+l}K(X_{jt}^{+}/b_{j})|\mathcal{F}_{t-1}) \\
\text{and}\quad & D_{jt}=\mathbb{E}(X_{jt}^{+l}K(X_{jt}^{+}/b_{j})|\mathcal{F%
}_{t-1})-\mathbb{E}(X_{jt}^{+l}K(X_{jt}^{+}/b_{j})).
\end{align*}%
Then the difference between $S_{jl,b}^{+}$ and its expectation can be
decomposed into two parts 
\begin{equation*}
\mathbb{E}_{0}S_{jl,b}^{+}=\sum_{t=1}^{T}M_{jt}+\sum_{t=1}^{T}D_{jt}=\mathrm{%
I}_{j1}+\mathrm{I}_{j2}.
\end{equation*}%
Note that for any $j,$ $M_{jt}$, $t\geq 1,$ are martingale differences with
respect to $\mathcal{F}_{t}$. Moreover, we have $M_{jt}\leq |K|_{\infty
}b_{j}^{l}$ and 
\begin{equation*}
\mathbb{E}M_{jt}^{2}\leq \int_{0}^{b_{j}}x^{2l}K^{2}(x/b_{j})g_{j}(x)dx\leq
b_{j}^{2l+1}\Big(g_{j}(0)\int_{0}^{1}K^{2}(y)dy+O(b_{j})\Big),
\end{equation*}%
uniformly over $j$. Thus by Freedman's inequality (\cite{freedman1975tail}%
), 
\begin{equation*}
\mathbb{P}\big(|\mathrm{I}_{j1}|\geq z\big)\lesssim \mathrm{exp}\big\{%
-z^{2}/(b_{j}^{l}z+Tb_{j}^{2l+1})\big\},
\end{equation*}%
where the constants in $\lesssim $ or $O(\cdot )$ here and all the
followings of this proof are independent of $N,T,j,b_{j}.$ That is with
probability greater than $1-(TN)^{-p}$, 
\begin{equation}
\max_{1\leq j\leq N}|\mathrm{I}_{j1}/(Tb_{j}^{l+1})|\lesssim \sqrt{\mathrm{%
log}(TN)/(T\underline{b})}.  \label{eq:ij1partlemma1}
\end{equation}%
For part $\mathrm{I}_{j2}$, by Taylor's expansion, we have 
\begin{equation*}
\mathbb{E}(X_{jt}^{+l}K(X_{jt}^{+}/b_{j})|\mathcal{F}_{t-1})=b_{j}^{l+1}%
\int_{0}^{1}y^{l}K(y)g_{j,t}(b_{j}y|\mathcal{F}_{t-1})dy=b_{j}^{l+1}\big(%
K_{l}^{+}g_{j,t}(0|\mathcal{F}_{t-1})+O(b_{j})\big),
\end{equation*}%
uniformly over $j$. By Theorem 6.2 in \cite{zhang2017gaussian}, for $%
z\gtrsim \sqrt{T\mathrm{log}(N)},$ 
\begin{equation*}
\mathbb{P}\bigg(\max_{1\leq j\leq N}\Big|\sum_{t=1}^{T}\mathbb{E}%
_{0}g_{j,t}(0|\mathcal{F}_{t-1})\Big|\geq z\bigg)\lesssim T^{(p/2-\alpha
p)\vee 1}\mathrm{log}(N)^{p/2}/z^{p}+e^{-z^{2}/T}.
\end{equation*}%
Hence with probability greater than $1-O(T^{-(\alpha p)\wedge (p/2-1)})$ we
have 
\begin{equation}
\max_{1\leq j\leq N}|\mathrm{I}_{j2}/(Tb_{j}^{l+1})|\lesssim 
\sqrt{\mathrm{log}(TN)/T}+\bar{b}.  \label{eq:lemma1ij2part}
\end{equation}%
Combining \eqref{eq:ij1partlemma1} and \eqref{eq:lemma1ij2part}, with
probability greater than $1-O(T^{-(\alpha p)\wedge (p/2-1)})$, we have 
\begin{equation*}
\max_{1\leq j\leq N}|\mathbb{E}_{0}S_{jl,b}^{+}|/(Tb_{j}^{l+1})\lesssim 
\sqrt{\mathrm{log}(TN)/(T\underline{b})}+\bar{b}.
\end{equation*}%
Lastly, we need to approximate the expectation part. Note that 
\begin{equation*}
\mathbb{E}(X_{jt}^{+l}K(X_{jt}^{+}/b_{j}))=%
\int_{0}^{b_{j}}x^{l}K(x/b_{j})g_{j}(x)dx=b_{j}^{l+1}K_{l}^{+}g_{j}(0)+O(b_{j}^{l+2}),
\end{equation*}%
uniformly over $j$. Hence \eqref{eq:Sjlplus} holds. Inequality %
\eqref{eq:witb} directly comes from \eqref{eq:Sjlplus}.

For \eqref{eq:bddforsumw}, note that $\sum_{t=1}^{T}|w_{jt,b}^{+}|\leq
(S_{j0}^{+}S_{j2}^{+}+S_{j1}^{+2})/(S_{j0}^{+}S_{j2}^{+}-S_{j1}^{+2}),$ thus
result follows from \eqref{eq:kernelcond}. ${\tiny \blacksquare }$

\bigskip
\noindent \textbf{Proof of Lemma \ref{lem:mdepend}. }We shall only work on
the set $\mathcal{A}_{T}$. Let $w_{jt,b}$ be as
defined in (\ref{eq:wjtandrjt}). Then $I_{\epsilon }$ can be rewritten as 
\begin{equation}
I_{\epsilon }\mathbf{1}_{\mathcal{A}_{T}}=\max_{1\leq j\leq N}\frac{%
(Tb_{j})^{1/2}}{v_{j}}\Big|\sum_{t=1}^{T}w_{jt,b}\big[\sigma
_{j}(X_{jt},U_{jt})\epsilon _{jt}+\varepsilon _{j}(X_{jt},U_{jt})\big]%
+\gamma _{j}\Big|.
\end{equation}%
Recall the definition of $\epsilon _{jt,m}$ in \eqref{eq:defofmdep} and $r_{j}(X_{jt},U_{jt})$ in (\ref{eq:wjtandrjt}), then 
\begin{align}
|I_{\epsilon }-I_{\epsilon ,m}|\mathbf{1}_{\mathcal{A}_{T}}\leq &
\max_{1\leq j\leq N}(Tb_{j})^{1/2}\Big|%
\sum_{t=1}^{T}w_{jt,b}r_{j}(X_{jt},U_{jt})(\epsilon _{jt}-\epsilon _{jt,m})%
\Big|  \notag \\
=& \max_{1\leq j\leq N}(Tb_{j})^{1/2}\Big|%
\sum_{t=1}^{T}w_{jt,b}r_{j}(X_{jt},U_{jt})\sum_{k\leq t-m}A_{t-k,j,\cdot
}^{\top }\eta _{k}\Big|  \notag \\
=& \max_{1\leq j\leq N}(Tb_{j})^{1/2}\Big|\sum_{k\leq T-m}\sum_{t=1\vee
(k+m)}^{T}w_{jt,b}r_{j}(X_{jt},U_{jt})A_{t-k,j,\cdot }^{\top }\eta _{k}\Big|,
\label{eq:bddiepandepm}
\end{align}%
where $A_{t,j,\cdot }$ is the $j$th row of matrix $A_{t}$. Let  
\begin{equation*}
\xi _{jk}=(Tb_{j})^{1/2}\sum_{t=1\vee
(k+m)}^{T}w_{jt,b}r_{j}(X_{jt},U_{jt})A_{t-k,j,\cdot}^\top\eta _{k}.
\end{equation*}%
Then conditional on $\mathcal{F}_{T}$ and $\tilde{\mathcal{F}}_{T}$, $\xi _{jk}$
are independent for different $k,$ and 
\begin{equation*}
|I_{\epsilon }-I_{\epsilon ,m}|\mathbf{1}_{\mathcal{A}_{T}}\leq \max_{1\leq
j\leq N}\Big|\sum_{k\leq T-m}\xi _{jk}\Big|.
\end{equation*}%
Let 
\begin{align}
\sigma ^{2}& =\max_{1\leq j\leq N}\sum_{k\leq T-m}%
\mathbb{E}(\xi _{jk}^{2}|\mathcal{F}_{T},\tilde{\mathcal{F}}_{T})  \notag \\
& \lesssim \max_{1\leq j\leq N}Tb_{j}\sum_{k\leq T-m}\Big(\sum_{t=1\vee
(k+m)}^{T}|w_{jt,b}r_j(X_{jt},U_{jt})A_{t-k,j,\cdot }|_{2}\Big)^{2}.  \label{eq:sumxiik2}
\end{align}%
Since kernel function $K(\cdot)$ is bounded, by Lemma \ref{lem:propwitb}, $%
\max_{t,j}(Tb_{j})|w_{jt,b}|=O(1)$ and $\max_{j}%
\sum_{t=1}^{T}|w_{jt,b}|=O(1),$ where the constants in $O(\cdot )$ and $\lesssim$ here and throughout this proof are independent of $N,T,b,j$.
Hence we have
\begin{equation}
\max_{j,k}(Tb_{j}) \sum_{t=1\vee (k+m)}^{T}|w_{jt,b}A_{t-k,j,\cdot
}|_{2}/v_j\lesssim m^{-\beta },  \label{eq:bddsummaxjkwra}
\end{equation}%
and 
\begin{equation}
\max_{j}\sum_{k\leq T-m}\sum_{t=1\vee (k+m)}^{T}|w_{jt,b}||A_{t-k,j,\cdot
}|_{2}/v_j=\max_{j}\sum_{t=1}^{T}\sum_{k\leq t-m}|w_{jt,b}||A_{t-k,j,\cdot
}|_{2}/v_j\lesssim m^{-\beta }.  \label{eq:bddmaxjsumwra}
\end{equation}%
It follows that 
\begin{equation}
\sigma ^{2}\lesssim m^{-2\beta }.  \label{eq:upbdfwjtb}
\end{equation}
By Lemma \ref{lem:freedman}, for any $z>0,$ 
\begin{align}
& \mathbb{P}\big(|I_{\epsilon }-I_{\epsilon ,m}|\mathbf{1}_{\mathcal{A}%
_{T}}\geq z\big|\mathcal{F}_{T},\tilde{\mathcal{F}}_{T}\big)  \notag \\
\lesssim & \sum_{k\leq T-m}\sum_{1\leq
j\leq N}\mathbb{P}\Big(|\xi _{jk}|\geq u|\mathcal{F}_{T},\tilde{\mathcal{F}}_{T}\Big)+2N%
\mathrm{exp}\Big(-\frac{z^{2}}{zu+m^{-2\beta}}\Big),  \label{eq:funlmd}
\end{align}
By Markov's inequality, we have 
\begin{align}
\sum_{k\leq T-m}\mathbb{P}\Big(|\xi _{jk}|\geq u\Big)& \leq \sum_{k\leq T-m}u^{-q}\EE\big(\EE(|\xi_{jk}|^q|\F_T,\tilde\F_T)\big) \notag \\
&\lesssim u^{-q} \sum_{k\leq T-m}(Tb_{j})^{q/2}
\EE\Big|\sum_{t=1\vee
(k+m)}^{T}w_{jt,b}r_{j}(X_{jt},U_{jt})A_{t-k,j,\cdot}^\top\Big|_2^q\notag\\
& \lesssim (T\underline{b})^{1-q/2}m^{-q\beta }u^{-q}.  \label{eq:bdforeta}
\end{align}%
We complete the proof by implementing 
\eqref{eq:bdforeta} into \eqref{eq:funlmd} with $u=z/\mathrm{log}(NT)$. 
${\tiny \blacksquare }$

\bigskip

\noindent \textbf{Proof of Lemma \ref{lem:mdepend_supexp}. }By %
\eqref{eq:bddiepandepm}, $|I_{\epsilon }-I_{\epsilon ,m}|$ can be bounded by 
\begin{equation*}
|I_{\epsilon }-I_{\epsilon ,m}|\mathbf{1}_{\mathcal{A}_{T}}\leq \max_{1\leq
j\leq N}\Big|\sum_{k\leq T-m}c_{jk}^{\top }\eta _{k}\Big|,
\end{equation*}%
where $c_{jk}=(Tb_{j})^{1/2}\sum_{t=1\vee
(k+m)}^{T}w_{jt,b}r_{j}(X_{jt},U_{jt})A_{t-k,j,\cdot }.$ Then by %
\eqref{eq:bddsummaxjkwra} and \eqref{eq:bddmaxjsumwra}, we have 
\begin{equation}
\max_{j,k}|c_{jk}|_{2}\lesssim m^{-\beta }(\underline{b}T)^{-1/2}\quad \text{%
and}\quad \max_{j}\sum_{k}|c_{jk}|_{2}^{2}\lesssim m^{-2\beta }.
\label{eq:cjki}
\end{equation}%
Conditional on $\mathcal{F}_{T}$ and $\tilde{\mathcal{F}}_{T}$, $%
c_{jk}^{\top }\eta _{k}$ are independent for different $k$, therefore for $%
\lambda |c_{jk}|_{2}\leq \lambda _{0}$, where $\lambda_0$ is defined in Assumption \ref{asmp:moment}. By Markov's inequality 
\begin{equation}
\mathbb{P}\Big(\sum_{k\leq T-m}c_{jk}^{\top }\eta _{k}\geq z\big|\mathcal{F}%
_{T},\tilde{\mathcal{F}}_{T}\Big)\leq e^{-\lambda z}\prod_{k\leq T-m}\mathbb{%
E}\big(e^{\lambda c_{jk}^{\top }\eta _{k}}\big|\mathcal{F}_{T},\tilde{%
\mathcal{F}}_{T}\big).  \label{eq:exp11}
\end{equation}%
Since $\mathbb{E}(\eta _{k}|\mathcal{F}_{T},\tilde{\mathcal{F}}_{T})=0$, we
have 
\begin{align*}
\mathbb{E}\big(e^{\lambda c_{jk}^{\top }\eta _{k}}\big|\mathcal{F}_{T},%
\tilde{\mathcal{F}}_{T}\big)& =1+\mathbb{E}\Big(e^{\lambda c_{jk}^{\top
}\eta _{k}}-1-\lambda c_{jk}^{\top }\eta _{k}\big|\mathcal{F}_{T},\tilde{%
\mathcal{F}}_{T}\Big) \\
& \leq 1+\mathbb{E}\bigg(\frac{e^{|\lambda c_{jk}^{\top }\eta
_{k}|}-1-|\lambda c_{jk}^{\top }\eta _{k}|}{\lambda ^{2}|c_{jk}|_{2}^{2}}%
\Big|\mathcal{F}_{T},\tilde{\mathcal{F}}_{T}\bigg)\lambda
^{2}|c_{jk}|_{2}^{2},
\end{align*}%
in view of $e^{x}-x\leq e^{|x|}-|x|$ for any $x.$ Note that for any fixed $%
x>0,$ $(e^{tx}-tx-1)/t^{2}$ is increasing on $t\in (0,\infty )$. Therefore 
\begin{align*}
\mathbb{E}\big(e^{\lambda c_{jk}^{\top }\eta _{k}}\big|\mathcal{F}_{T},%
\tilde{\mathcal{F}}_{T}\big)& \leq 1+\mathbb{E}\bigg(\frac{e^{\lambda
_{0}|c_{jk}^{\top }\eta _{k}|/|c_{jk}|_{2}}-1-\lambda _{0}|c_{jk}^{\top
}\eta _{k}|/|c_{jk}|_{2}}{\lambda _{0}^{2}}\Big|\mathcal{F}_{T},\tilde{%
\mathcal{F}}_{T}\bigg)\lambda ^{2}|c_{jk}|_{2}^{2} \\
& \leq 1+a_{0}'\lambda ^{2}|c_{jk}|_{2}^{2} \\
& \leq e^{a_{0}'\lambda ^{2}|c_{jk}|_{2}^{2}},
\end{align*}%
where the last inequality is due to $1+x\leq e^{x}$ and $a_{0}'=a_0/\lambda_0^2$ with $a_0$ defined in Assumption \ref{asmp:moment}. Combining the above results with %
\eqref{eq:exp11}, we have 
\begin{equation*}
\mathbb{P}\Big(\sum_{k\leq T-m}c_{jk}^{\top }\eta _{k}\geq z\big|\mathcal{F}%
_{T},\tilde{\mathcal{F}}_{T}\Big)\leq e^{-\lambda z}\mathbb{E}\Big(%
e^{a_{0}'\lambda ^{2}\sum_{k\leq T-m}|c_{jk}|_{2}^{2}}\big|\mathcal{F}_{T},%
\tilde{\mathcal{F}}_{T}\Big)\leq e^{-\lambda z}e^{c\lambda^ 2 m^{-2\beta
}},
\end{equation*}%
where the last inequality is due to \eqref{eq:cjki} and $c>0$
is some constant. The same argument can be applied for $\sum_{k\leq
T-m}c_{jk}^{\top }\eta _{k}<-z,$ then we have 
\begin{equation*}
\mathbb{P}\big(|I_{\epsilon }-I_{\epsilon ,m}|\mathbf{1}_{\mathcal{A}%
_{T}}\geq z\big)
\leq 2Ne^{-\lambda z+c\lambda^2 m^{-2\beta }}
\lesssim N%
\mathrm{exp}\Big\{-c'\frac{z^{2}}{m^{-\beta
}(\underline bT)^{-1/2}z+m^{-2\beta }}\Big\},
\end{equation*}%
where $c'>0$ is some constant and we complete the proof. 
${\tiny \blacksquare }$

\subsection{Proofs of the Technical Lemmas in Section \protect\ref{SecA.3} 
\label{SecB.3}}

\noindent \textbf{Proof of Lemma \ref{lem:mdependQ_homo}. } $(i)$ Recall the
definition of $a_{jt,b}$ in \eqref{eq:ajtb}. We have 
\begin{equation*}
\sum_{t=1}^{T}a_{jt,b}(\epsilon _{jt}-\epsilon _{jt,m})=\sum_{k=-\infty
}^{T-m}\sum_{t=1\vee (k+m)}^{T}a_{jt,b}A_{t-k,j,\cdot }^{\top }\eta _{k}.
\end{equation*}%
Denote 
\begin{equation*}
\xi _{jk}=(Tb_{j})^{1/2}\sum_{t=1\vee (k+m)}^{T}\Big(a_{jt,b}A_{t-k,j,\cdot}-%
\sum_{l=1}^{N}a_{lt,b}A_{t-k,l,\cdot}/N\Big)^\top\eta _{k}/\tilde v_j.
\end{equation*}%
Then $\mathcal{Q}_{\epsilon }-\mathcal{Q}_{\epsilon ,m}$ can be rewritten
into 
\begin{equation*}
|\mathcal{Q}_{\epsilon }-\mathcal{Q}_{\epsilon ,m}|\leq \max_{1\leq j\leq N}%
\Big|\sum_{k\leq T-m}\xi _{jk}\Big|.
\end{equation*}%
Given $\mathcal{F}_{T},\tilde{\mathcal{F}}_{T},$ $\xi _{jk}$ are
independent for different $k.$ By the same argument as used to obtain %
\eqref{eq:funlmd}, we have 
\begin{align}
& \mathbb{P}\big(|\mathcal{Q}_{\epsilon }-\mathcal{Q}_{\epsilon ,m}|\geq z|%
\mathcal{F}_{T},\tilde{\mathcal{F}}_{T}\big)  \notag \\
\leq & \sum_{k\leq T-m}\mathbb{P}\Big(\max_{1\leq
j\leq N}|\xi _{jk}|\geq u\big|\mathcal{F}_{T},\tilde{\mathcal{F}}_{T}\Big)%
+2Ne^{-z^{2}/(2zu+2v)},  \label{eq:funlmd11}
\end{align}%
where $v$ is the upper bound for 
\begin{align*}
\sigma ^{2}=& \max_{1\leq j\leq N}\sum_{k\leq T-m}
\mathbb{E}\big(\xi _{jk}^{2}|\mathcal{F}_{T},\tilde{\mathcal{F}}_{T}\big) \\
=& \max_{1\leq j\leq N}\sum_{k\leq T-m}Tb_{j}\Big|
\sum_{t=1\vee (k+m)}^{T}\Big(a_{jt,b}A_{t-k,j,\cdot}-%
\sum_{l=1}^{N}a_{lt,b}A_{t-k,l,\cdot}/N\Big)\Big|_2^{2}/\tilde v_j^2 \\
\leq & \max_{1\leq j\leq N}Tb_{j}\sum_{k\leq T-m}\Big(\sum_{t=1\vee
(k+m)}^{T}\Big(|a_{jt,b}||A_{t-k,j,\cdot
}|_{2}+\sum_{l=1}^{N}|a_{lt,b}||A_{t-k,l,\cdot }|_{2}/N\Big)\Big)^{2}/\tilde v_j^2.
\end{align*}%
As in the proof of \eqref{eq:upbdfwjtb}, we have 
\begin{equation}
\sigma ^{2}\lesssim (\bar{b}/\underline{b})m^{-2\beta }\lesssim m^{-2\beta }.
\label{eq:boundforsigma2heto}
\end{equation}%
Since $\sum_{k\geq
m}|A_{k,j,\cdot }|_{2}\lesssim m^{-\beta },$  
by the same argument as used to derive \eqref{eq:bdforeta}, we have 
\begin{equation}
\sum_{k\leq T-m}\sum_{1\leq j\leq
N}\mathbb{P}\Big(|\xi _{jk}|\geq u\Big)\lesssim N(T\underline{b})^{-q/2+1}m^{-q\beta
}u^{-q}.  \label{eq:uselaterheto}
\end{equation}%
Take $u=z/\mathrm{log}(NT)$. Substituting \eqref{eq:boundforsigma2heto} and %
\eqref{eq:uselaterheto} into \eqref{eq:funlmd11}, we complete the proof of
part $(i)$.

$(ii)$ Note that $|\mathcal{Q}_{\epsilon }-\mathcal{Q}_{\epsilon ,m}|$ can be
bounded by 
\begin{equation*}
|\mathcal{Q}_{\epsilon }-\mathcal{Q}_{\epsilon ,m}|\leq \max_{1\leq j\leq N}%
(Tb_{j})^{1/2}\Big|\sum_{k\leq T-m}c_{jk}^{\top }\eta _{k}\Big|/\tilde v_j,
\end{equation*}%
where 
\begin{equation*}
c_{jk}=\sum_{t=1\vee (k+m)}^{T}\Big(a_{jt,b}A_{t-k,j,i}-%
\sum_{l=1}^{N}a_{lt,b}A_{t-k,l,i}/N\Big).
\end{equation*}%
By arguments as used to obtain \eqref{eq:bddsummaxjkwra} and %
\eqref{eq:bddmaxjsumwra}, we have the same bound of $c_{jk}$ as in %
\eqref{eq:cjki}. Hence the desired result follows from Lemma \ref%
{lem:mdepend_supexp}. ${\tiny \blacksquare }$\bigskip

\noindent \textbf{Proof of Lemma \ref{lem:gapprox_homo}} 
$(i)$
Recall the definition of $a_{jt,b}$ in \eqref{eq:ajtb}, then $\mathcal{Q}%
_{\epsilon ,m}$ can be rewritten as 
\begin{align}
\mathcal{Q}_{\epsilon ,m}\mathbf{1}_{\mathcal{A}_{T}}& =\max_{1\leq j\leq
N}(Tb_{j})^{1/2}\Big|\sum_{t=1}^{T}\Big(a_{jt,b}\epsilon
_{jt,m}-\sum_{l=1}^{N}a_{lt,b}\epsilon _{lt,m}/N+w_{jt,b}\varepsilon
_{j}(X_{jt},U_{jt})  \notag \\
&\quad -\sum_{l=1}^{N}w_{lt,b}\varepsilon _{l}(X_{lt},U_{lt})/N\Big)+(\gamma _{j}-%
\bar{\gamma})\Big|/\tilde v_j  \notag \\
& =\max_{1\leq j\leq N}(Tb_{j})^{1/2}\Big|\sum_{k=2-m}^{T}c_{jk}^{\top }\eta
_{k}+\sum_{t=1}^{T}\xi _{jt}+(\gamma _{j}-\overline{{\gamma}})\Big|/\tilde v_j,
\end{align}%
where 
\begin{equation*}
c_{jk}=\sum_{t=1\vee k}^{T\wedge (k+m-1)}\Big(a_{jt,b}A_{t-k,j,\cdot
}-\sum_{l=1}^{N}a_{lt,b}A_{t-k,l,\cdot }/N\Big)
\end{equation*}%
and 
\begin{equation*}
\xi _{jt}=w_{jt,b}\varepsilon
_{j}(X_{jt},U_{jt})-\sum_{l=1}^{N}w_{lt,b}\varepsilon _{l}(X_{lt},U_{lt})/N.
\end{equation*}
Decompose $a_{jt,b}$ into two parts, 
\begin{equation}
M_{jt}=a_{jt,b}-\mathbb{E}(a_{jt,b}|\mathcal{F}_{t-1},\tilde{\mathcal{F}}_{t-1})\quad 
\text{and}\quad D_{jt}=\mathbb{E}(a_{jt,b}|\mathcal{F}_{t-1},\tilde{\mathcal{F}}%
_{t-1}).  \label{eq:mdjthomo}
\end{equation}%
Let $\tilde{\sigma}_{j}(X_{jt})=\mathbb{E}(\sigma
_{j}(X_{jt},U_{jt})|X_{jt}).$ Since density function $g_{j}$ is bounded, 
\begin{align*}
|D_{jt}|& \lesssim \big|\mathbb{E}(\tilde{\sigma}%
_{j}(X_{jt})K(X_{jt}/b_{j})/(b_{j}T)|\mathcal{F}_{t-1})\big| \\
& \lesssim \Big|\int T^{-1}K(x)\tilde{\sigma}_{j}(b_{j}x)g_{j}(b_{j}x|%
\mathcal{F}_{t-1})dx\Big|\lesssim T^{-1},
\end{align*}%
where the constants in $\lesssim $ and $O(\cdot )$ here and all the
followings of this proof are independent of $j,t,N,T.$ In the following we
shall firstly show that with probability greater than $1-O(T^{-(\alpha
p)\wedge (p/2-1)})$, 
\begin{equation}
\min_{1\leq j\leq N}(Tb_{j})\sum_{k=2-m}^{T}\mathbb{E}\big((c_{jk}^{\top
}\eta _{k})^{2}|\mathcal{F}_{T},\tilde{\mathcal{F}}_{T}\big)\geq c,
\label{eq:lowerbddcjketa}
\end{equation}%
where $c>0$ is some constant. Note that 
\begin{equation}
\sum_{k=2-m}^{T}\mathbb{E}\big((c_{jk}^{\top }\eta _{k})^{2}|\mathcal{F}_{T},%
\tilde{\mathcal{F}}_{T}\big)\geq \sum_{k=1}^{T-m}\Big|\sum_{t=k}^{k+m-1}\Big(%
a_{jt,b}A_{t-k,j,\cdot }-\sum_{l=1}^{N}a_{lt,b}A_{t-k,l,\cdot }/N\Big)\Big|%
_{2}^{2}.  \label{eq:lower4varheto}
\end{equation}%
Denote $M_{jt}$ and $D_{jt}$ as in \eqref{eq:mdjthomo}, and let 
\begin{equation*}
Y_{jk}=\Big|\sum_{t=k}^{k+m-1}\Big(M_{jt}A_{t-k,j,\cdot
}-\sum_{l=1}^{N}M_{lt}A_{t-k,l,\cdot }/N\Big)\Big|_{2}^{2}.
\end{equation*}%
Since $|D_{jt}|\lesssim 1/T,$ similarly to \eqref{eq:xijk12lbdyjk}, %
\eqref{eq:lower4varheto} can be further lower bounded by 
\begin{equation*}
\sum_{k=2-m}^{T}\mathbb{E}((c_{jk}^{\top }\eta _{k})^{2}|\mathcal{F}_{T},%
\tilde{\mathcal{F}}_{T})\geq \sum_{k=1}^{T-m}\Big(Y_{jk}^{1/2}-O(1/T)\Big)^{2}\geq
\sum_{k=1}^{T-m}Y_{jk}-O\Big(\Big(\sum_{k=1}^{T-m}Y_{jk}/T\Big)^{1/2}\Big).
\end{equation*}%
By Assumption \ref{asmp:bddxj1j2density}, for $j\neq l$, we have $\mathbb{E}%
(M_{jt}M_{lt})=O(T^{-2})$ and thus 
\begin{align}
&\sum_{k=1}^{T-m}\mathbb{E}(Y_{jk})\notag\\
=& \sum_{k=1}^{T-m}\sum_{t=k}^{k+m-1}%
\mathbb{E}\Big|M_{jt}A_{t-k,j,\cdot }-\sum_{l=1}^{N}M_{lt}A_{t-k,l,\cdot }/N%
\Big|_{2}^{2}  \notag \\
=& \sum_{k=1}^{T-m}\sum_{t=k}^{k+m-1}\Big(\mathbb{E}%
(M_{jt}^{2})(1-1/N)^{2}|A_{t-k,j,\cdot }|_{2}^{2}+\sum_{\{l:l\neq j\}}%
\mathbb{E}(M_{lt}^{2})|A_{t-k,l,\cdot }|_{2}^{2}/N^{2}\Big)+O(1/T)  \notag \\
\gtrsim & (\underline{b}T)^{-1}.  \label{eq:eyjk_homo}
\end{align}%
We shall bound the difference between $Y_{jk}$ and $\mathbb{E}Y_{jk}$. Let $%
L_{jtk}=M_{jt}A_{t-k,j,\cdot }-\sum_{l=1}^{N}M_{lt}A_{t-k,l,\cdot }/N$ and
denote 
\begin{equation*}
\tilde{\xi}_{jt}=\sum_{k=1\vee (t-m+1)}^{(T-m)\wedge t}\Big(%
|L_{jtk}|_{2}^{2}+2L_{jtk}^{\top }\sum_{k\leq s<t}L_{jsk}\Big).
\end{equation*}%
Then the difference can be decomposed into two parts, 
\begin{equation*}
\sum_{k=1}^{T-m}\mathbb{E}_{0}Y_{jk}=\sum_{t=1}^{T}(\tilde{\xi}_{jt}-\mathbb{%
E}(\tilde{\xi}_{jt}|\mathcal{F}_{t-1},\tilde{\mathcal{F}}_{t-1}))+%
\sum_{t=1}^{T}(\mathbb{E}(\tilde{\xi}_{jt}|\mathcal{F}_{t-1},\tilde{\mathcal{%
F}}_{t-1})-\mathbb{E}\tilde{\xi}_{jt})=\mathrm{I}_{j1}+\mathrm{I}_{j2},
\end{equation*}%
Similarly to the argument in \eqref{eq:ij1apart}, we have $|\tilde{\xi}%
_{jt}|=O((b_{j}T)^{-2})$, $\sum_{t=1}^{T}\mathbb{E}(\tilde{\xi}_{jt}^{2}|%
\mathcal{F}_{t-1},\tilde{\mathcal{F}}_{t-1})=O((b_{j}T)^{-3})$ and thus 
\begin{equation}
\mathbb{P}\Big(\max_{1\leq j\leq N}|\mathrm{I}_{j1}|\geq z/(b_{j}T)^{2}\Big)%
\lesssim N\mathrm{exp}\Big\{-\frac{z^{2}}{z+b_{j}T}\Big\}.
\end{equation}%
Hence with probability greater than $1-T^{-p}$, we have $\max_{1\leq j\leq
N}(b_{j}T)|\mathrm{I}_{j1}|=\sqrt{\mathrm{log}(NT)/(\underline{b}T)}.$

For $\mathrm{I}_{j2}$ part, similar to \eqref{eq:ij2parta}, we have 
\begin{equation*}
\mathbb{E}(M_{jt}^{2}|\mathcal{F}_{t-1},\tilde{\mathcal{F}}_{t-1})=\int
b_{j}(b_{j}T)^{-2}K^{2}(x)\tilde{\sigma}_{j}^{2}(c_{j0})g_{j}(c_{j0}|\mathcal{F%
}_{t-1})\d x(1+O(b_{j})),
\end{equation*}%
and that 
\begin{equation*}
\mathbb{E}(\tilde{\xi}_{jt}|\mathcal{F}_{t-1},\tilde{\mathcal{F}}%
_{t-1})=\sum_{k=1\vee (t-m+1)}^{(T-m)\wedge t}\mathbb{E}(|L_{jtk}|_{2}^{2}|%
\mathcal{F}_{t-1},\tilde{\mathcal{F}}_{t-1}).
\end{equation*}%
Hence by Theorem 6.2 in \cite{zhang2017gaussian}, 
\begin{equation*}
\mathbb{P}\Big(\max_{1\leq j\leq N}(b_{j}T^{2})|\mathrm{I}_{j2}|\geq z+O(%
\bar{b})\Big)\lesssim z^{-p}T^{(p/2-\alpha p)\vee 1}\mathrm{log}%
(N)^{p/2}+e^{-z^{2}/T}.
\end{equation*}%
Hence combining $\mathrm{I}_{j1}$ and $\mathrm{I}_{j2},$ with probability
greater than $1-O(T^{-(\alpha p)\wedge (p/2-1)}),$ 
\begin{equation}
b_{j}T\Big|\sum_{k=1}^{T-m}\mathbb{E}_{0}Y_{jk}\Big|\lesssim \sqrt{\mathrm{%
log}(NT)/(\underline{b}T)}+\bar{b}.  \label{eq:bdde0yjk_homo}
\end{equation}%
This, in conjunction with \eqref{eq:eyjk_homo}, completes the proof of %
\eqref{eq:lowerbddcjketa}. \ 

We shall then show that with probability greater than $1-O(T^{-(\alpha
p)\wedge (p/2-1)})$, for $B_{T}=c\underline b^{-1/2},$ some constant $c>0$ large
enough, $l=1,2,$ we have 
\begin{equation}
\max_{1\leq j\leq N}b_{j}^{(2+l)/2}T^{1+l}\sum_{k=2-m}^{T}\mathbb{E}%
(|c_{jk}^{\top }\eta _{k}|^{2+l}|\mathcal{F}_{T},\tilde{\mathcal{F}}%
_{T})\leq B_{T}^{l}.  \label{eq:bdd2l_homo}
\end{equation}%
Since $\eta _{ik}$ are independent for different $i,k$, 
\begin{equation}
\sum_{k=2-m}^{T}\mathbb{E}(|c_{jk}^{\top }\eta _{k}|^{2+l}|\mathcal{F}_{T},%
\tilde{\mathcal{F}}_{T})\lesssim \sum_{k=2-m}^{T}|c_{jk}|_{2}^{2+l}\lesssim
(b_{j}T)^{-l}\sum_{k=2-m}^{T}Y_{jk}+T^{-(1+l)}.  \label{eq:bddcjketak_homo}
\end{equation}%
By a similar argument as used to derive \eqref{eq:eyjk_homo}, we have $%
b_{j}T\sum_{k=2-m}^{T}\mathbb{E}Y_{jk}=O(1).$ Hence together with %
\eqref{eq:bdde0yjk_homo}, with probability greater than $1-O(T^{-(\alpha
p)\wedge (p/2-1)}),$ 
\begin{equation}
b_{j}T\sum_{k=2-m}^{T}Y_{jk}=O(1).  \label{eq:bdde0yjk_homo_2}
\end{equation}%
Combining \eqref{eq:bddcjketak_homo} and \eqref{eq:bdde0yjk_homo_2}, we
complete the proof of \eqref{eq:bdd2l_homo}. Let $\tilde{Z}_{j,1}$, $1\leq
j\leq N$, be Gaussian distributed random variables with the same conditional
covariance structure as $(Tb_{j})^{1/2}\sum_{k=2-m}^{T}c_{jk}^{\top }\eta
_{k}$ given $\mathcal{F}_{T}$ and $\tilde{\mathcal{F}}_{T}$. For any $j$
conditional on $\mathcal{F}_{T}$ and $\tilde{\mathcal{F}}_{T},$ $%
c_{jk}^{\top }\eta _{k}$ are independent for different $k$. Thus by
Proposition 2.1 in \cite{MR3693963}, 
\begin{align*}
& \sup_{u\in \mathbb{R}}\bigg|\mathbb{P}\big(\mathcal{Q}_{\epsilon ,m}%
\mathbf{1}_{\mathcal{A}_{T}}\leq u\big)-\mathbb{P}\Big(\max_{1\leq j\leq N}%
\Big|\tilde{Z}_{j,1}+(Tb_{j})^{1/2}\Big(\sum_{t=1}^{T}\xi _{jt}+(\gamma _{j}-%
\bar{\gamma})\Big)\Big|/\tilde v_j\leq u\Big)\bigg| \\
\lesssim & (\underline{b}T)^{-1/6}\mathrm{log}^{7/6}(NT)+(T^{2/q}/(%
\underline{b}T))^{1/3}\mathrm{log}(NT).
\end{align*}
Define $$\tilde{v}_{j,1}^2=v_j^2\big\{(1-1 / N)^2 v_{j, 1}^2+\sum_{l=1} v_{l, 1}^2 / N^2\big\},$$
where $v_{j,1}$ defined in \eqref{eq:defvj12}.
Note that 
\begin{equation*}
\mathbb{E}(\mathrm{Var}(\tilde{Z}_{j,1}|\mathcal{F}_{T},\tilde{\mathcal{F}}%
_{T}))=\tilde{v}_{j,1}^{2}+O(b_{j}+m^{-\beta }).
\end{equation*}%
And for any $j_{1}\neq j_{2}$
we have 
\begin{equation*}
\mathbb{E[}\mathrm{Cov}(\tilde{Z}_{j_{1},1},\tilde{Z}_{j_{2},1}|\mathcal{F}%
_{T},\tilde{\mathcal{F}}_{T})]=O(\bar{b}).
\end{equation*}%
Let $Z_{j,1}$ be i.i.d. from $N(0,\tilde{v}_{j,1}^{2})$ and is independent of $%
\mathcal{F}_{T}$ and $\tilde{\mathcal{F}}_{T}$. With probability greater
than $1-O(T^{-(\alpha p)\wedge (p/2-1)}),$ the maximum difference of $%
\mathrm{Var}(\tilde{Z}_{j,1}|\mathcal{F}_{T},\tilde{\mathcal{F}}_{T})$ and $%
\mathrm{Cov}(\tilde{Z}_{j_{1},1},\tilde{Z}_{j2,1}|\mathcal{F}_{T},\tilde{%
\mathcal{F}}_{T})$ from their expectation are bounded by $\bar{b}+\sqrt{%
\mathrm{log}(NT)/(\underline{b}T)}.$ Hence by Lemma \ref{lem:comparison}, we
further have 
\begin{align*}
& \sup_{u\in \mathbb{R}}\bigg|\mathbb{P}\big(\mathcal{Q}_{\epsilon ,m}%
\mathbf{1}_{\mathcal{A}_{T}}\leq u\big)-\mathbb{P}\Big(\max_{1\leq j\leq N}%
\Big|Z_{j,1}+(Tb_{j})^{1/2}\Big(\sum_{t=1}^{T}\xi _{jt}+(\gamma _{j}-\bar{%
\gamma})\Big)\Big|/\tilde{v}_j\leq u\Big)\bigg| \\
\lesssim & (\underline{b}T)^{-1/6}\mathrm{log}^{7/6}(NT)+(T^{2/q}/(%
\underline{b}T))^{1/3}\mathrm{log}(NT)+(\bar{b}+m^{-\beta })^{1/3}\mathrm{log%
}^{2/3}(N).
\end{align*}
Random variable $Z_{j,1}=(Tb_{j})^{1/2}\sum_{t=1}^{T}\tilde{\xi}_{jt,1},$
where $\tilde{\xi}_{jt,1}$ are i.i.d. from $N(0,\tilde{v}%
_{j,1}^{2}/(T^{2}b_{j}))$ and independent from $\mathcal{F}_{T},\tilde{\mathcal{F}}_{T}$. Then 
\begin{equation*}
Z_{j,1}+(Tb_{j})^{1/2}\Big(\sum_{t=1}^{T}\xi _{jt}+(\gamma _{j}-\bar{\gamma})%
\Big)=(Tb_{j})^{1/2}\Big(\sum_{t=1}^{T}(\tilde{\xi}_{jt,1}+\xi
_{jt})+(\gamma _{j}-\bar{\gamma})\Big).
\end{equation*}

Define $Z_{j}^{\diamond }$s as independent Gaussian random variable with mean $0$ and variance $\tilde{v}_j^2$.
Given $\mathcal{F}_{T}$, $\tilde{\xi}_{jt,1}+\xi _{jt,2}$ are independent
for different $t.$ Similar to the argument in the proof of %
\eqref{eq:bdd4iemandzj}, we have 
\begin{align}
& \sup_{u\in \mathbb{R}}\bigg|\mathbb{P}\big(\mathcal{Q}_{\epsilon ,m}%
\mathbf{1}_{\mathcal{A}_{T}}\leq u\big)-\mathbb{P}\Big(\max_{1\leq j\leq N}%
\big|Z_{j}^{\diamond }+(Tb_{j})^{1/2}(\gamma _{j}-\bar{\gamma})\big|/\tilde{v}_j\leq u%
\Big)\bigg|  \notag \\
\lesssim & (\underline{b}T)^{-1/6}\mathrm{log}^{7/6}(NT)+(T^{2/q}/(%
\underline{b}T))^{1/3}\mathrm{log}(NT)+(\bar{b}+m^{-\beta })^{1/3}\mathrm{log%
}^{2/3}(N).
\end{align}

Proof of $(ii)$. Since $\max_{j,k}|c_{jk}|_{2}=O(1/(\underline bT)),$ for $B_{T}=c\underline b^{-1/2}$
some constant $c>0$ large enough, we have $Tb_{j}^{1/2}|c_{jk}|_{2}/B_{T}%
\leq \lambda _{0}\mathrm{log}_{(a_{0}\vee 2)}(2)$ and thus 
\begin{equation*}
\mathbb{E}\big(\mathrm{exp}\{Tb_{j}^{1/2}|c_{jk}^{\top }\eta _{k}|/B_{T}\}%
\big)\leq 2.
\end{equation*}%
Then $(ii)$ follows from Proposition 2.1 in \cite{MR3693963}. 
${\tiny \blacksquare }$